\documentclass[pdflatex,sn-mathphys-ay]{sn-jnl}

\usepackage{graphicx}%
\usepackage{multirow}%
\usepackage{amsmath,amssymb,amsfonts}%
\usepackage{amsthm}%
\usepackage{mathrsfs}%
\usepackage[title]{appendix}%
\usepackage{xcolor}%
\usepackage{textcomp}%
\usepackage{manyfoot}%
\usepackage{booktabs}%
\usepackage{algorithm}%
\usepackage{algorithmicx}%
\usepackage{algpseudocode}%
\usepackage{listings}%

\newcommand{\bx}{\mathbf{x}}
\newcommand{\by}{\mathbf{y}}

\newcommand{\bu}{\mathbf{u}}

\newcommand{\bC}{\mathbf{C}}

\newcommand{\R}{\mathbb R}
\newcommand{\E}{\mathbb E}

\usepackage{subcaption}
\usepackage[export]{adjustbox}  
\usepackage{soul}
\usepackage{comment}

\newif\ifshowchanges          
\showchangestrue              

\ifshowchanges
    {\egroup\endgroup}                    
\else
  \excludecomment{strikeblock}            
\fi

\newtheorem{assumption}{Assumption Set}[section]

\begin{document}

\title[Exploring the Generalizability of the 0.234 Acceptance Rate]{Exploring the generalizability of the optimal 0.234 acceptance rate in random-walk Metropolis and parallel tempering algorithms}

\author*[1]{\fnm{Aidan} \sur{Li}}\email{aidan.li@umontreal.ca\footnote{for correspondence only, work done while at the University of Toronto}}

\author[1]{\fnm{Liyan} \sur{Wang}}

\author[1]{\fnm{Tianye} \sur{Dou}}

\author[1]{\fnm{Jeffrey S.} \sur{Rosenthal}}

\affil*[1]{\orgdiv{Department of Statistical Sciences}, \orgname{University of Toronto}}

\abstract{
For random-walk Metropolis (RWM) and parallel tempering 
(PT) algorithms, an asymptotic acceptance rate of around 0.234 is known to be optimal in certain high-dimensional limits. 
However, its practical relevance is uncertain due to restrictive derivation conditions. 
We synthesise previous theoretical advances in extending the 0.234 acceptance rate to more general settings, and demonstrate its applicability with a comprehensive empirical simulation study on examples examining how acceptance rates affect Expected Squared Jumping Distance (ESJD).
Our experiments show the optimality of the 0.234 acceptance rate for RWM is surprisingly robust even in lower dimensions across various non-spherically symmetric proposal distributions, multimodal target distributions that may not have an i.i.d.\ product density, and curved Rosenbrock target distributions with nonlinear correlation structure.
Parallel tempering experiments also show that the idealized 0.234 spacing of inverse temperatures may be approximately optimal for low dimensions and non i.i.d.\ product target densities, and that constructing an inverse temperature ladder with spacings given by a swap acceptance of 0.234 is a viable strategy. 

}


\maketitle

\section{Introduction}\label{sec:intro}

Markov chain Monte Carlo (MCMC) methods, such as the random-walk Metropolis algorithm, are used to draw samples from complex high-dimensional target probability distributions. They enjoy strong theoretical asymptotic guarantees of accuracy, converging to the target distribution in stationarity \citep{brooks_handbook_2011, robert_monte_2004}. However, they may also be inefficient and 
prohibitively slow
to provide a good approximation of the target distribution they are sampling \citep{rosenthal_minorization_1995}. Therefore, the analysis of their running time is an important practical issue to consider. 

\cite{gelman_gilks_roberts_1997} proved an important result for independently and identically distributed (i.i.d.) product target densities: in the state space high-dimensional limit $d \to \infty$, the asymptotic acceptance rate of a Metropolis algorithm is central to its efficiency. Under their assumptions, an asymptotic acceptance rate of approximately 0.234 maximises the efficiency of a random-walk Metropolis (RWM) algorithm with a Gaussian proposal distribution starting at stationarity. 
However, their assumptions on the target and proposals are very restrictive. Many practical scenarios often involve finite and relatively small dimensions. Additionally, many probability distributions do not have densities of the i.i.d.\ product form, and for those that do, sampling a one-dimensional target is much simpler than using RWM. 
A burgeoning active research area is dedicated to showing that this figure still applies beyond these assumptions, and much of this focuses on weakening the i.i.d.\ product assumption.

Many papers have shown that the optimal average acceptance rate of 0.234 for RWM holds in greater generality than the assumptions used in \citet{gelman_gilks_roberts_1997}. 
For more general theoretical results, \citet{sherlock_optimal_2013} replaces the restrictive i.i.d. product target and Gaussian proposal setting of \citet{gelman_gilks_roberts_1997} with a set of ``shell'', relative variability, and eccentricity conditions on the target, eccentricity condition on the jump distance matrix, and shell condition on the proposal. This framework does not require a product form and is based on the geometric properties of the (log) target distribution in high dimensions.
\citet{jun_yang_2020} makes the 0.234 result applicable to targets with sparse dependency structures, such as those arising from Bayesian graphical models, using a set of sufficient conditions based on local dependencies and bounded asymptotic behaviour of derivatives. We expand on these results in Section \ref{sec:rwm_background}.
The 0.234 acceptance rate in optimal scaling has also been shown to apply to the spacing and swapping between inverse temperatures in the parallel tempering MCMC method \citep{atchade_towards_2011, roberts_rosenthal_2014_temperature}; we describe this in greater detail in Section \ref{sec:pt_background}.

That being said, there are some cases in which 0.234 is not optimal, such as in the case of discontinuous targets \citep{Neal2012Discontinuous} where the optimal acceptance rate is approximately 0.1353. Additionally, these results of optimality are still theoretical ideals, with required assumptions that could be further relaxed, and in the limit of $d \to \infty$, where the Markov chain converges to a diffusion process.

Other relaxations of the 0.234 assumptions include: non-Gaussian proposals \citep{neal_2011_nongaussian}, discrete hypercube target distributions \citep{Roberts1998hypercube}, Gibbs random fields \citep{breyer_metropolis_2000}, perturbations from product targets using pre-conditioned proposals \citep{beskos_optimal_2009},
independent target components with inhomogeneous scaling \citep{bedard_noniid_2007, bedard_general_2008, roberts_rosenthal_2001}, partial updates where not all components are updated at once \citep{neal_roberts_2006_optimal_scaling}, elliptical symmetric unimodal target densities \citep{sherlock_elliptical_2009}, and infinite-dimensional target distributions with non-trivial dependence structures \citep{mattingly_diffusion_2012}.

These results are of great interest to practitioners who would like to understand how to tune their MCMC algorithms to maximise their efficiency.
Previous simulation studies have been carried out on more general target distributions like products of independent, non-identical components \citep{bedard_2008_applications, roberts_rosenthal_2001}, and 
for the Markov modulated Poisson process \citep{sherlock_theoryandpractice_2010}, but there is still a lack of thorough experimentation and guidance for how applicable the 0.234 rule is on realistic target distributions that may have lower or higher dimensionality and/or multimodality.
Our research examines how necessary these assumptions are for the theoretical result to remain relevant: where can we relax some assumptions in lower dimensions and still have the optimal acceptance rate of approximately 0.234, and where can we not do this? And in the end, how applicable is the 0.234 figure for the practitioner, who may want to use MCMC for complicated target distributions beyond the necessary assumptions in finite dimensions?
We thoroughly dissect various aspects of the Metropolis algorithm and experiment with them, and in doing so, we show empirically where the theoretical ideal value can still align with more realistic scenarios.

We begin our paper with a description of the RWM and its optimal scaling framework, as well as formally stating sufficient theoretical conditions for the 0.234 rule that are most relevant to our work in Section \ref{sec:rwm_background}.
In Section \ref{sec:rwm-simulations}, we describe RWM experiments in low dimensions that show where an acceptance rate of 0.234 can be optimal in the sense that it maximises the Expected Squared Jumping Distance (ESJD) for various proposal and target densities which are not necessarily all i.i.d.\ product forms. 
Next, we describe the parallel tempering method and its own optimal scaling framework in Section \ref{sec:pt_background},
and show using experiments in Section \ref{sec:pt-simulations} that the 0.234 swap acceptance rate in parallel tempering may also be optimal in lower dimensions for the multivariate Gaussian target density and a Gaussian mixture target density which is not of an i.i.d.\ product form. 
Lastly, we discuss the implications of our findings in Section \ref{sec:discussion}.

\section{Random-Walk Metropolis Background}
\label{sec:rwm_background}

We first introduce the Metropolis algorithm and the Optimal Scaling framework. We then discuss measures to evaluate the efficiency of MCMC algorithms, such as the expected squared jumping distance, and explain how the asymptotic acceptance rate of a MCMC algorithm is tied to the algorithm's efficiency. Finally, we discuss some of the conditions for the 0.234 acceptance rate to be optimal that are most relevant to our work.

\subsection{Metropolis Algorithm}\label{subsec:background-mh-stepbystep}

A Metropolis algorithm \citep{metropolis_equation_1953} constructs a Markov chain $\bx^{(0)}, \bx^{(1)}, \dots, \bx^{(t)}$. If the chain is constructed to be irreducible and aperiodic, it is guaranteed to have the target distribution $\pi$ as its unique stationary distribution. This means that as the number of steps $t$ becomes large, the distribution of $\bx^{(t)}$ converges to $\pi$. Therefore, after an initial burn-in period, the subsequent states of this single, long chain can be used as a sample from the target distribution $\pi$.  
When taking each step in the Markov chain, we generate a new state $\by$ from a proposal distribution $Q(\cdot | \bx^{(t)})$ that is conditional on the current state $\bx^{(t)}$ and is symmetric, meaning $Q(\by | \bx^{(t)}) = Q(\bx^{(t)} | \by)$. We accept this new state with probability $\alpha(\bx, \by)$ given by $\alpha = \min\left\{1, \dfrac{\pi(\by)}{\pi(\bx)}\right\}$, where, by a common abuse of notation, we let $\pi(\cdot)$ also denote the probability density function of the target distribution $\pi$.
If the proposed value is accepted, we set the next state $\bx^{(t+1)} = \by$. Otherwise, we set $\bx^{(t+1)} = \bx^{(t)}$.

\subsection{Optimal Scaling Framework}\label{subsec:optimal-scaling-framework}

Optimal Scaling \citep{gelman_gilks_roberts_1997, roberts_rosenthal_2001} is one of the most successful frameworks for performing asymptotic analysis of high-dimensional MCMC methods, and provides mathematically-grounded guidance on how to best optimise MCMC performance by tuning the ``scaling'' parameter(s) of the proposal distribution. 
Typically, a multivariate Gaussian distribution centred at the current state is used as the proposal distribution.
The scaling of the Gaussian proposal is then given by a scaling factor $\sigma$, and this is often used as a variance $\sigma^2$ that is applied to the $d \times d$ identity matrix $I_d$. So, $\epsilon \sim \mathcal{N}(0, \sigma^2 I_d)$.
If the proposal's variance is too small, many proposed steps are accepted by the algorithm, but each step does not explore the state space much. Vice versa, if the proposal's variance is too large, too few proposed steps are accepted, leading to slow exploration of the state space since the algorithm stays at a state for too long. 
Therefore, the proposal's variance is crucial to the performance of the algorithm; the optimal scaling should balance the exploration of new areas in the state space with exploiting high-density areas of the target distribution.

We next discuss how to evaluate the efficiency of an MCMC algorithm for determining optimal scaling.
A very popular measure of efficiency both theoretically (e.g. \citet{sherlock_optimal_2013, roberts_rosenthal_2014_temperature,  jun_yang_2020}) and in practice (e.g. \citet{ pasarica_adaptively_2010,sherlock_efficiency_2015}) is the Euclidean \textit{Expected Squared Jumping Distance} (ESJD) metric which we use for our experiments.
The expected squared jumping distance measures how far, in expectation, the MCMC chain moves in a single iteration. For the standard random-walk Metropolis algorithm, we define this as 
\begin{equation}\label{eq:ESJD}
    \E\left[\left\|\bx^{(t+1)} - \bx^{(t)}\right\|^2\right] \approx \frac{1}{n-1}\sum_{i=1}^{n-1} \left\|\bx^{(t+1)} - \bx^{(t)}\right\|^2
\end{equation}
where $n$ is the total number of iterations of the algorithm.
Maximising the ESJD aligns with minimizing the first-order auto-correlation of the Markov chain and subsequently maximises efficiency if each higher order auto-correlation is a monotonically increasing function of the first-order auto-correlation \citep{pasarica_adaptively_2010, jun_yang_2020}. It is worth mentioning that ESJD primarily measures local convergence performance. Because heavy-tailed targets disrupt the monotonic relationship between first-order and higher-order autocorrelations, achieving global convergence for heavy-tailed targets requires maximising an adapted version of ESJD \citep{Kamatani_2020}. However, we do not consider heavy-tailed targets in this work, so the working definition of ESJD is sufficient.

There are other notions of efficiency of a Markov chain, but in the high-dimensional limit $d \to \infty$, if the moment conditions in Assumption \ref{assump:rgr97} hold, the chain converges to a diffusion process, and all efficiency measures are effectively equivalent \citep{gelman_gilks_roberts_1997, roberts_rosenthal_2001}.
A key result of these referenced papers is that, given a Metropolis algorithm with a Gaussian proposal distribution $\mathcal{N}(\bx^{(t)}, \frac{\ell^2}{d} I_d)$ where $\ell>0$ is a fixed scaling constant and $I_d$ is the identity matrix, maximising the diffusion's speed measure $h(\ell)$, which is a function of the scaling constant $\ell$, yields the most efficient asymptotic diffusion. 
Furthermore, the speed measure has a clear relation to a much simpler quantity to tune for: the \textit{asymptotic acceptance rate} of the proposed new states (moves) of the algorithm, defined as
$$a = \lim_{n\rightarrow \infty}\frac{\text{\# accepted moves}}{n}.$$ 
Both the speed measure $h(\ell)$ and asymptotic acceptance rate $a(\ell)$ are functions of the scaling constant $\ell$.
Given the restrictive conditions mentioned in Section \ref{subsec: background-assumptions}, the scaling constant $\ell$ that maximises the speed measure $h(\ell)$ corresponds to an asymptotic acceptance rate $a(\ell)$ of approximately 0.234. Hence, an asymptotic acceptance rate of approximately 0.234 should necessarily optimise a measure of efficiency such as the ESJD of the algorithm.

\subsection{Theoretical Assumptions for Optimality of a 0.234 Acceptance Rate} \label{subsec: background-assumptions}

Certain work has shown that under restrictive assumptions, a 0.234 acceptance rate is asymptotically optimal for random-walk Metropolis algorithms in certain high-dimensional limits. We briefly list the theoretical assumptions of these works that are most relevant to our simulations.
We first describe the seminal result of \citet{gelman_gilks_roberts_1997} alluded to in the introduction in full. 

\begin{assumption}
\label{assump:rgr97} 
(\citet{gelman_gilks_roberts_1997})
Let $\bx$ be a random variable from the target density $\pi$. The target distribution has an i.i.d. product density $\pi(\bx) = \prod_{i=1}^d f(x_i)$.
The component density $f: \mathbb{R} \to \mathbb{R}^+$ satisfies the following regularity conditions:
\begin{enumerate}
    \item $f$ is positive and twice continuously differentiable on its support.
    \item The derivative of its logarithm, $f'/f$, is Lipschitz continuous.
    \item The following moment conditions hold:
    \begin{equation} \label{eq:rgr97moments}
       \mathbb{E}_f\left[\left(\frac{f'(X)}{f(X)}\right)^8\right] < \infty \quad \text{and} \quad \mathbb{E}_f\left[\left(\frac{f''(X)}{f(X)}\right)^4\right] < \infty
    \end{equation}
\end{enumerate}
    
\end{assumption}

Under the conditions in Assumption \ref{assump:rgr97}, \citet{gelman_gilks_roberts_1997} showed that for an RWM algorithm with a Gaussian proposal distribution starting in stationarity, when the whole process is sped-up by a factor of $d$, the first component of the process converges to a limiting Langevin diffusion. Maximizing the speed of this limiting process, in turn, corresponds to tuning the algorithm to an asymptotic acceptance rate of approximately 0.234.

The \citet{gelman_gilks_roberts_1997} result works because an i.i.d. product target looks locally the same no matter which coordinate you observe. Then, the analysis of \citet{sherlock_optimal_2013} shows that a similar homogeneity emerges for a much wider class of targets once the dimension is large.

\begin{assumption}
\label{assump:sherlock2013}
(\citet{sherlock_optimal_2013})
Let $\bx$ be a random variable from the target density $\pi$, $M(\bx) = ||\nabla\log\pi(\bx)||$ denote the norm of the gradient of the log target, $H(\bx) = -\nabla^2\log\pi(\bx)$ denote the negative Hessian matrix of the log target, and $\Delta(\bx, \bu) = H(\bx + \bu) - H (\bx)$ denote the matrix of perturbations of the negative Hessian. The primary conditions are:
\begin{enumerate}
    \item \textbf{Target weak shell.} There exist sequences $\{M^{(d)}\}$ and $\{H^{(d)}\}$, as $d \to \infty$:
    \begin{equation} \label{eq:sherlock_target_shell}
        \frac{M^{(d)}(\bx^{(d)})}{M^{(d)}} \xrightarrow{p} 1 \quad \text{and} \quad \frac{H^{(d)}(\bx^{(d)})}{H^{(d)}} \xrightarrow{p} 1 
    \end{equation}
    \item \textbf{Hessian relative variability.} The Hessian is locally stable, meaning it does not vary significantly on the scale of a typical proposal jump. Formally, for any fixed $\mu > 0$ and $\delta > 0$, and for a standard Gaussian random vector $\mathbf{z}^{(d)} \sim \mathcal{N}(\mathbf{0}, \mathbf{I}_d)$ independent of the target state $\bx^{(d)}$, we require that as $d \to \infty$,
\begin{equation}\label{eq:sherlock_hessian}
    \mathbb{P}_{\mathbf{x,z}}\left( \frac{1}{H^{(d)}} \left| \left(\mathbf{z}^{(d)}\right)^T \Delta\left(\bx^{(d)}, t \mu \frac{M^{(d)}}{H^{(d)}} \mathbf{z}^{(d)}\right) \mathbf{z}^{(d)} \right| < \delta \quad \text{for all } t \in [0,1] \right) \to 1.
\end{equation}

    \item \textbf{Target eccentricity.} The curvature is not disproportionately concentrated in any single direction. Denoting the eigenvalues of $\mathbf{H}(\bx)$ by $\beta_i(\bx)$:
    \begin{equation}\label{sherlock_target_eccentricity}
        \frac{\max_i |\beta_i(\bx^{(d)})|}{\sum_{j=1}^d \beta_j(\bx^{(d)})} \xrightarrow{p} 0.
    \end{equation}

    \item \textbf{Proposal shell.} There exists a sequence $\{k_u^{(d)}\}$ such that the sequence of $d$-dimensional spherically symmetric proposals $\{Q^{(d)}\}$ satisfies this as $d \to \infty$:
    \begin{equation}\label{sherlock_proposal_shell}
        \frac{||Q^{(d)}||}{k_u^{(d)}} \xrightarrow{m.s.} 1.
    \end{equation}
    \item \textbf{Jump distance matrix eccentricity.} If the ESJD is defined with respect to a positive definite symmetric matrix $\mathbf{T}^{(d)}$ as $\mathbb{E}[(\mathbf{Y}^{(d)})^T \mathbf{T}^{(d)} \mathbf{Y}^{(d)}]$, then the eigenvalues of this \textit{jump distance matrix} must be similarly well-behaved. Let $\tau_i(d)$ be the eigenvalues of $\mathbf{T}^{(d)}$. When $d \to \infty$: 
    \begin{equation}\label{sherlock_jumping}
        \frac{\max_{i} \tau_i(d)}{\sum_{j=1}^d \tau_j(d)} \to 0.
    \end{equation}
\end{enumerate}
\end{assumption}

Put together, these geometric regularities recover the same result of the optimality of the 0.234 acceptance rate as \citet{gelman_gilks_roberts_1997} but with weaker assumptions on the target and proposal.

\section{Random-Walk Metropolis Simulations} \label{sec:rwm-simulations}

In this section, we explore the applicability of the 0.234 acceptance rate figure for RWM algorithms using several practical RWM examples in lower dimensions. We begin with simple examples where the target densities meet all the key conditions in \citet{gelman_gilks_roberts_1997} (provided as Assumption \ref{assump:rgr97}) required for the 0.234 acceptance rate to be optimal. Next, we investigate examples that do not use Gaussian proposal distributions, and then we explore other target densities that may lack these assumptions, such as smoothness or the i.i.d.\ product form. We estimate the Expected Squared Jumping Distance (ESJD) of the algorithm as a metric of the algorithm's efficiency as mentioned in Section \ref{subsec:optimal-scaling-framework}, and aim to see under what conditions an acceptance rate of approximately 0.234 maximises the ESJD for various target and proposal distributions in lower dimensions.

Our experimental procedure is as follows: we run RWM simulations using (except for Section \ref{subsec:non-gaussian-proposal}) a Gaussian proposal distribution
    \begin{equation}
Q(\by \mid \bx) = \mathcal{N}(\by \mid \bx, \sigma^2 I_d)
\label{eq:gaussian_proposal}
\end{equation}
where $\sigma$ is a proposal scaling factor, or $\sigma^2$ can be thought of as a variance, and $I_d$ is the $d$-dimensional identity matrix. We run 40 different RWM simulations where each simulation has a different scaling factor, and each simulation has a burn-in of at least 1,000 steps before taking at least 200,000 steps. It is important to note that functions of acceptance rates are somewhat flat around the optimal values and that the values and figures reported are still prone to random error inherent in a Monte Carlo simulation. Therefore, we repeat this procedure over at least 20 computer seeds and average the results to reduce random error. This procedure is applied in each subsection and specific simulation details may be mentioned in a subsection if the experiment used more than these minimum values.

\subsection{Simpler Examples: Gamma and Beta i.i.d.\ Targets}

We begin with simpler examples demonstrating that the 0.234 acceptance rate is still roughly optimal for lower dimensions with any i.i.d.\ product target distribution and a Gaussian proposal. 
As mentioned in Section \ref{sec:rwm_background}, \cite{gelman_gilks_roberts_1997} proved the 0.234 optimal acceptance rate for i.i.d.\ product target distributions with a standard Gaussian proposal distribution in the infinite-dimensional limit, and \cite{roberts_rosenthal_2001} and \cite{bedard_2008_applications} extended this with MCMC simulations showing that with a i.i.d.\ product target distribution and Gaussian proposal, the 0.234 acceptance rate seems approximately optimal for dimensions as low as 10. Our experiment provides further evidence for this being the case, and we extend this with simulations for $d=2,5$.
We present two i.i.d.\ target densities $\pi_1, \pi_2$ with different single-dimension component densities $f_1, f_2$:
\begin{align}
    \pi_1(\bx) &= \prod_{i=1}^n f_1(x_i), \quad f_1(x) = \text{Gamma}(x \mid 3, 2), \\
    \pi_2(\bx) &= \prod_{i=1}^n f_2(x_i), \quad f_2(x) = \text{Beta}(x \mid 3, 2)
    \label{eq:targets_gamma_beta}
\end{align}

The first example sets $f_1$ with shape parameter 3 and scale parameter 2, and the second example sets $f_2$ with shape parameters 3 and 2. 
We use the multivariate Gaussian proposal distribution defined in \eqref{eq:gaussian_proposal} in both examples. We run this experiment in dimensions $d \in \{2, 5, 10, 30, 50, 100\}$.

\begin{table}[htbp]
\centering
\caption{Empirical optimal RWM acceptance rates (i.e., maximizing ESJD) for i.i.d.\ product targets with a Gaussian proposal.}
\label{tab:rwm_iid_targets}
\begin{tabular}{lcccccc}
\toprule
\textbf{Target Density} & \(d=2\) & \(d=5\) & \(d=10\) & \(d=30\) & \(d=50\) & \(d=100\) \\
\midrule
$\pi_1$: i.i.d.\ Gamma(3, 2) & 0.3036 & 0.2378 & 0.2199 & 0.2101 & 0.2141 & 0.2140 \\
$\pi_2$: i.i.d.\ Beta(3, 2)  & 0.3903 & 0.2937 & 0.2561 & 0.2319 & 0.2248 & 0.2159 \\
\bottomrule
\end{tabular}
\end{table}

\begin{figure}[htbp]
\centering

\begin{subfigure}[t]{0.32\textwidth}
    \includegraphics[width=\linewidth]{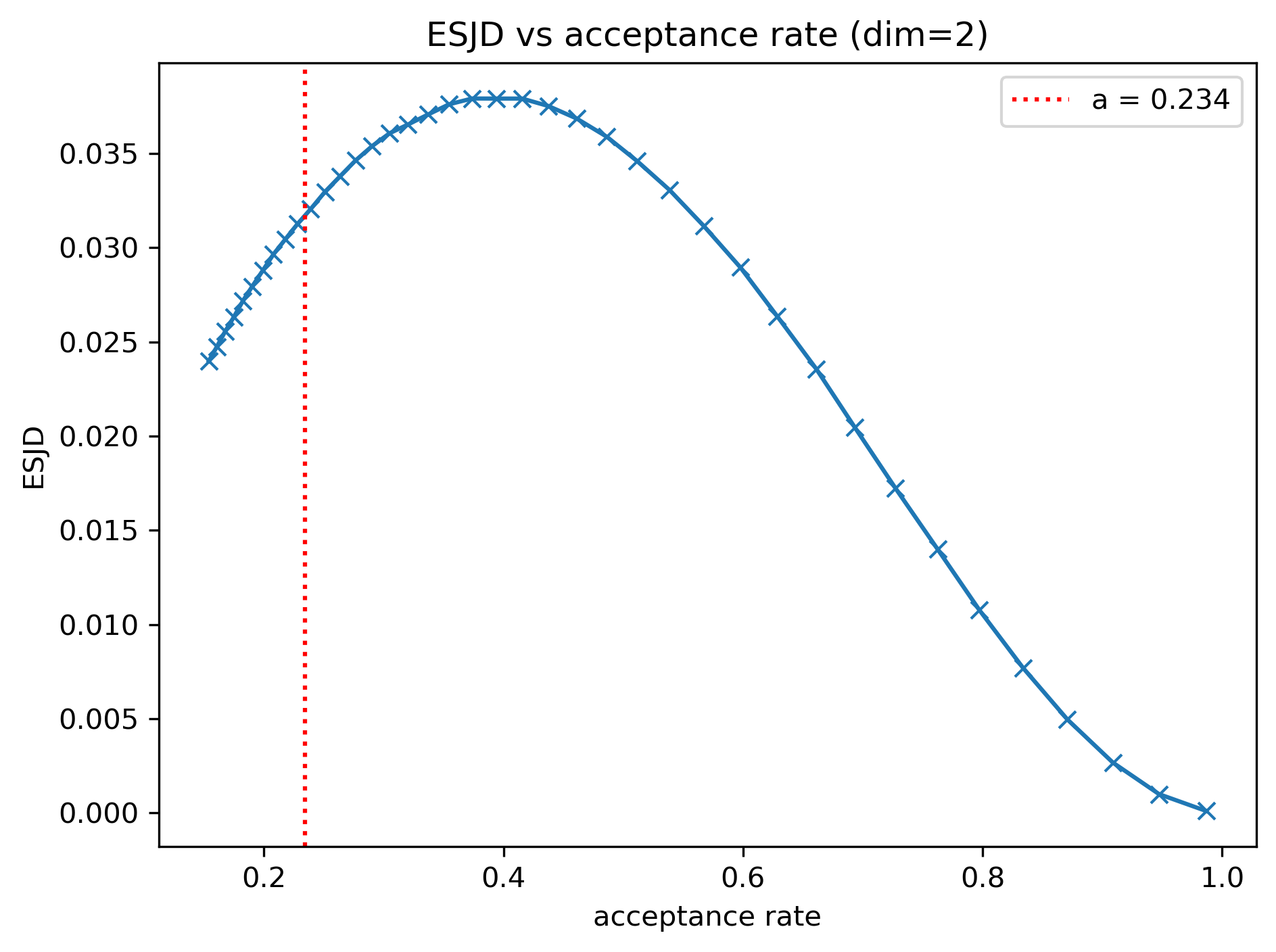}
\end{subfigure}
\hfill
\begin{subfigure}[t]{0.32\textwidth}
    \includegraphics[width=\linewidth]{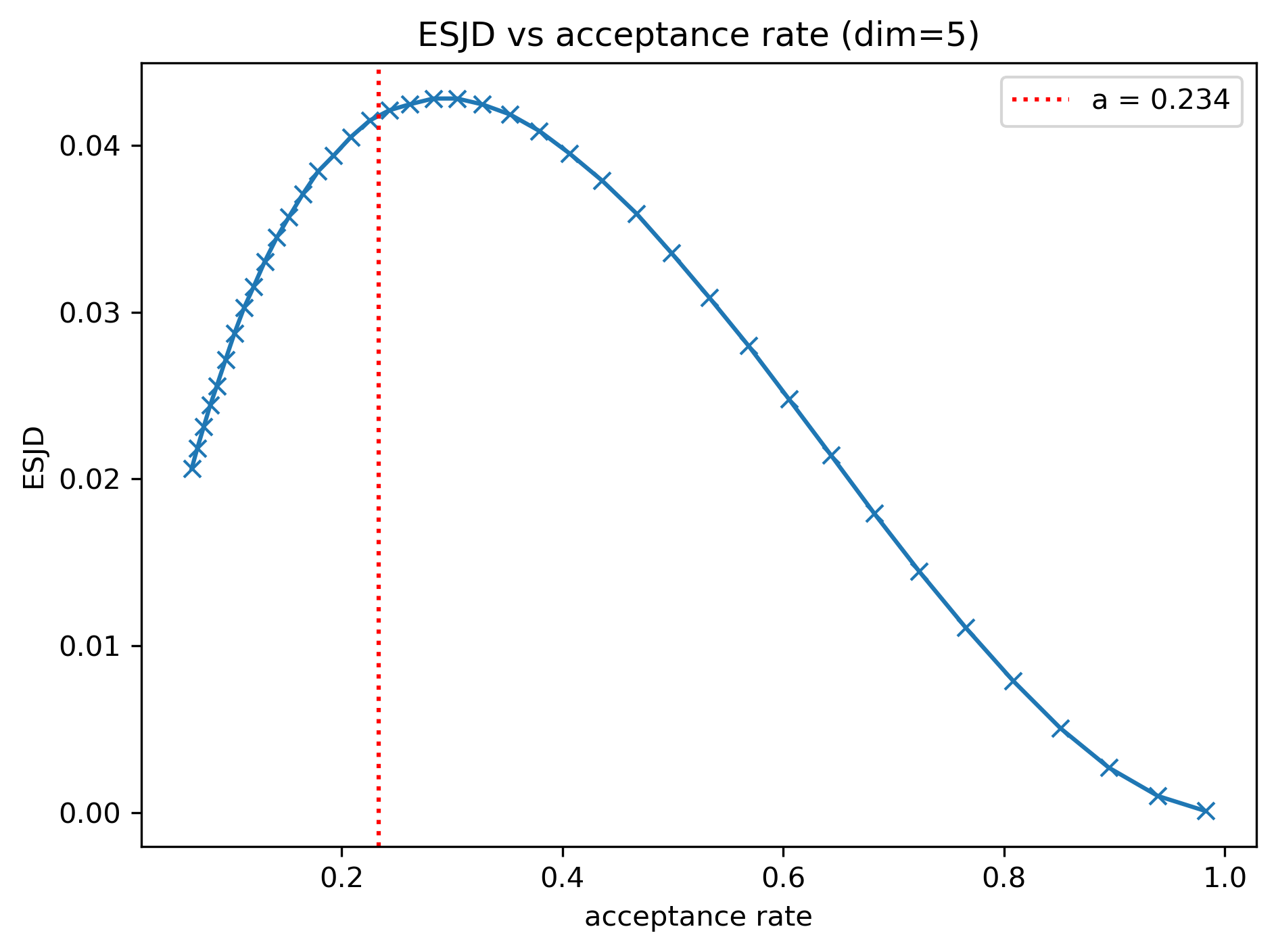}
\end{subfigure}
\hfill
\begin{subfigure}[t]{0.32\textwidth}
    \includegraphics[width=\linewidth]{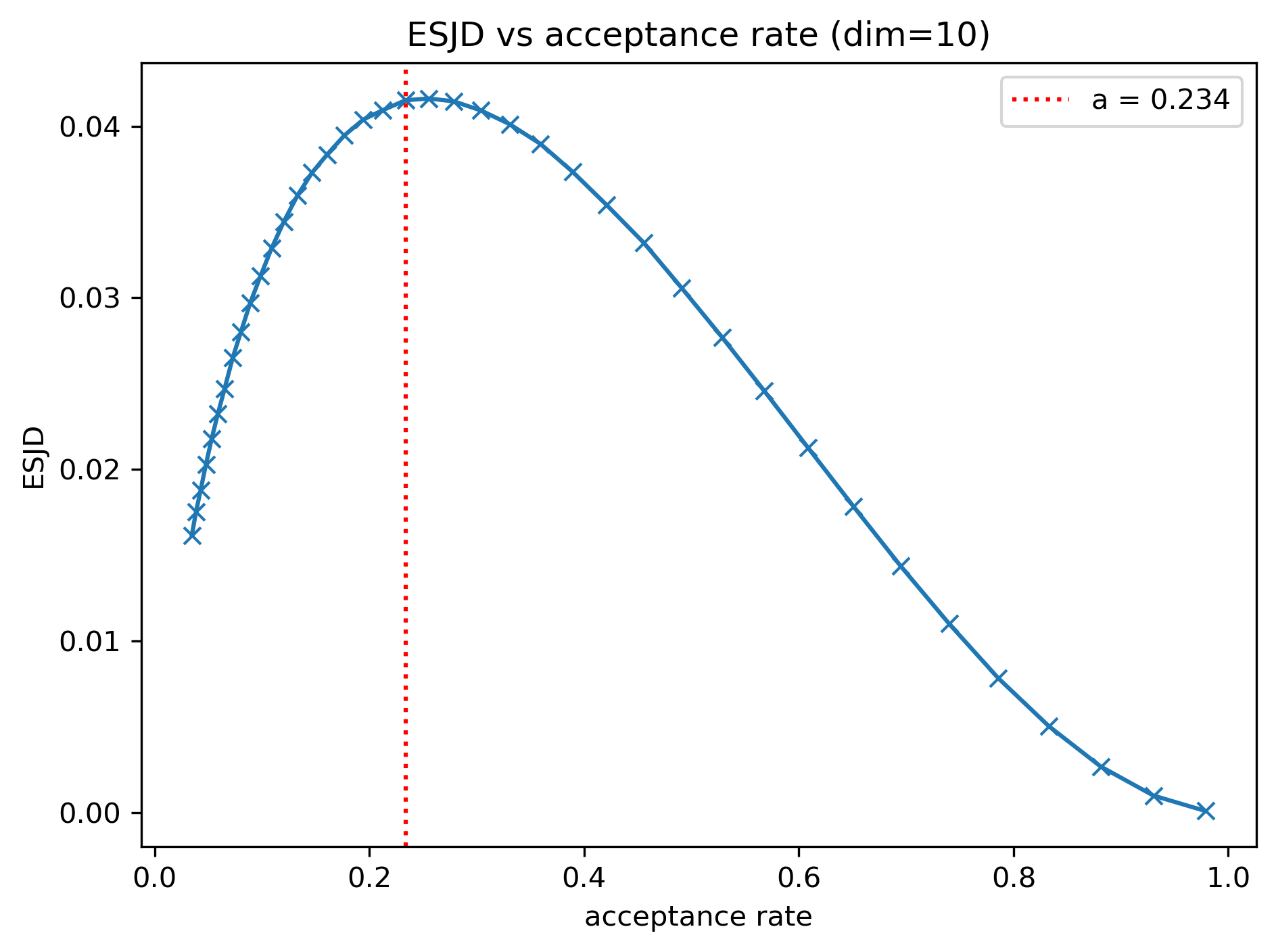}
\end{subfigure}

\vspace{0.5em}

\begin{subfigure}[t]{0.32\textwidth}
    \includegraphics[width=\linewidth]{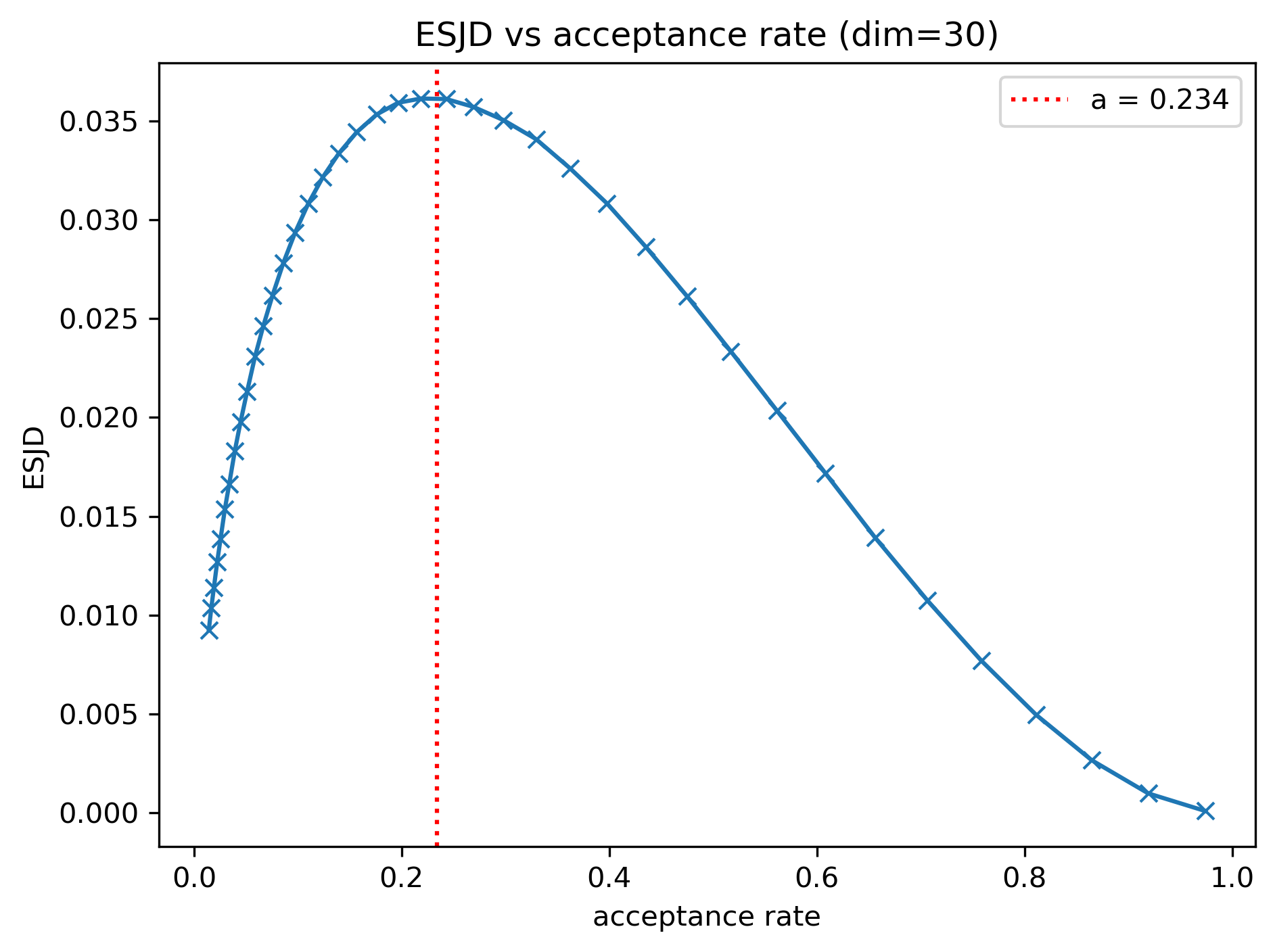}
\end{subfigure}
\hfill
\begin{subfigure}[t]{0.32\textwidth}
    \includegraphics[width=\linewidth]{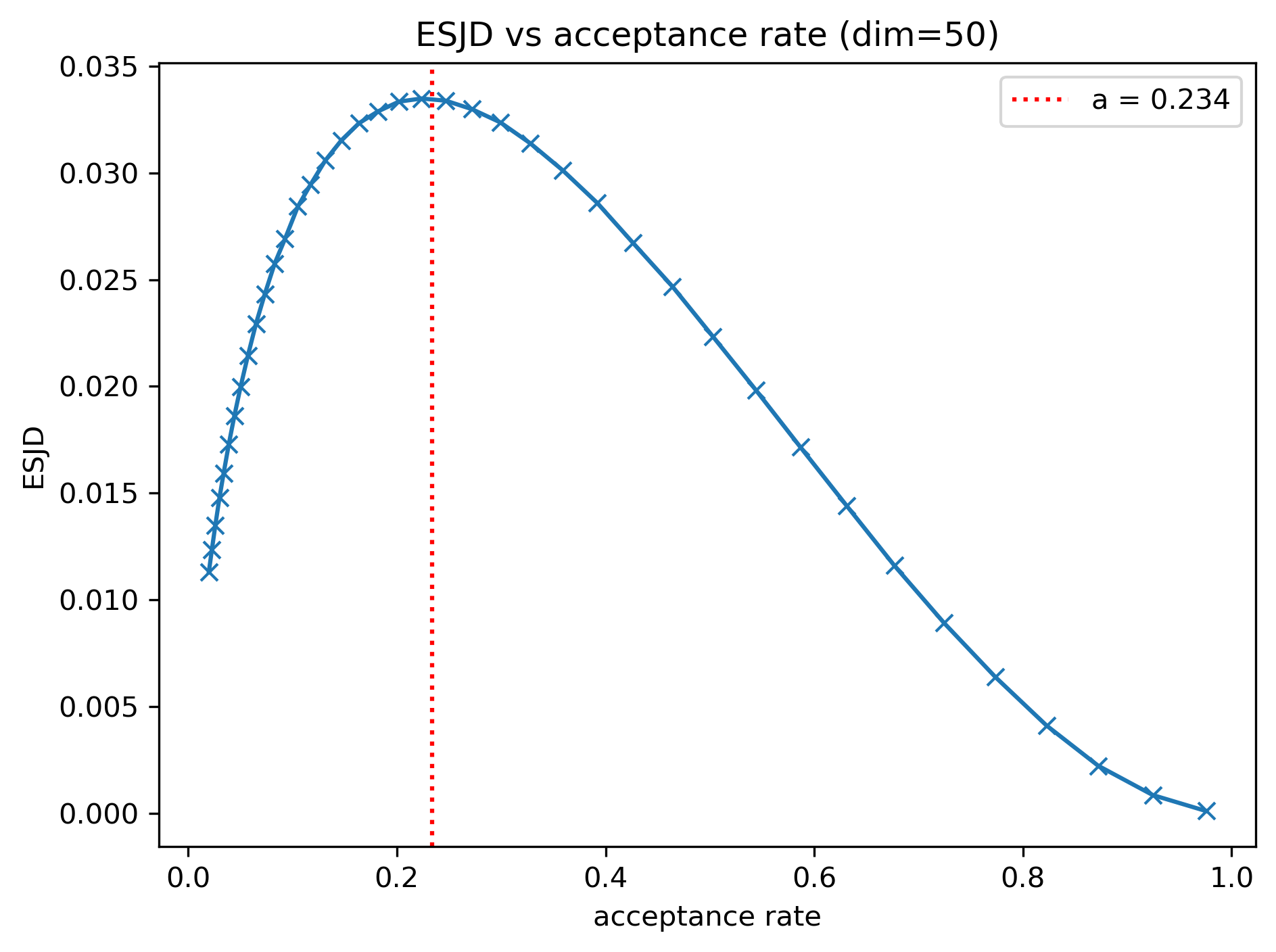}
\end{subfigure}
\hfill
\begin{subfigure}[t]{0.32\textwidth}
    \includegraphics[width=\linewidth]{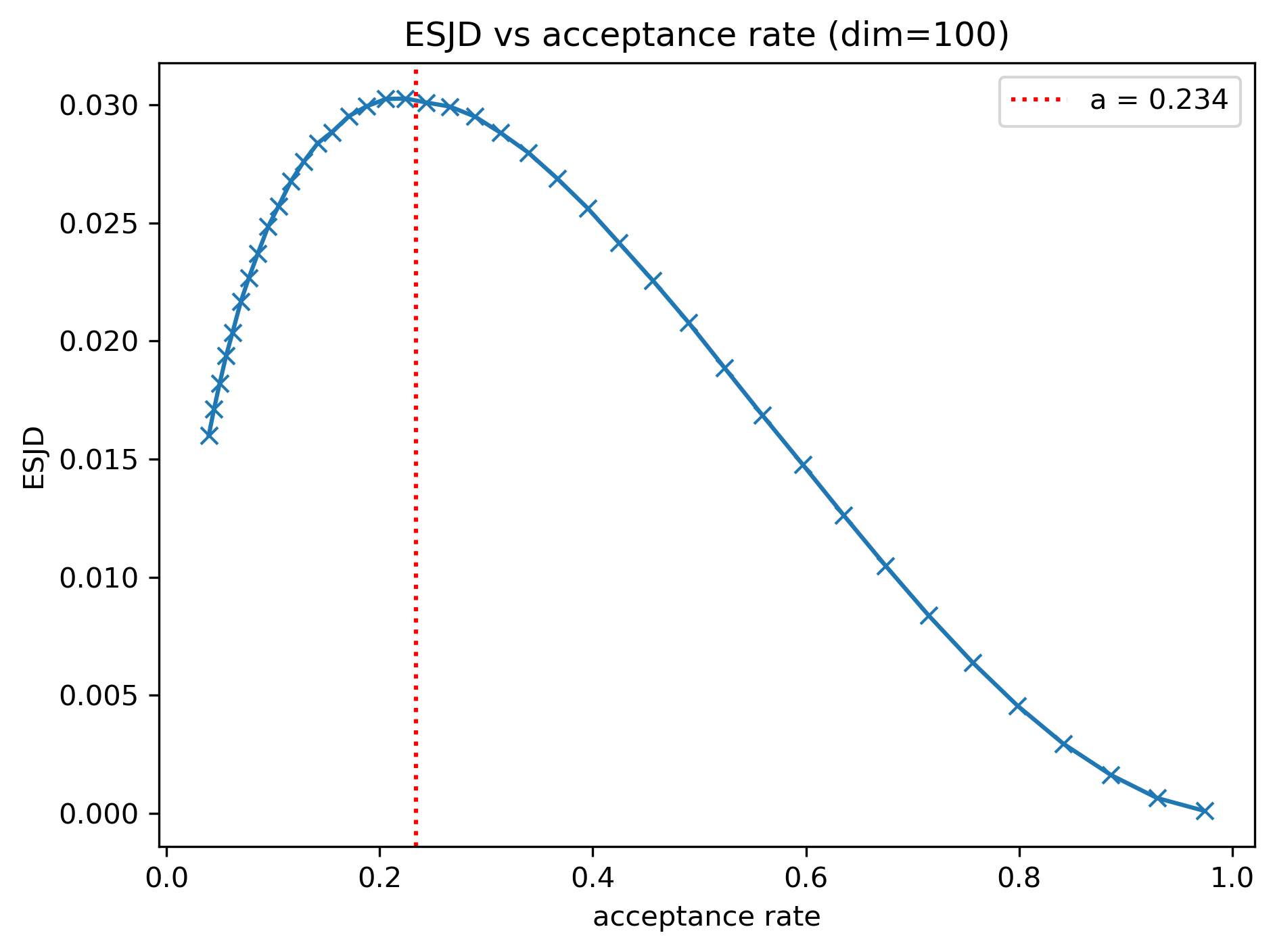}
\end{subfigure}

\caption{ESJD vs.\ acceptance rate for the i.i.d.\ Beta(3, 2) target distribution $\pi_2$ under RWM with a Gaussian proposal in dimensions $d \in \{2, 5, 10, 30, 50, 100\}$ from top-left to bottom-right. Red dotted line indicates an acceptance rate of 0.234.}
\label{fig:esjd_beta_iid}
\end{figure}

The resulting ESJD-maximising acceptance rates are summarised in Table \ref{tab:rwm_iid_targets} and the i.i.d. Beta target has its trends of ESJD versus acceptance rate with different dimensions displayed in Figure \ref{fig:esjd_beta_iid}.
For the Gamma target, the optimal acceptance rate drops from 0.304 at $d=2$ to around 0.21 once $d \ge 10$; for the Beta target they fall from 0.390 to around 0.22 over the same range. Figure \ref{fig:esjd_beta_iid} confirms that the ESJD curves have a moving plateau towards a centre slightly below the classical 0.234 value as the dimension grows. Our simulation provides further evidence that the optimal acceptance rate for maximizing ESJD is approximately 0.234 even in low dimensions.

Since most ESJD versus acceptance rate plots look fairly similar and have a similar progression as dimension changes, the full set of curves for ESJD versus acceptance rate plots for other subsections are in Appendix \ref{app:plots}, including the results for the i.i.d. Gamma in Figure \ref{fig:esjd_gamma_iid}. 
As an aside, in all dimensions, there is a relatively broad plateau where the
acceptance rate is approximately optimal; thus, the coarse 40-point grid pinpoints the optimum to approximately $\pm 0.01$. 

\subsection{Non-Gaussian Proposal Densities}\label{subsec:non-gaussian-proposal}

Next, we investigate the efficacy of different proposal distributions for the Metropolis algorithm. One of the conditions assumed by the 0.234 theorem of \citet{gelman_gilks_roberts_1997} is the use of a multivariate Gaussian proposal distribution $Q(\by | \bx) = \mathcal{N}(\by | \bx, \sigma^2I_d)$. 
The additional results in \cite{roberts_rosenthal_2001} depend on the assumptions of light-tailed proposals. In contrast, \cite{Jarner_Roberts_2002} extend this analysis by examining the behaviour of the algorithms with heavy-tailed proposals. 
They find that, unlike the diffusion behaviour seen with light-tailed proposals, heavy-tailed proposals lead to a different dynamic characterized by abrupt movements followed by periods of inactivity.
Furthermore, \cite{neal_2011_nongaussian} investigated optimal scaling results for heavy-tailed proposal distributions such as the Cauchy distribution, and spherical proposal distributions of fixed radius. 
In their RWM simulation study with three different continuous i.i.d.\ product target densities, both the Gaussian and Cauchy proposals had an estimated asymptotically optimal acceptance rate of approximately 0.234 in $d=100$, and a spherical proposal with a fixed radius of 2.38 achieved an optimal acceptance rate of about 0.234 in $d=2$ and $d \geq 10$. Finally, \citet{sherlock_optimal_2013} proved that spherically symmetric proposal distributions following a geometric ``shell'' condition from Assumption \ref{assump:sherlock2013} \eqref{sherlock_proposal_shell} should yield an optimal acceptance rate of approximately 0.234 when sampling from a suitable target.

We experiment with two different proposal distributions in $d\in \{2, 5, 10, 20, 50, 100\}$ and set the target density $\pi_3$ to a standard multivariate Gaussian:

\begin{equation}
\begin{aligned}
&\pi_3(\bx) = \mathcal{N}(\bx \mid \mathbf{0}, I_d) \\[-0.3em]
&Q_L(\by \mid \bx) \propto \prod_{i=1}^d \exp\left(-\frac{|y_i - x_i|}{\sigma}\right), \quad 
Q_U(\by \mid \bx) = \prod_{i=1}^d \frac{\mathbb{I}\left[ |y_i - x_i| \le \tfrac{b}{2} \right]}{b} 
\end{aligned}
\label{eq:gaussian_target_laplace_uniform_proposals}
\end{equation}

The first proposal distribution we consider is the multivariate Laplace distribution, which may also be referred to as the double exponential distribution. We set the location parameter $\mu = \bx$  and scale parameter $\sigma$ to be chosen to influence the acceptance rate of the algorithm. We also consider an example with the uniform proposal distribution within the closed interval $\left[x_i - \frac{b}{2}, x_i + \frac{b}{2}\right]$ for each component $i$, where $b > 0$ is the interval length.

\begin{table}[htbp]
\centering
\caption{Empirical optimal RWM acceptance rates across dimensions for different proposal distributions targeting a standard multivariate Gaussian.}
\label{tab:rwm_proposals}
\begin{tabular}{lcccccc}
\toprule
\textbf{Proposal} & \(d=2\) & \(d=5\) & \(d=10\) & \(d=20\) & \(d=50\) & \(d=100\) \\
\midrule
$Q_L$: Laplace           & 0.3780 & 0.3036 & 0.2841 & 0.2570 & 0.2429 & 0.2377 \\
$Q_U$: Uniform           & 0.3194 & 0.2516 & 0.2391 & 0.2368 & 0.2365 & 0.2316 \\
\bottomrule
\end{tabular}
\end{table}

As observed in Table \ref{tab:rwm_proposals} and the corresponding Figures \ref{fig:esjd_prop_laplace}, \ref{fig:esjd_prop_uniform} in Appendix \ref{app:plots}, the optimal acceptance rates for both the Laplace and Uniform proposals exhibit a clear trend. In low dimensions, the optimal rate is significantly higher than 0.234, but converges towards 0.234 from above as the dimension increases. For the Uniform proposal, the optimal rate is already approximately 0.239 at $d=10$, and for both proposals, the rates are close to the high-dimensional limit for $d \ge 20$.

These empirical findings provide strong validation for the generalized theory of optimal scaling. While the foundational result of \citet{gelman_gilks_roberts_1997} was derived for Gaussian proposals, the more general criteria from \citet{sherlock_optimal_2013}, summarized in Assumption \ref{assump:sherlock2013}, extend this theory. The key requirement on the proposal is the shell condition \eqref{sherlock_proposal_shell} that requires the norm of the proposal vector to concentrate around a constant in high dimensions. 
Interestingly, although both the multivariate Laplace and Uniform distributions used here are not spherically symmetric as required by \eqref{sherlock_proposal_shell}, they still satisfy the spirit of the shell condition mentioned in the previous sentence. If these more relaxed conditions still satisfy the necessary requirements for the theory to work, the theory predicts that the 0.234 acceptance rate should indeed be asymptotically optimal when using these proposals on a suitable target. Our results confirm this prediction, and characterize the finite-dimensional behaviour, demonstrating that convergence to the 0.234 optimum occurs from above for these proposals. The lack of spherical symmetry in our proposals also suggests that weaker conditions on the proposal than spherical symmetry may be sufficient for the 0.234 rate to be optimal.   

\subsection{Multimodal Target Densities}\label{subsec:multimodal}
\begin{figure}[ht]
    \centering
    \begin{subfigure}{0.4\textwidth}
        \includegraphics[width=\linewidth, valign=t]{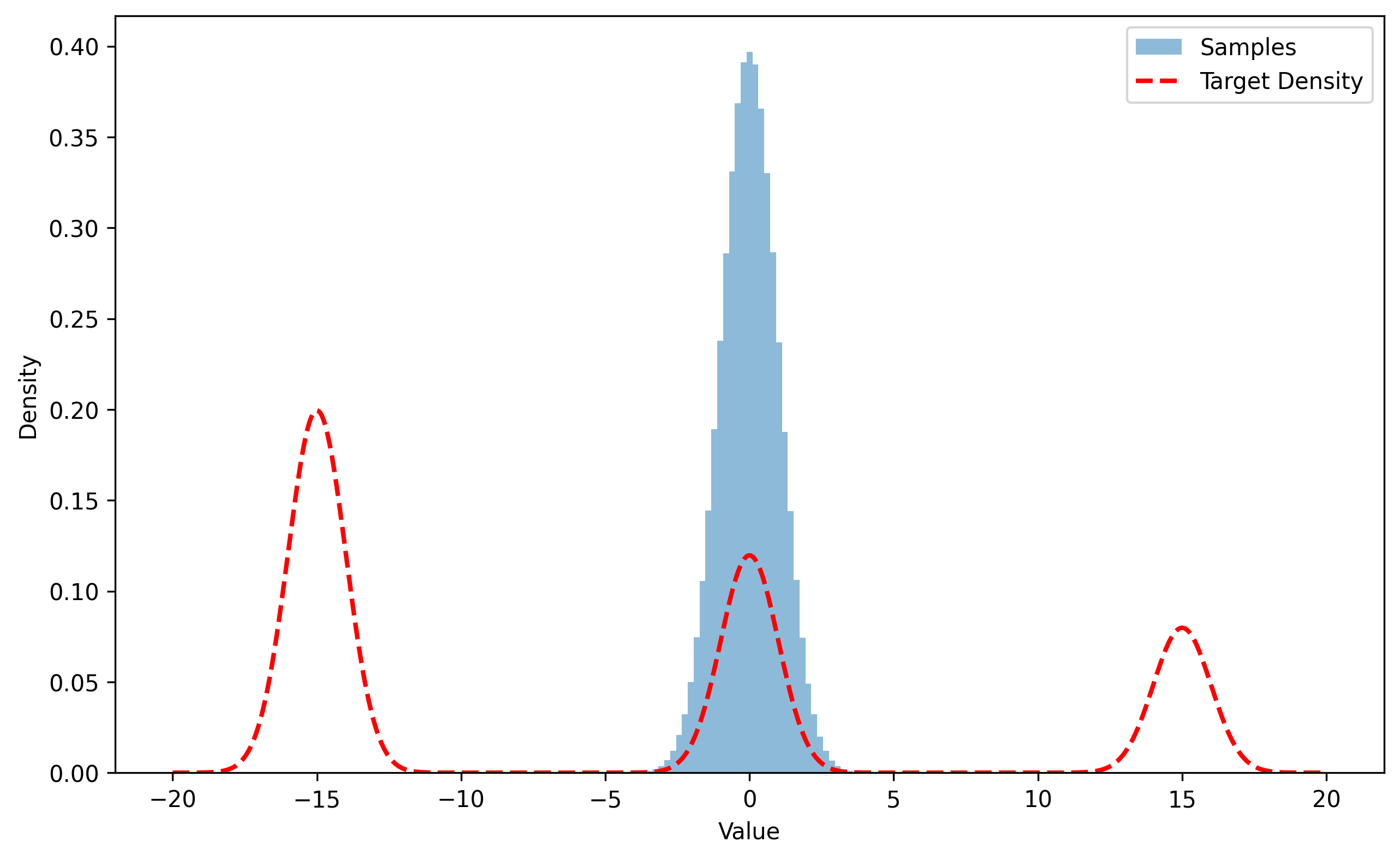}
    \end{subfigure}
    \begin{subfigure}{0.4\textwidth}
        \includegraphics[width=\linewidth, valign=t]{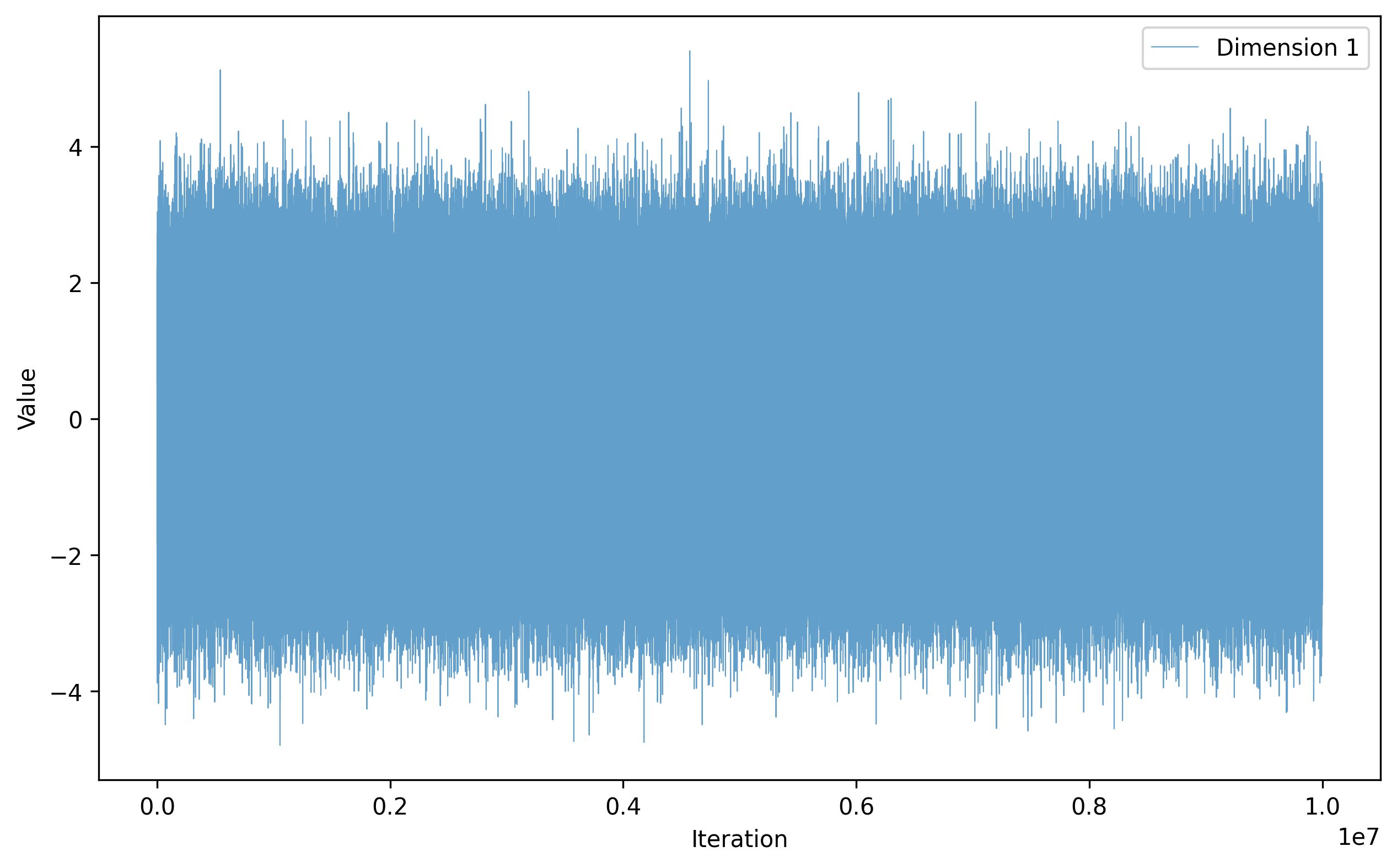}
    \end{subfigure}
    \caption{Histogram and traceplot of 10 million samples drawn from a multimodal single-dimensional target density. The true density values are traced with the dotted lines.}
    \label{fig:10m no escape}
\end{figure}

In this subsection, we investigate target distributions with multimodal densities. Many real-world probability and data distributions are multimodal, and standard RWM algorithms may struggle to explore all the modes of a multimodal distribution in a reasonable amount of time particularly when the modes are far apart. Additionally, multimodal densities may not necessarily fall under the i.i.d.\ product umbrella of target densities, and even if they do, it is debatable how relevant the 0.234 rule is in these cases. For example, a RWM algorithm may achieve a 0.234 acceptance rate in its run over many iterations and yet completely fail to escape a mode to explore other modes. We illustrate this with Figure \ref{fig:10m no escape}, where we attempt to draw 10 million samples from a multimodal distribution with three modes but fail to escape the central mode.

When we have widely-separated modes, a RWM with the Gaussian proposal distribution struggles to reach the other modes in finite time regardless of the proposal scaling and the dimension of the target distribution. 
Therefore, for these multimodal target density experiments, we deliberately set the modes to be close enough to each other so that a RWM still mixes between the modes reasonably often.
We list out the various target densities before detailing the experimental results.

\subsubsection{The ``Rough Carpet'' Target Density}\label{sec:rough-carpet}

\begin{figure}[h]
    \centering
    \includegraphics[width=0.6\textwidth]{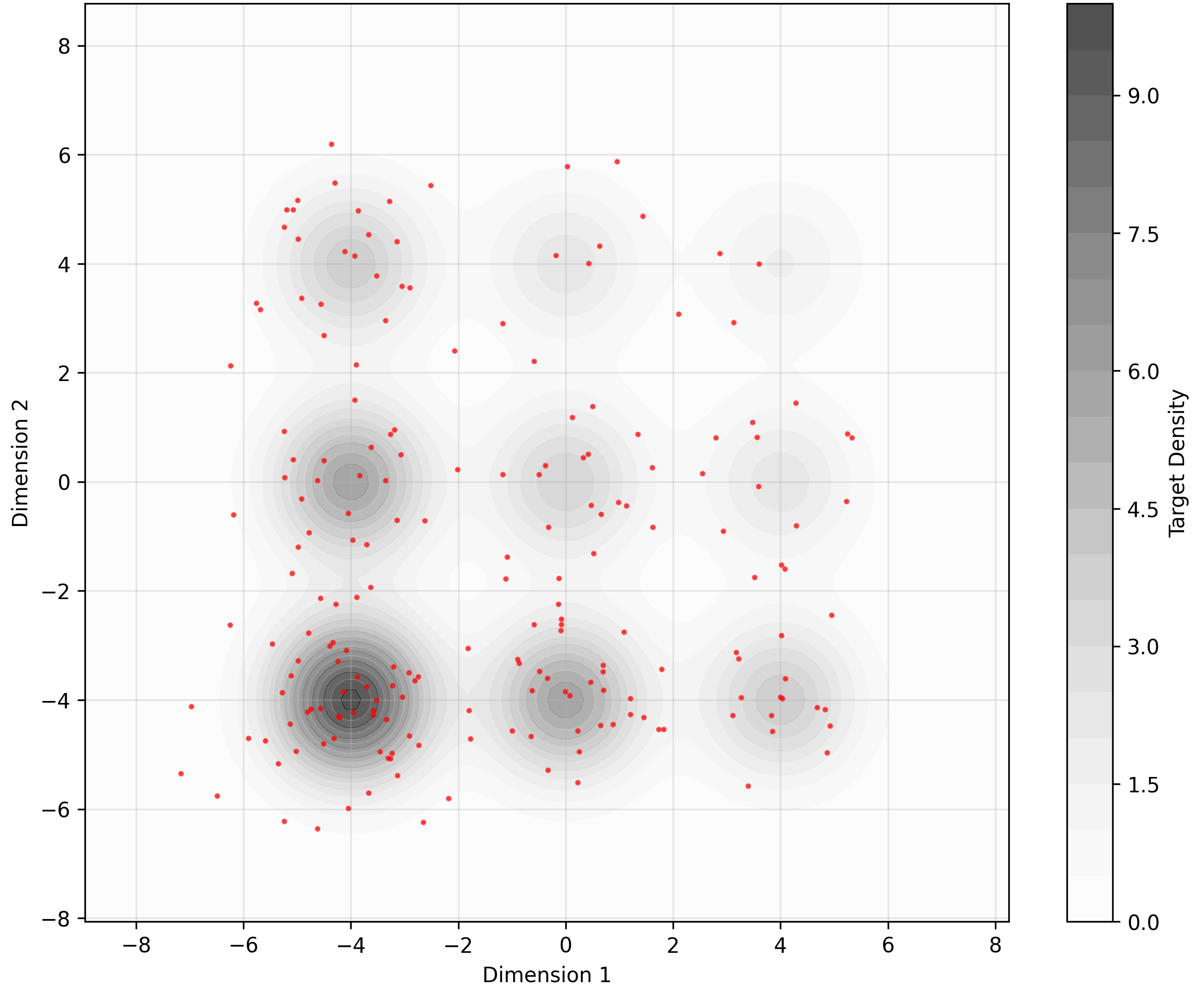}
    \caption{Density of the first two dimensions of the ``rough carpet'' target distribution with MCMC samples from one RWM simulation plotted in red. One out of every 200 MCMC samples drawn are plotted.}
    \label{fig:rough_carpet_density}
\end{figure}

First, we examine a simpler i.i.d.\ product density. Our first multimodal distribution has the form 
\begin{equation} \label{eq:target-rough-carpet}
\pi_4(\bx) = \prod_{i=1}^d f_4(x_i), \quad f_4(x) = 0.5 \, \mathcal{N}(x | \mu_1, 1) + 0.3 \, \mathcal{N}(x| \mu_2, 1) + 0.2 \, \mathcal{N}(x| \mu_3, 1)
\end{equation}
where the single component density $f_4$ is a one-dimensional density with three modes at $(\mu_1, \mu_2, \mu_3) = (-5, 0, 5)$ and $\mathcal{N}(x|\mu_i, 1)$ indicates the density of the univariate Gaussian distribution at point $x$ with mean $\mu_i$ and variance 1. This forms a ``rough carpet''-like appearance for the target distribution when we take the product of this single-component density over many dimensions (see the visualization of the first two components in Figure \ref{fig:rough_carpet_density} for an example). Since we have $3^d$ modes, the density values at each mode may not be very high. Yet, since this is a i.i.d.\ product form, the theoretical result of 0.234 being the optimal acceptance rate should also apply. 

\subsubsection{The ``Rough Carpet'' Target Density: Inhomogeneously Scaled Components}
Next, we examine a multimodal distribution with a density of the form 
\begin{equation} \label{eq:target-rough-carpet-scaled}
\pi_5(\bx) =\prod_{i=1}^d C_i f_5(C_i x_i), ~~ f_5(x) = 0.5 \, \mathcal{N}(x | \mu_1, 1) + 0.3 \, \mathcal{N}(x| \mu_2, 1) + 0.2 \, \mathcal{N}(x| \mu_3, 1).
\end{equation}
Note that $f_5 = f_4$, so this density $\pi_5$ is a more general form of $\pi_4$ by including inhomogeneous component-wise scaling factors $C_i$, where $C_i > 0$ and $\E(C_i) = 1$ and $\text{Var}(C_i) < \infty$. This is a special case of a class of inhomogeneously-scaled i.i.d.\ product target densities discussed by \cite{roberts_rosenthal_2001}, which proved the extension of the 0.234 asymptotically optimal acceptance rate to this class. We independently sample the component scaling factors $C_i \sim \text{Uniform}[0.02,1.98]$.

\subsubsection{A Tale of Three Mixtures}\label{sec:3mode}
The above ``rough carpet'' distributions are multimodal, but notice that they still follow the general product form $\pi(\bx) = \prod_{i=1}^d f(x_i)$. Here, we examine a distribution that does not have this product assumption at all; instead, we have a weighted sum of densities as our target density. 
We examine a mixture of three Gaussians; a multimodal distribution with just three modes regardless of the number of dimensions. The density is 
\begin{equation} \label{eq:3mode}
\pi_6(\bx) = w_1 \mathcal{N}(\bx | \boldsymbol{\mu}_1, \Sigma_1) + w_2 \mathcal{N}(\bx | \boldsymbol{\mu}_2, \Sigma_2) + w_3 \mathcal{N}(\bx | \boldsymbol{\mu}_3, \Sigma_3)
\end{equation}
 where we have non-negative scalar weights $w_i \geq 0$ such that $w_1 + w_2 + w_3 = 1$, and $\mathcal{N}$ is the density function of a multivariate Gaussian with mean $\boldsymbol{\mu}_i \in \R^d$ and covariance matrix $\Sigma_j \in \R^{d \times d}$.

\begin{figure}[ht!]
    \centering
    \begin{subfigure}{0.4\textwidth}
        \includegraphics[width=\linewidth, valign=t]{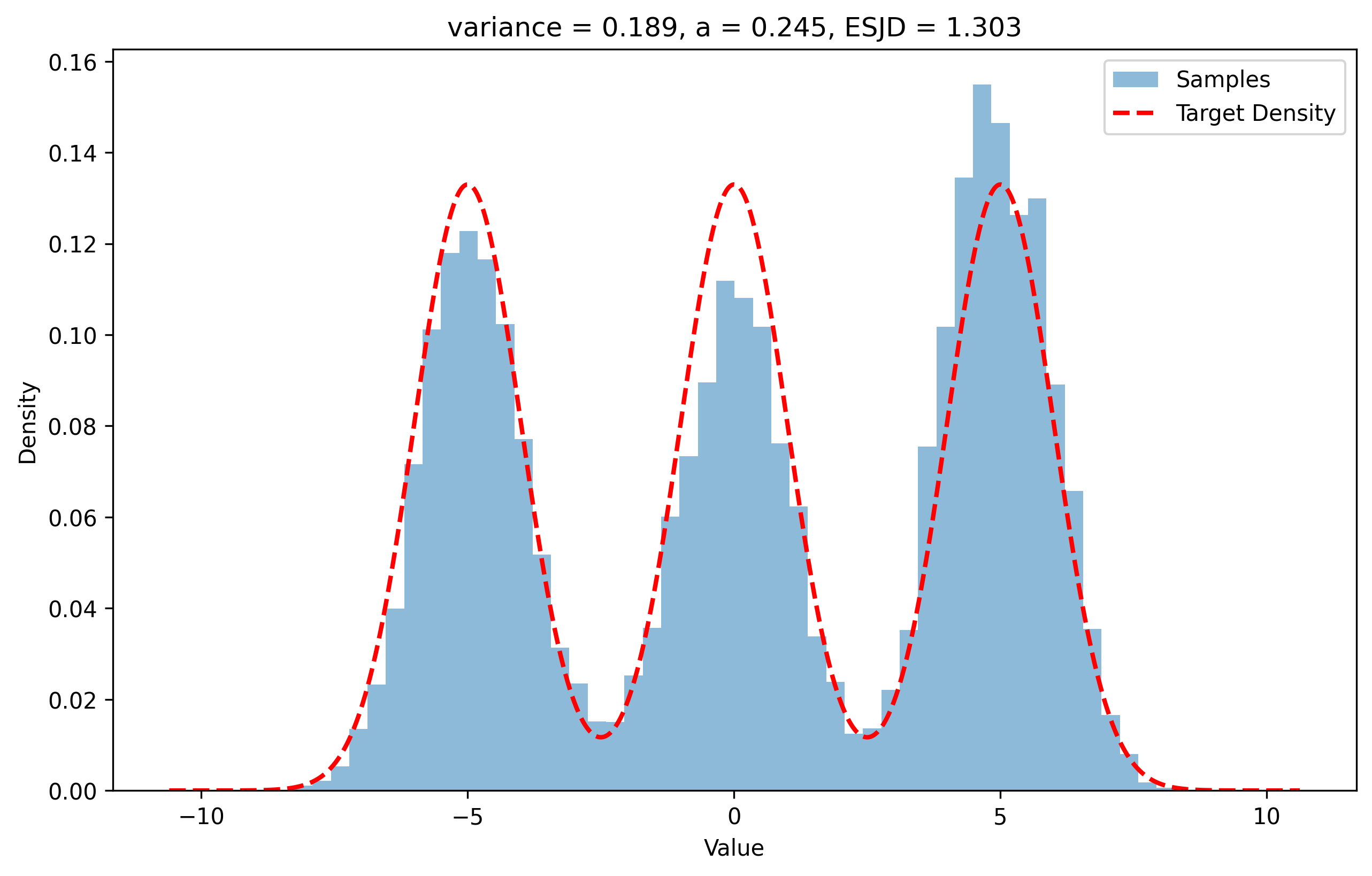}
    \end{subfigure}
    \begin{subfigure}{0.4\textwidth}
        \includegraphics[width=\linewidth, valign=t]{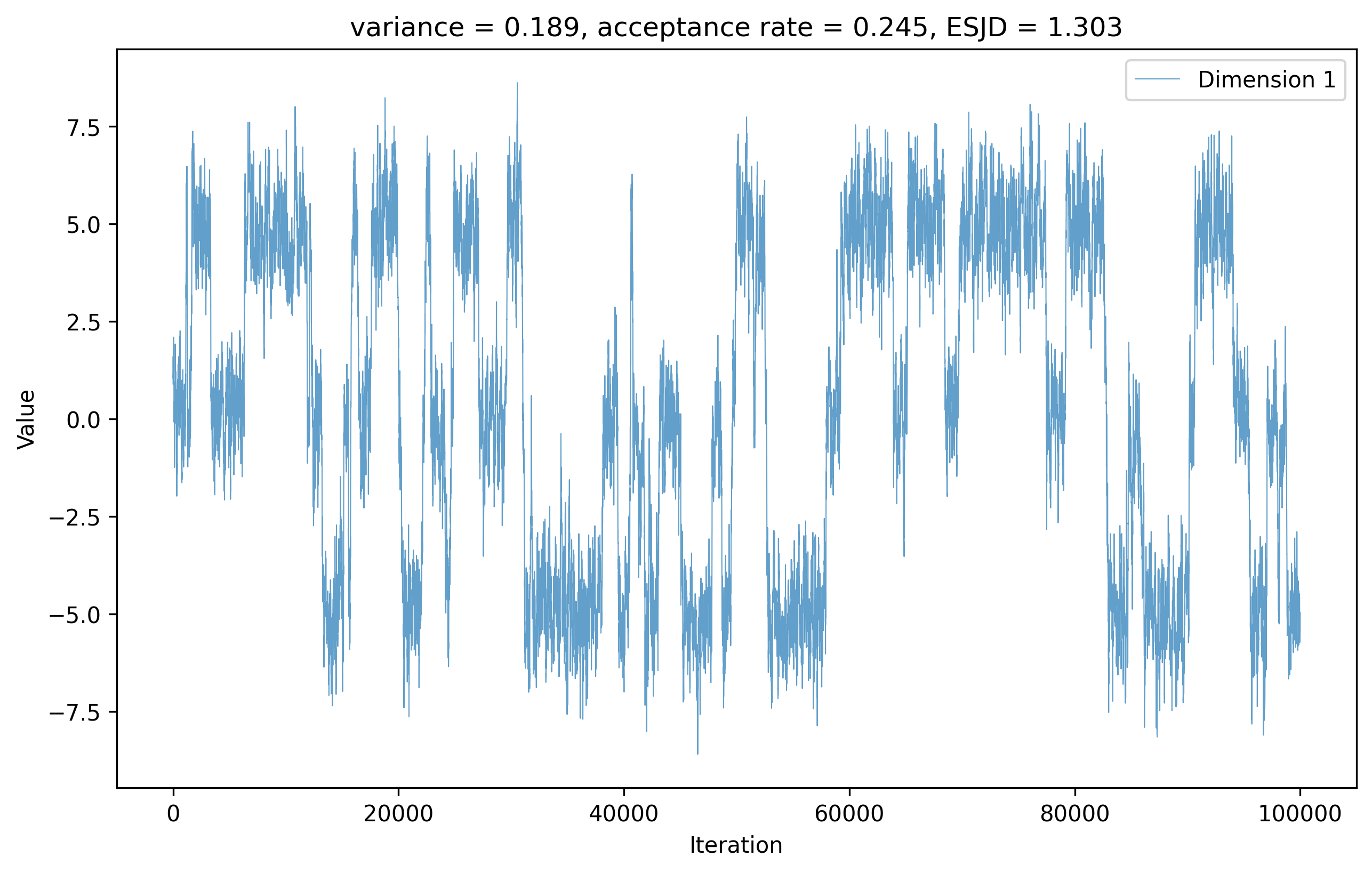}
    \end{subfigure}
    \caption{A histogram and traceplot of the first component from one of the simulations in the three mixture target distribution experiment after 100,000 iterations.}
    \label{fig:3mode_rwm_hist_new}
\end{figure}

For these experiments, we set the weights $w_1=w_2=w_3 = \frac{1}{3}$, the covariances $\Sigma_1 = \Sigma_2 = \Sigma_3 = I_d$, and set the means $\boldsymbol{\mu}_1 = (\epsilon, 0, 0, \dots, 0), \boldsymbol{\mu}_2 = (0, 0, \dots, 0), \boldsymbol{\mu}_3 = (-\epsilon, 0, 0, \dots, 0)$ for some constant $\epsilon \in \R$, i.e. just varying the first component of the means and setting the other components of the means to 0. 
As mentioned (refer to Figure \ref{fig:10m no escape}), we choose the constant $\epsilon = 5.0$ in the first dimension after observing via traceplots that jumping between modes still happens frequently enough. We refer the reader to Figure \ref{fig:3mode_rwm_hist_new} for an example histogram and traceplot to illustrate how the Markov chain explores the three modes.

\subsubsection{A Tale of Three Mixtures: Inhomogenous Scaling Factors}

We repeat our experiment with inhomogeneous scaling factors applied to each Gaussian mode. Concretely, we instantiate our target density as
\begin{equation} \label{eq:3mode-scaled}
\pi_7(\bx) = \sum_{k=1}^3 w_k \cdot ( \prod_{j=1}^d C_j ) \cdot \mathcal{N}(\bx \odot \bC | \mu_k, I_d)
\end{equation}
where $C_i \sim \mathrm{Uniform}[0.2,1.8]$ and $\bx \odot \bC$ denotes the Hadamard (element-wise) product $(x_1 C_1, ..., x_d C_d)$. This allows each Gaussian mode to follow the functional form defined in \citet{roberts_rosenthal_2001} Section 6; each mode has its own density of the functional form $\pi(\bx) = \prod_{i=1}^d C_i f(C_i x_i)$, where $\{C_i\}$ are i.i.d.\ with $\E(C_i) = 1$ and $\E(C_i^2) < \infty$.

\subsubsection{Multimodal Experiments and Results}

We conduct our experiments in $d \in \{20, 30, 50\}$ using the default experimental procedure described. We summarize our results in Table \ref{tab:rwm_multimodal}, and refer the reader to Appendix \ref{app:plots} Figures \ref{fig:esjd_rough_carpet}, \ref{fig:esjd_rough_carpet_scaled}, \ref{fig:esjd_3mix}, and  \ref{fig:esjd_3mix_scaled} for plots of the trend of ESJD with acceptance rate. 

\begin{table}[htbp]
\centering
\caption{Empirical optimal RWM acceptance rates for multimodal target distributions.}
\label{tab:rwm_multimodal}
\begin{tabular}{lccccc}
\toprule
\textbf{Target} & \(d=20\) & \(d=30\) & \(d=50\) \\
\midrule
$\pi_4$: Rough Carpet (homogeneous)         & 0.2133 & 0.2202 & 0.2291 \\
$\pi_5$: Rough Carpet (inhomogeneous)       & 0.2009 & 0.2174 & 0.2351 \\
$\pi_6$: 3-Mixture (homogeneous)            & 0.2489 & 0.2454 & 0.2394 \\
$\pi_7$: 3-Mixture (inhomogeneous)           & 0.2360 & 0.2334 & 0.2371 \\
\bottomrule
\end{tabular}
\end{table}

The results for our multimodal experiments provide strong empirical support for the generalizability of the 0.234 acceptance rate heuristic. We observe that for all four target distributions, the optimal acceptance rate converges towards 0.234 as the dimension increases, even for those that lack an i.i.d. product structure. Their optimal acceptance rates are already very close to 0.234 even in low dimensions ($d=10$).
Theoretically, the 0.234 value is likely to be justified by the general criteria of \citet{sherlock_optimal_2013} that are stated in Assumption \ref{assump:sherlock2013}. The three mixture targets ($\pi_6, \pi_7$), while not product densities, are constructed from isotropic Gaussians. In high dimensions, they form a well-behaved geometric shell with low eccentricity, thus satisfying the target conditions for the 0.234 rule to apply. 
The rough carpet targets $\pi_4, \pi_5$ are i.i.d. products of a more complex multimodal density. The component-wise multimodality creates a more intricate target landscape, which may require higher dimensions before the geometric properties average out to satisfy the shell and stability conditions as effectively; this may explain why their optimal rates approach 0.234 from below.

There is an interesting note that for both families of targets, introducing inhomogeneous scaling  consistently pushes the optimal acceptance rate lower in smaller dimensions. This could be explained by the Target Eccentricity condition of \citet{sherlock_optimal_2013} (mentioned in Assumption \ref{assump:sherlock2013} \eqref{sherlock_target_eccentricity}). The inhomogeneous scaling factors introduce anisotropy into the target's curvature, creating directions with much higher curvature than others. With a standard isotropic proposal such as the multivariate Gaussian, the optimal proposal scale must shrink to avoid constant rejections in these high-curvature directions. This makes the sampler inefficient in the less-curved directions, degrading the algorithm's overall performance. This less efficient dynamic alters the trade-off between proposal size and acceptance probability, shifting the peak of the ESJD curve to the left. Our results show that for such low-dimensional inhomogeneously scaled targets, the optimal balance is achieved at a lower acceptance rate than the 0.234 ideal. Then, as $d$ grows, the law of large numbers mitigates the effect of any single scaling factor, the target becomes effectively less eccentric, and the optimal acceptance rate recovers to 0.234.

\subsection{Rosenbrock Target Densities}\label{subsec:rosenbrock}

The multimodal experiments in Section \ref{subsec:multimodal} demonstrated the effectiveness of the 0.234 acceptance rate for multimodal distributions, where the main challenge is mode isolation. In practice, however, many posterior landscapes are not only multi-peaked but also have nonlinear correlation structure, providing challenges for local proposals even when there is only a single global mode. We examine this setting by sampling three different extensions of the two-dimensional Rosenbrock target density to $d$-dimensions \citep{goodman_ensemble_2010, dixon_effect_1994, pagani_n-dimensional_2022}, each of which have different curvature and therefore different difficulty from a MCMC perspective. 

\citet{rosenbrock_automatic_1960} originally introduced the Rosenbrock function as an optimization problem: a non-convex function which has a global minimum inside a long, narrow parabolic canyon. The Rosenbrock function can then be converted into a probability density that preserves these properties. Because the local correlation structure changes rapidly along the ridge, step sizes that work near the mode are either rejected or become inefficient elsewhere. Consequently, Rosenbrock targets have become common benchmarks to test MCMC algorithms \citep{haario_heikki_saksman_2001, huijser_properties_2022, hunt-smith_accelerating_2024, militzer_ensemble_2025}.
Including these targets in our empirical study extends our findings to targets with curved correlation structure. 

We first describe the unnormalized Rosenbrock target densities, then describe our experimental procedure and results. We use the Rosenbrock kernels described by \citet{pagani_n-dimensional_2022}.

\subsubsection{Full Rosenbrock}

The $d$-dimensional Full Rosenbrock distribution \citep{goodman_ensemble_2010} has the following kernel for $\bx \in \R^d$:
\begin{equation}
\pi_8(\mathbf{x}) \propto \exp\left\{ -\sum_{i=1}^{d-1} \left[ 100(x_{i+1} - x_i^2)^2 + (1 - x_i)^2 \right]/20 \right\}
\label{eq:full_rosenbrock}
\end{equation}
The original 2-dimensional Rosenbrock resembles a long banana-shaped valley whose floor follows the parabola $x_2=x_1^{2}$. As stated in \citet{pagani_n-dimensional_2022}, as $d$ increases, the long, narrow ridge that makes sampling challenging becomes more concentrated around the mode, decreasing the problem's difficulty. However, the variance of $x_d$ increases steeply as each new random variable is directly dependent on the squared value of the
previous variable.

\subsubsection{Even Rosenbrock}

The $d$-dimensional Even Rosenbrock distribution extended from the optimization function in \citep{dixon_effect_1994} has the following kernel for $\bx, \boldsymbol{\mu} \in \R^d$, where $d$ must be even:

\begin{equation}
\pi_9(\mathbf{x}) \propto \exp\left\{ -\sum_{i=1}^{d/2} \left[ (x_{2i-1} - \mu_{2i-1})^2 + 100(x_{2i} - x_{2i-1}^2)^2 \right]/20 \right\}
\label{eq:even_rosenbrock}
\end{equation}

The Even Rosenbrock partitions the state into $\tfrac{d}{2}$ independent 2-dimensional Rosenbrock blocks.  Because the blocks do not interact, the global target is the product of smaller bananas; the joint density still exhibits the same curved valleys within each block but lacks the long-range, serial dependence of the Full Rosenbrock. Consequently, we expect higher ESJD (and hence slightly lower optimal acceptance rates) than in the fully coupled case. 

Our experiments only use a single scaling factor applied to the whole covariance matrix of the proposal, but it could be useful for a practitioner to note that it is possible to locally tune the proposal for each 2-dimensional Rosenbrock block to make the RWM algorithm  more efficient on this distribution. For example, when using a Gaussian proposal, the covariance matrix could be set as a block diagonal matrix where each 2x2 block on the diagonal of the Gaussian covariance matrix is tailored to and achieves optimality on its own 2-dimensional Rosenbrock.

\subsubsection{Hybrid Rosenbrock}

The $d$-dimensional Hybrid Rosenbrock distribution introduced by \citet{pagani_n-dimensional_2022} has the following kernel for $\bx \in \R^d$, where $\mu, x_{j,i} \in \R; a, b_{j,i} \in \R^+ (\forall j,i)$:
\begin{equation}
\pi_{10}(\mathbf{x}) \propto \exp\left\{ -a(x_1 - \mu)^2 - \sum_{j=1}^{n_2} \sum_{i=2}^{n_1} b_{j,i}(x_{j,i} - x_{j,i-1}^2)^2 \right\}
\label{eq:hybrid_rosenbrock}
\end{equation}
The integers $n_1$ and $n_2$ describe a grid of blocks: there are $n_2$ vertical stacks, each containing $n_1\!-\!1$ Rosenbrock-type links and a shared root coordinate $x_1$.  The total dimension is therefore $d=(n_1-1)n_2+1$.
Within every column $j$, the curvature coefficients $b_{j,i}$ can vary with depth $i$, producing steeper or flatter valleys as we move away from the root.  Horizontally, however, columns are independent, so the Hybrid form interpolates between the serial dependence of the Full Rosenbrock (one long chain: $n_2=1$) and the independent blocks of the Even Rosenbrock ($n_1=3$, $n_2=d/2$). This allows the difficulty of the distribution to be modified by setting $n_1, n_2$; larger $n_1$ increases non-linearity within a column, while larger $n_2$ raises overall dimension without adding extra curvature.

\subsubsection{Simulations and Results}

For our Rosenbrock simulations, we found that our initial settings of 200,000 Markov chain steps and 20 seeds were insufficient to produce meaningful trends, so we ran more simulations for more steps with the Rosenbrock targets. In this section, we continue using a Gaussian proposal with 40 different variance values. However, we run each simulation for 1,000,000 steps and we repeated each experiment with 30 different seeds instead of the usual 20. For the Full Rosenbrock, we select $d \in \{5,10,20,30\}$. For the Even Rosenbrock, we select $d \in \{4,10,20,30\}$. For the Hybrid Rosenbrock, we select $(n_1, n_2) \in \{(3,2), (5,2), (7,3), (8,4)\}$ which gives total dimension $d \in \{5,9,19,29\}$ respectively. 

We also provide a plot of the ESJD versus acceptance rate trend for the 2-dimensional Rosenbrock. Note that for $d=2$, each of the three Rosenbrocks reduce to the standard Rosenbrock 2-dimensional density, assuming $\mu_1 = 1$ for the Even Rosenbrock and $n_1=2, n_2=1, a = \frac{1}{20}, b = \frac{100}{20}, \mu = 1$ for the Hybrid Rosenbrock. All other plots of ESJD versus acceptance rate trends for these distributions are in Appendix \ref{app:plots} Figures \ref{fig:esjd_rosenbrock_full}, \ref{fig:esjd_rosenbrock_even}, and  \ref{fig:esjd_rosenbrock_hybrid} for plots of the trend of ESJD with acceptance rate. 

\begin{table}[htbp]
\centering
\caption{Empirical optimal RWM acceptance rates for three different $d$-dimensional Rosenbrock target distributions.}
\label{tab:rwm_rosenbrock}
\begin{tabular}{lcccc}
\toprule
\textbf{Target} & \(d=4\, \text{or} \,5\) & \(d=9\, \text{or} \,10\) & \(d=19\, \text{or} \,20\) & \(d=29\, \text{or} \,30\) \\
\midrule
$\pi_8$: Full Rosenbrock   & 0.2047 & 0.2359 & 0.2384 & 0.2337 \\
$\pi_9$: Even Rosenbrock   & 0.0646 & 0.1430 & 0.2003 & 0.2107 \\
$\pi_{10}$: Hybrid Rosenbrock & 0.1394 & 0.2272 & 0.2349 & 0.2340 \\
\bottomrule
\end{tabular}
\end{table}

Our experiments with the Rosenbrock family of targets, presented in Table 4, demonstrate that the 0.234 heuristic remains effective even for distributions with non-linear dependencies and curvature. For the Full Rosenbrock ($\pi_8$) and Hybrid Rosenbrock ($\pi_{10}$) targets, the optimal acceptance rate robustly converges to approximately 0.234 as the dimension increases. Although these targets are not i.i.d. and possess a highly anisotropic banana-shaped geometry, they should satisfy the general criteria from \citet{sherlock_optimal_2013} (stated as Assumption \ref{assump:sherlock2013}) in the high-dimensional limit. As $d$ grows, the complex local correlations become less influential relative to the global structure, allowing the target's geometric properties to stabilize in the manner required by Assumption \ref{assump:sherlock2013}.
On the other hand, the Even Rosenbrock ($\pi_9$) has a significantly lower optimal acceptance rate in low dimensions than the other Rosenbrocks, and the optimal acceptance rate increases slowly towards 0.234 with increasing dimension. This seems reasonable and the analysis of \citet{beskos_asymptotic_2018} is useful here.

The Even Rosenbrock is a product of independent 2-dimensional Rosenbrock blocks, and for a small number of blocks, the standard high-dimensional scaling theory is less applicable. 
We elaborate on this in the next paragraph about the 2-dimensional Rosenbrock, but the key relevant message from \citet{beskos_asymptotic_2018} is that in low dimensions, large proposed jumps and an acceptance rate closer to zero are most efficient, explaining the low optimal acceptance rate.
As the dimension $d$ and thus the number of independent blocks grow, the dynamics transition from a Markov jump process towards a limiting diffusion. While the theory for such targets suggests a position-dependent diffusion speed and thus a ``local'' 0.234 rule \citep[see point (iv), p. 2968]{beskos_asymptotic_2018}, the growing number of independent blocks and therefore the growing number of ``large'' components may induce a law of large numbers effect where position-dependent variations in diffusion speed average out across the many blocks. 
This would result in an approximately constant global diffusion speed and therefore a global 0.234 acceptance rate being approximately optimal in high dimensions for the Even Rosenbrock. 

\begin{figure}[h]
    \centering
    \includegraphics[width=0.5\textwidth]{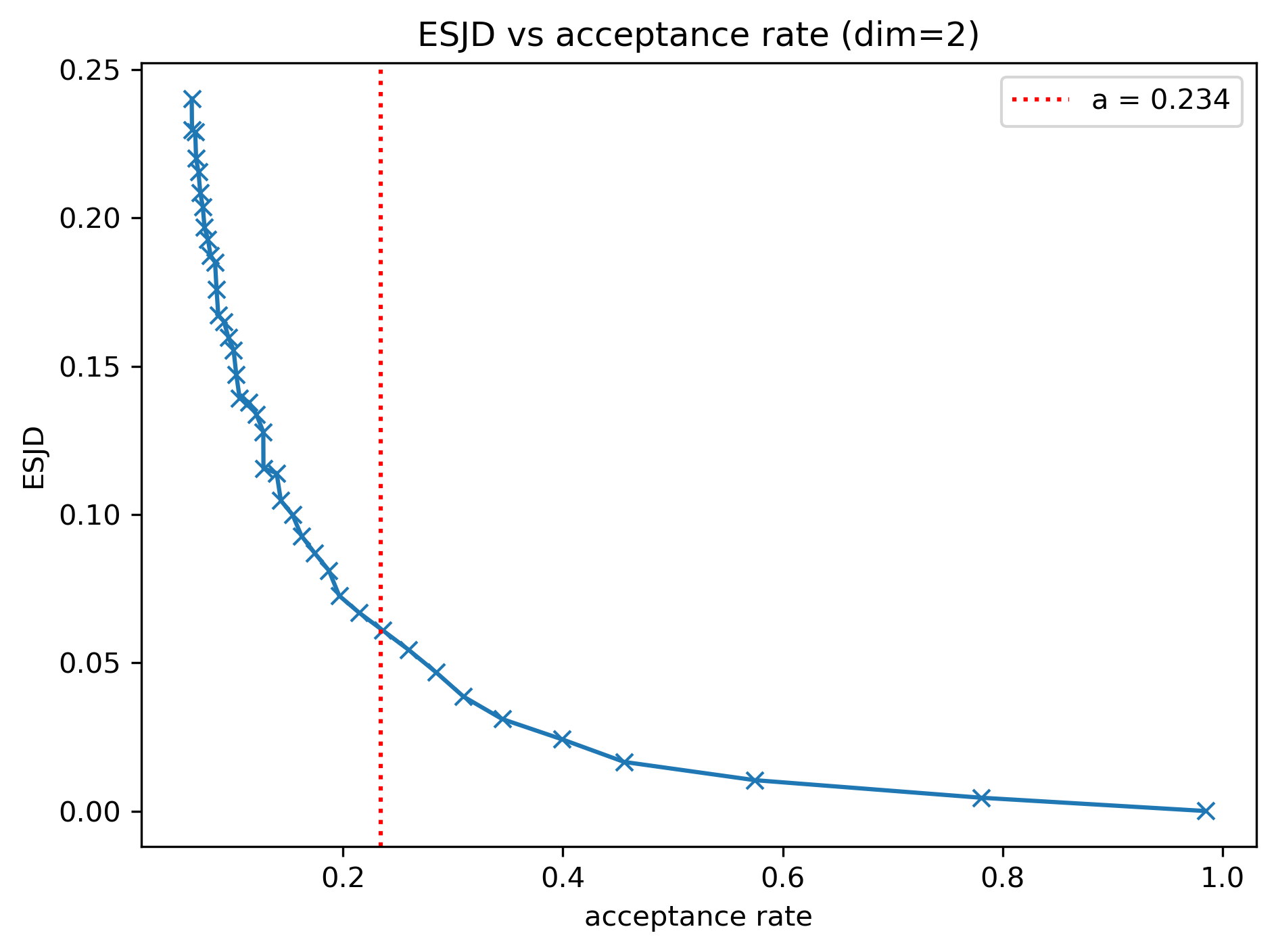}
    \caption{ESJD vs.\ acceptance rate for the 2-dimensional Rosenbrock target distribution $\pi_8$ under RWM with a Gaussian proposal. Red dotted line indicates an acceptance rate of 0.234.}
    \label{fig:rosenbrock-2D}
\end{figure}

The result for the 2-dimensional Rosenbrock target, shown in Figure \ref{fig:rosenbrock-2D}, provides empirical evidence for the low-dimensional sampling phenomena analyzed by \citet{beskos_asymptotic_2018}.
Unlike the higher-dimensional cases, the ESJD does not exhibit a clear peak around a non-zero acceptance rate. Instead, it appears to increase monotonically as the acceptance rate approaches zero.

As theoretically established by \citet{beskos_asymptotic_2018},  the optimal sampling strategy for this low-dimensional ridged target is not a diffusion process. 
They state that efficiency is maximized by proposing extremely large jumps using a large proposal scaling $\sigma$ and allowing the acceptance rate to go to zero. 
A heuristic explanation for this is that in this case, the squared distance of a proposal increases proportionally to $\sigma^2$ but the fraction of accepted proposals decreases proportionally to $\frac{1}{\sigma}$, and therefore ESJD increases proportionally to $\sigma$ until the proposal scale becomes comparable to the characteristic scale of the target distribution itself, at which point the acceptance probability begins to decay more rapidly.
Taking the limit, as the thickness (scale $\varepsilon$ in \citet{beskos_asymptotic_2018}) of the support of the target distribution goes to zero, the algorithm's dynamics converge to a continuous-time Markov jump process rather than a diffusion. 
Our observation for the 2-dimensional Rosenbrock target, and by extension the low-dimensional Even Rosenbrock target (e.g., for $d=4$, which consists of two independent 2-dimensional blocks), is an instance of this theoretically-predicted behaviour.

\subsection{Non-smooth Target Densities: Continuous Hypercube}\label{subsec:hypercube}

Now, consider a RWM using a Gaussian proposal and a target density still of the product form $\pi(\bx) = \prod_{i=1}^d f(x_i)$, but where the individual component density $f$ is Uniform$[a, b]$. 
\begin{equation} \label{eq:hypercube}
\pi_{11}(\bx) = \prod_{i=1}^d f_{11}(x_i), \quad f_{11}(x) = \frac{\mathbb{I}\left[a \le x \le b \right]}{b - a}
\end{equation}

Unlike the previous examples, the component density $f$ and its logarithm are not continuously differentiable on $\R$ due to the discontinuities at the boundaries $x_i=a$ and $x_i=b$. Since this is one of the assumptions in Assumption \ref{assump:rgr97}, the asymptotic results where the optimal acceptance rate is 0.234 may not apply. This discontinuous target density case has been explored by \cite{Neal2012Discontinuous}, which found that for product densities with a discontinuity at the boundary, such as the unit hypercube $[0,1]^d$, the efficiency of the RWM is maximised with an asymptotic optimal acceptance rate of 0.1353 as $d \to \infty$. 
We provide simulation studies in various dimensions to investigate the applicability of this particular 0.1353 acceptance rate figure in lower dimensions on the hypercube $[0,1]^d$. We use 300,000 iterations instead of the typical 200,000 for this target to make the plots sufficiently smooth.

\begin{table}[htbp]
\centering
\caption{Empirical optimal RWM acceptance rates for the continuous hypercube target.}
\label{tab:rwm_hypercube}
\begin{tabular}{lccccccc}
\toprule
\textbf{Target} & \(d=2\) & \(d=5\) & \(d=10\) & \(d=20\) & \(d=30\) & \(d=50\) & \(d=100\) \\
\midrule
$\pi_{11}$: Hypercube [0,1]\textsuperscript{d} & 0.4316 & 0.2767 & 0.2027 & 0.1670 & 0.1525 & 0.1439 & 0.1423 \\
\bottomrule
\end{tabular}
\end{table}

\begin{figure}[htbp]
\centering

\begin{subfigure}[t]{0.32\textwidth}
    \includegraphics[width=\linewidth]{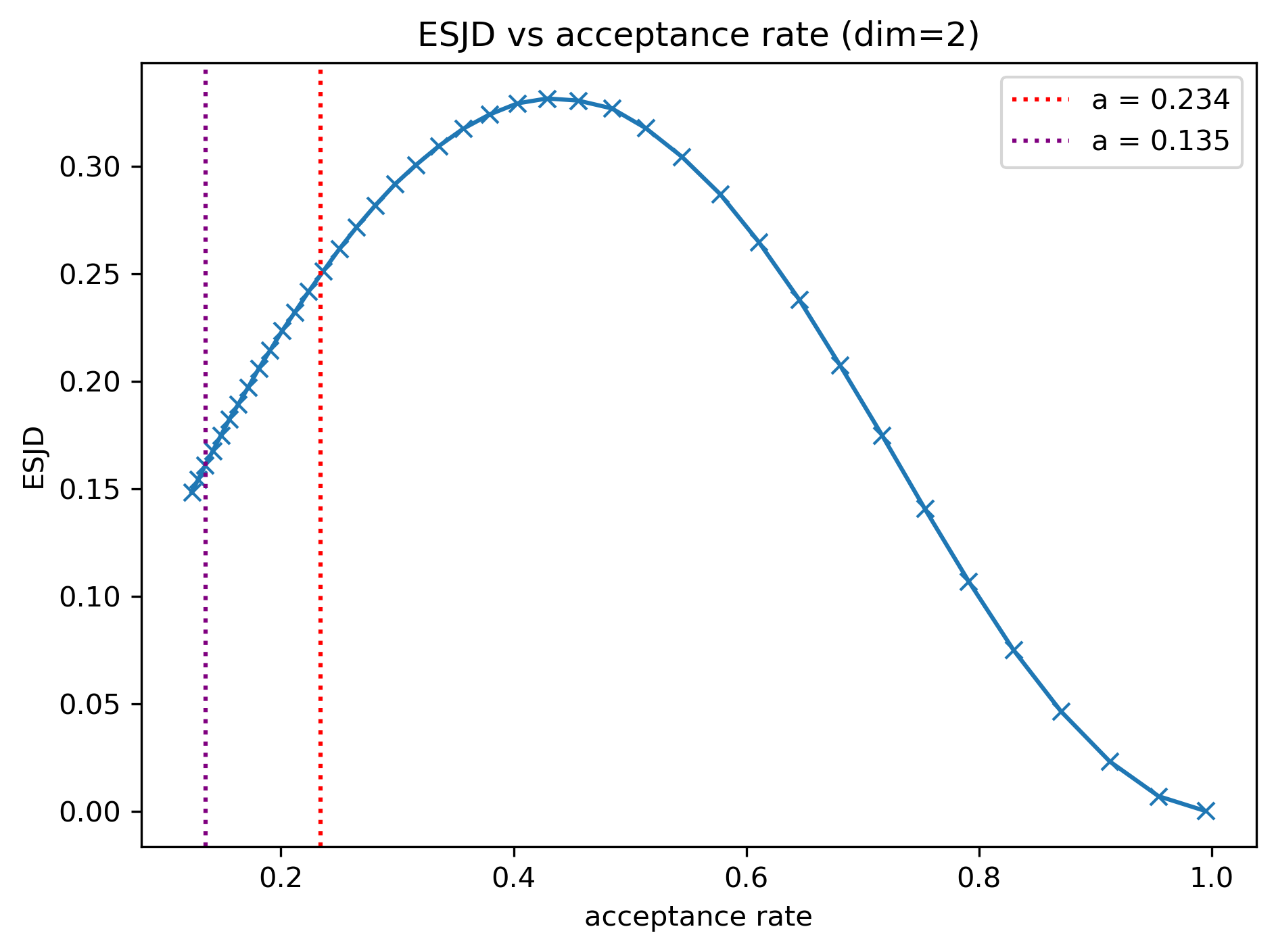}
\end{subfigure}
\hfill
\begin{subfigure}[t]{0.32\textwidth}
    \includegraphics[width=\linewidth]{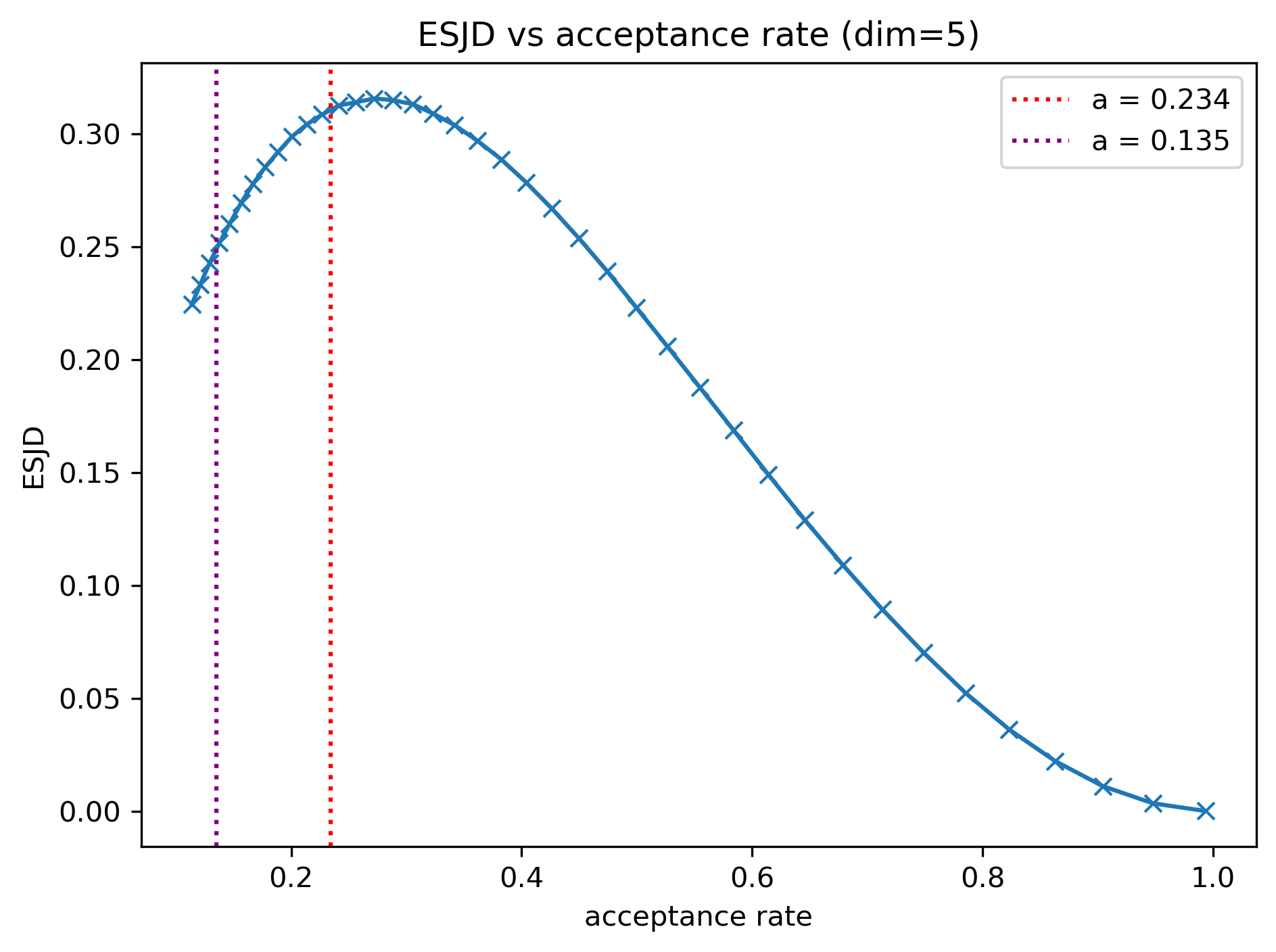}
\end{subfigure}
\hfill
\begin{subfigure}[t]{0.32\textwidth}
    \includegraphics[width=\linewidth]{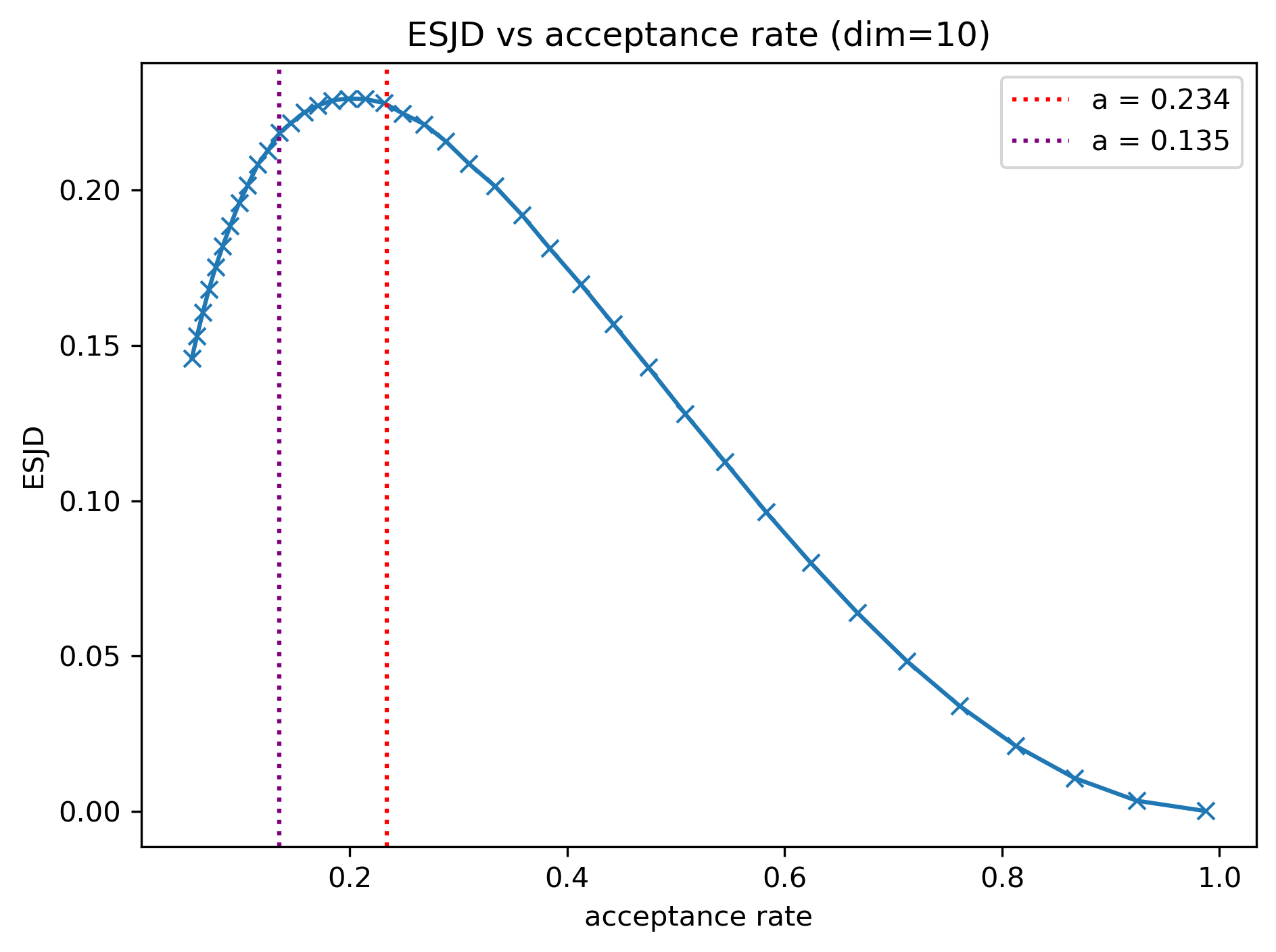}
\end{subfigure}

\vspace{0.5em}

\begin{subfigure}[t]{0.32\textwidth}
    \includegraphics[width=\linewidth]{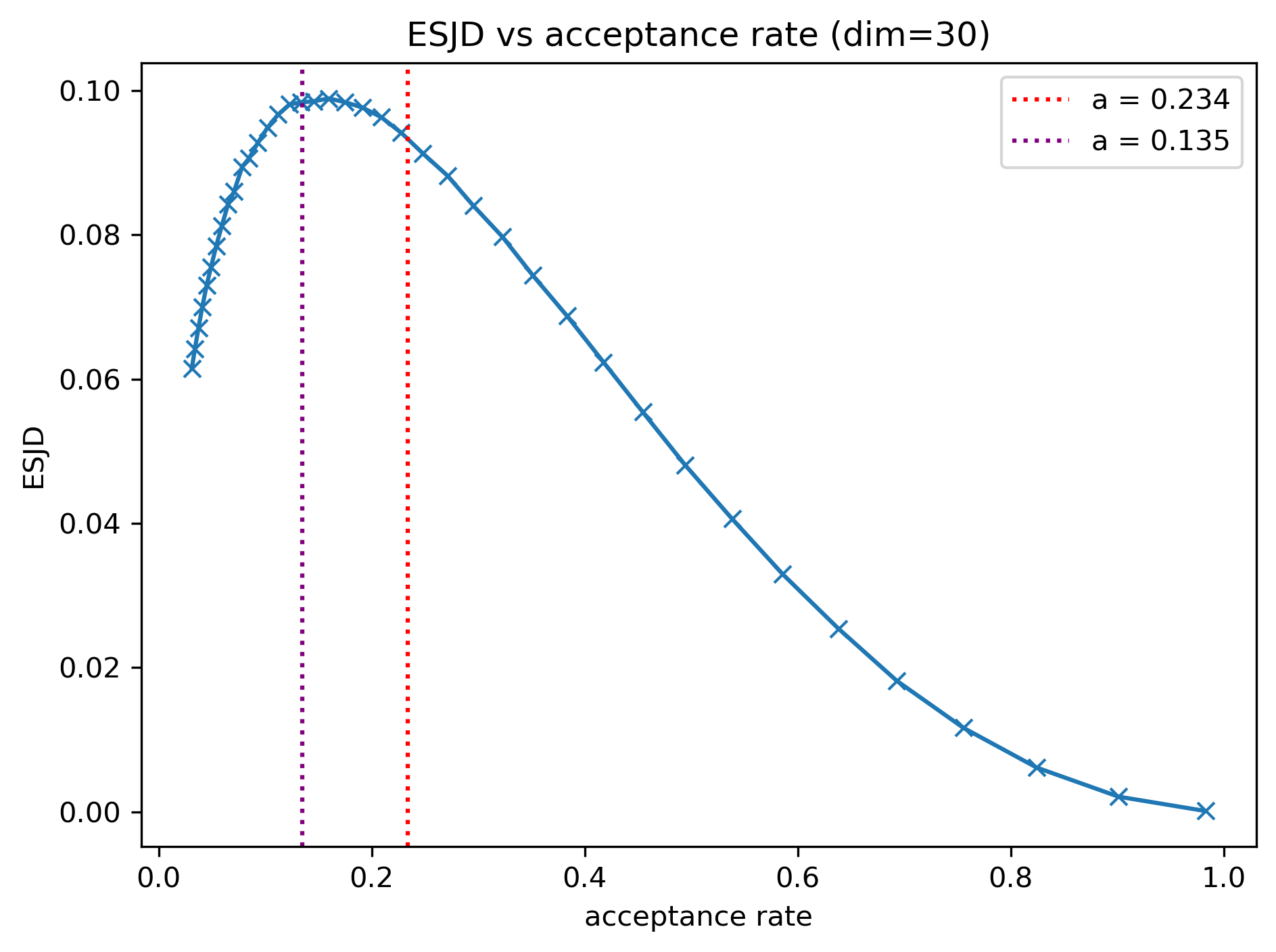}
\end{subfigure}
\hfill
\begin{subfigure}[t]{0.32\textwidth}
    \includegraphics[width=\linewidth]{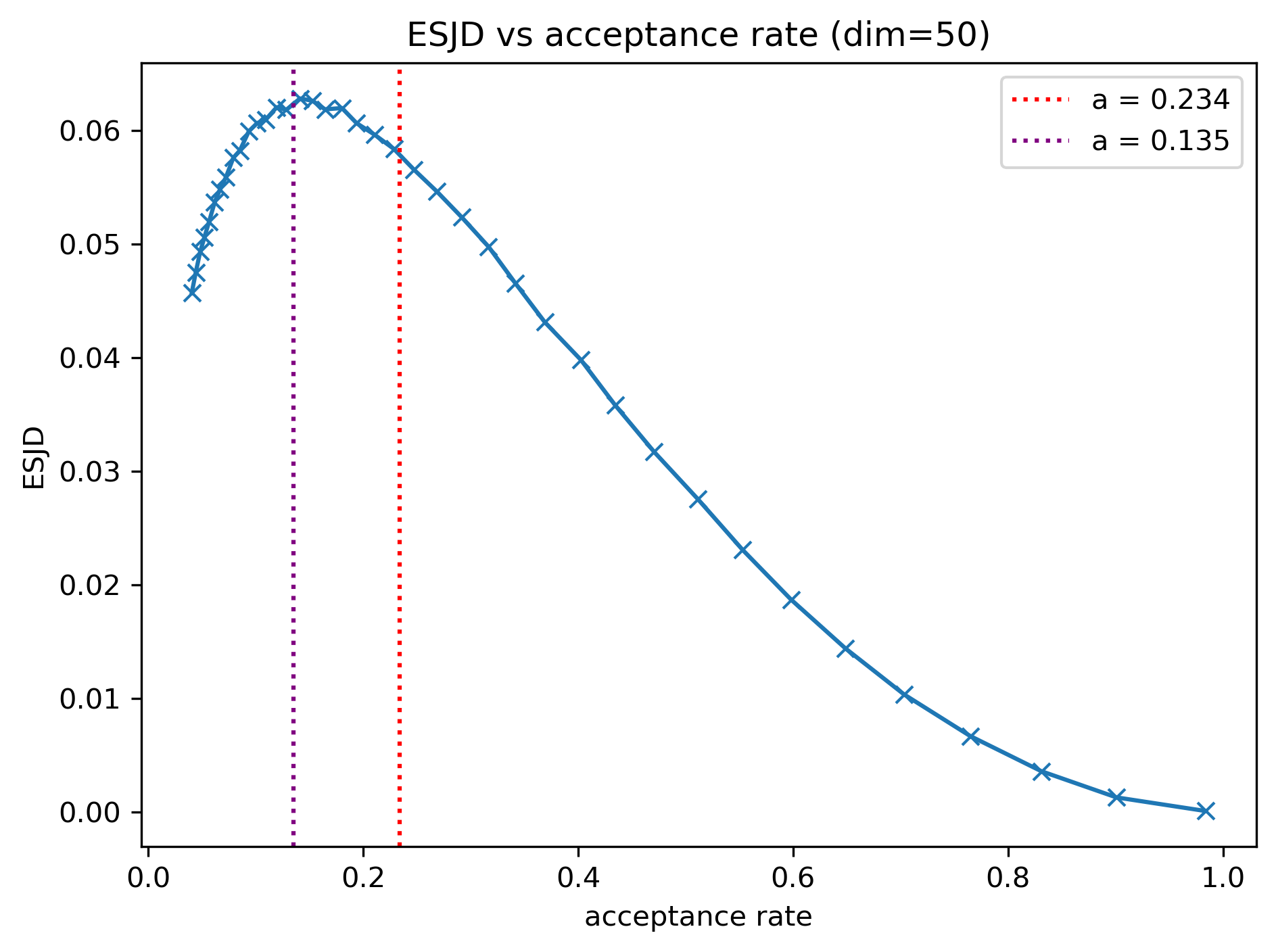}
\end{subfigure}
\hfill
\begin{subfigure}[t]{0.32\textwidth}
    \includegraphics[width=\linewidth]{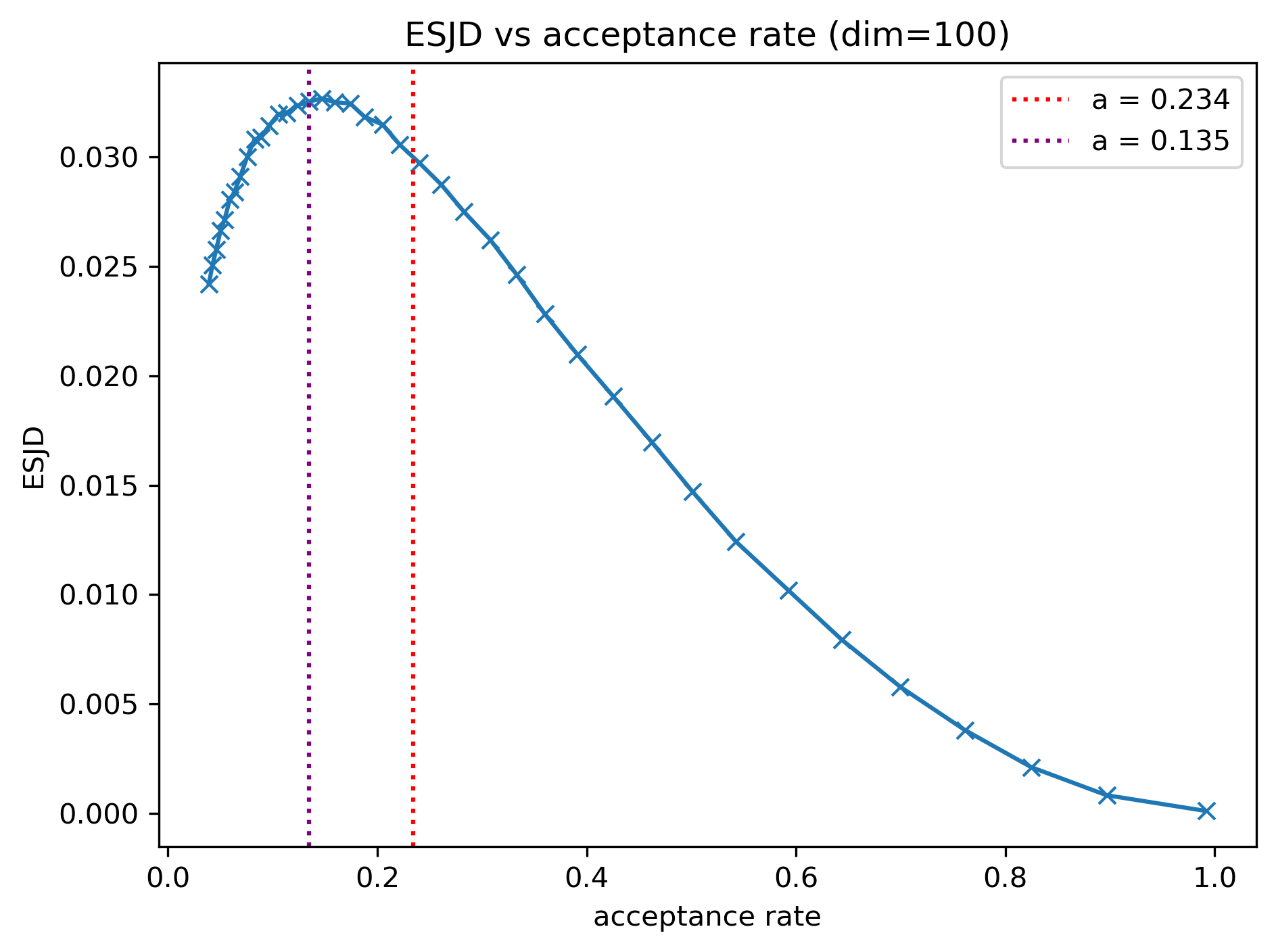}
\end{subfigure}

\caption{ESJD vs.\ acceptance rate for the i.i.d.\ Hypercube $[0,1]^d$ target distribution $\pi_{11}$ under RWM with a Gaussian proposal in dimensions $d \in \{2, 5, 10, 30, 50, 100\}$ from top-left to bottom-right. Red dotted line indicates an acceptance rate of 0.234. Purple dotted line indicates an acceptance rate of 0.135.}
\label{fig:esjd_hypercube}
\end{figure}

 Compared to the previous examples (see Appendix \ref{app:plots} for their ESJD versus acceptance rate curves, which tend to be smooth and gentle at the plateau), which seem to be fairly consistently optimal with an acceptance rate of approximately 0.234, we find (refer to Table \ref{tab:rwm_hypercube} and Figure \ref{fig:esjd_hypercube}) that the optimal acceptance rate for the hypercube target changes significantly as the dimension increases, and notice that the plateau of the curve tends to get quite sharp as the dimension increases. This is because discontinuities introduce additional complexity in the target distribution, affecting the behaviour and performance of the RWM algorithm.

As the dimension $d$ increases, the algorithm's performance converges to the theoretical 0.1353 value. 
In this specific case, the shrinking optimal acceptance rate might be because of the unique zero or non-zero form of the target density beyond the hypercube boundary. As discussed by \cite{Neal2012Discontinuous}, whether a proposed move is accepted depends completely on whether the proposed state is inside the hypercube. Subsequently, when using a Gaussian proposal, the probability that at least one proposed component will lie outside the boundary increases as dimension increases, leading to a higher probability of instant rejection when the proposal scaling factor independent of dimension is kept constant.

The trends in Figure \ref{fig:esjd_hypercube} offer practical guidance for tuning. In high dimensions (e.g., $d=100$), the ESJD curve peaks sharply near the theoretical optimum of 0.1353, and tuning to 0.234 would result in a significant loss of efficiency. However, in lower dimensions (e.g., $d=2, 5, 10$), the optimal acceptance rate is much closer to 0.234 than to 0.1353. The ESJD curve is also flatter, meaning the penalty for slight mistuning is less severe. This suggests a mini-max strategy for practitioners unsure of the dimensionality effects: tuning to an acceptance rate of approximately 0.234 is a robust choice across all dimensions. While it could be suboptimal for very high dimensions, it is nearly empirically optimal for low dimensions, whereas tuning to 0.1353 in these instances would be highly inefficient even though theory supports it over the 0.234 rate for this specific target distribution.

\subsection{Neal's Funnel Target Density} \label{subsec:funnel}

To probe the limits of the 0.234 heuristic, we now examine a target distribution that is canonical for its challenging geometry: the funnel distribution introduced by \citet{neal_slice_2003}. Unlike the Rosenbrock target, which has localized curvature, Neal's funnel possesses a global, non-linear dependency structure that presents a severe challenge to standard MCMC algorithms. In this distribution, a single variable controls the variance of all other variables, creating an extreme, state-dependent anisotropy. This structure should violate the geometric regularity conditions outlined in Assumption~\ref{assump:sherlock2013}, particularly the target eccentricity and Hessian stability.

The $d$-dimensional Neal's funnel target density, which we denote $\pi_{12}$, is defined for $\mathbf{x} = (x_1, x_2, \dots, x_d)$. The density is given by the product of a marginal density for the first component $x_1$ and conditional densities for the other components:
\begin{equation}
\label{eq:funnel}
\pi_{12}(\mathbf{x}) = \mathcal{N}(x_1 \mid 0, 3^2) \prod_{i=2}^{d} \mathcal{N}(x_i \mid 0, \exp(x_1))
\end{equation}
 The key feature is that the variance of each $x_i$ after $x_1$ is $\exp(x_1)$. This creates a funnel shape in the state space: when $x_1$ is large and negative, the funnel is a very narrow neck where the $x_i$'s are tightly constrained; when $x_1$ is positive, the funnel opens into a wide mouth where the $x_i$'s can have enormous variance. 
 
For our experiments, we follow the standard parameterization from \citet{neal_slice_2003}. We test this target in dimensions $d \in \{5, 10, 20, 30\}$. Due to the difficulty of sampling this target, we increased our simulation budget significantly, using $1{,}000{,}000$ iterations per simulation (across 40 seeds, 40 million total Metropolis steps), a burn-in of $50{,}000$ steps, and averaging the results over $30$ random seeds for each data point. We continue using the Gaussian proposal with variance as described in \eqref{eq:gaussian_proposal}.

\begin{figure}[htbp]
\centering

\begin{subfigure}[t]{0.4\textwidth}
    \includegraphics[width=\linewidth]{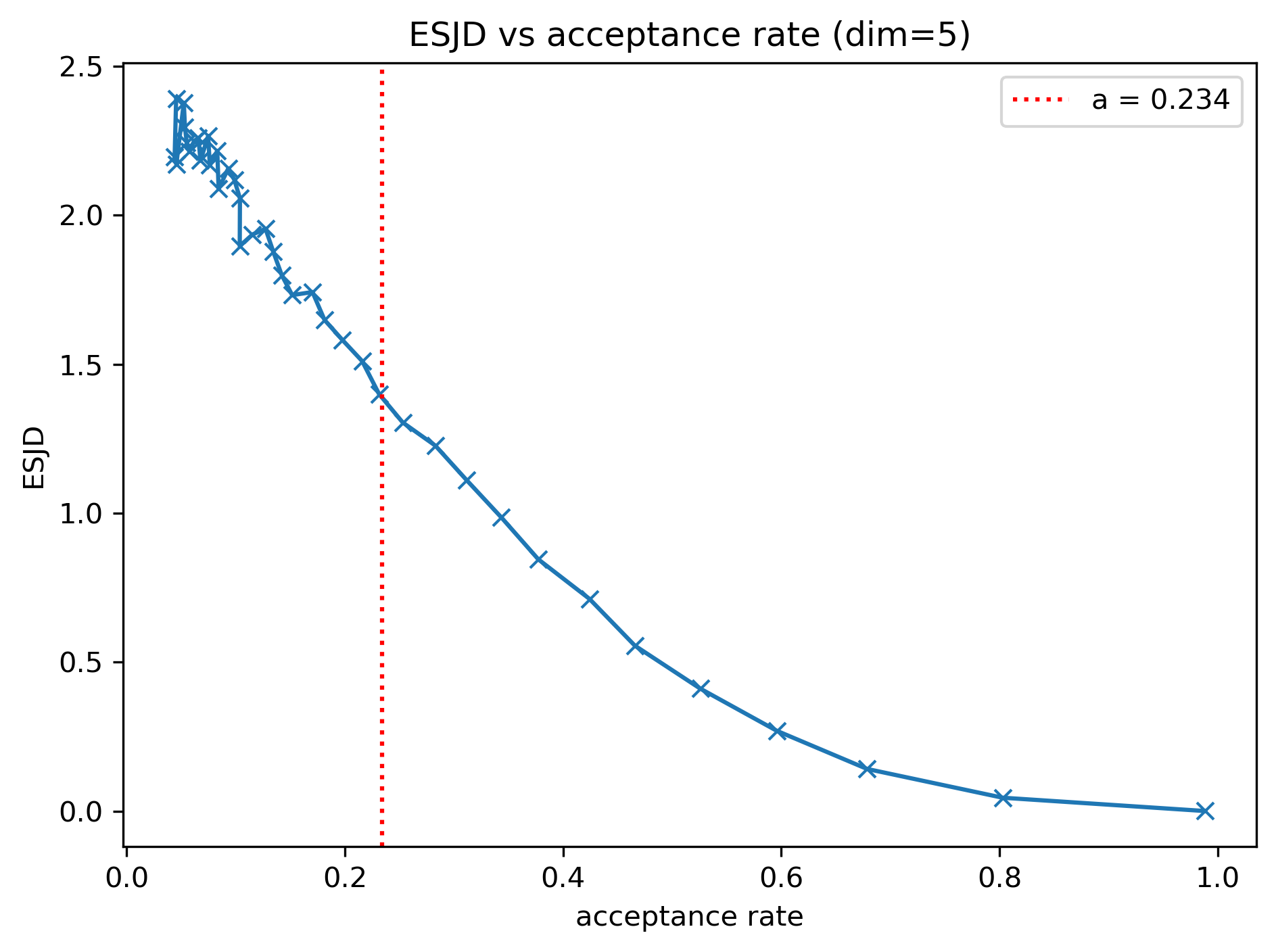}
\end{subfigure}
\begin{subfigure}[t]{0.4\textwidth}
    \includegraphics[width=\linewidth]{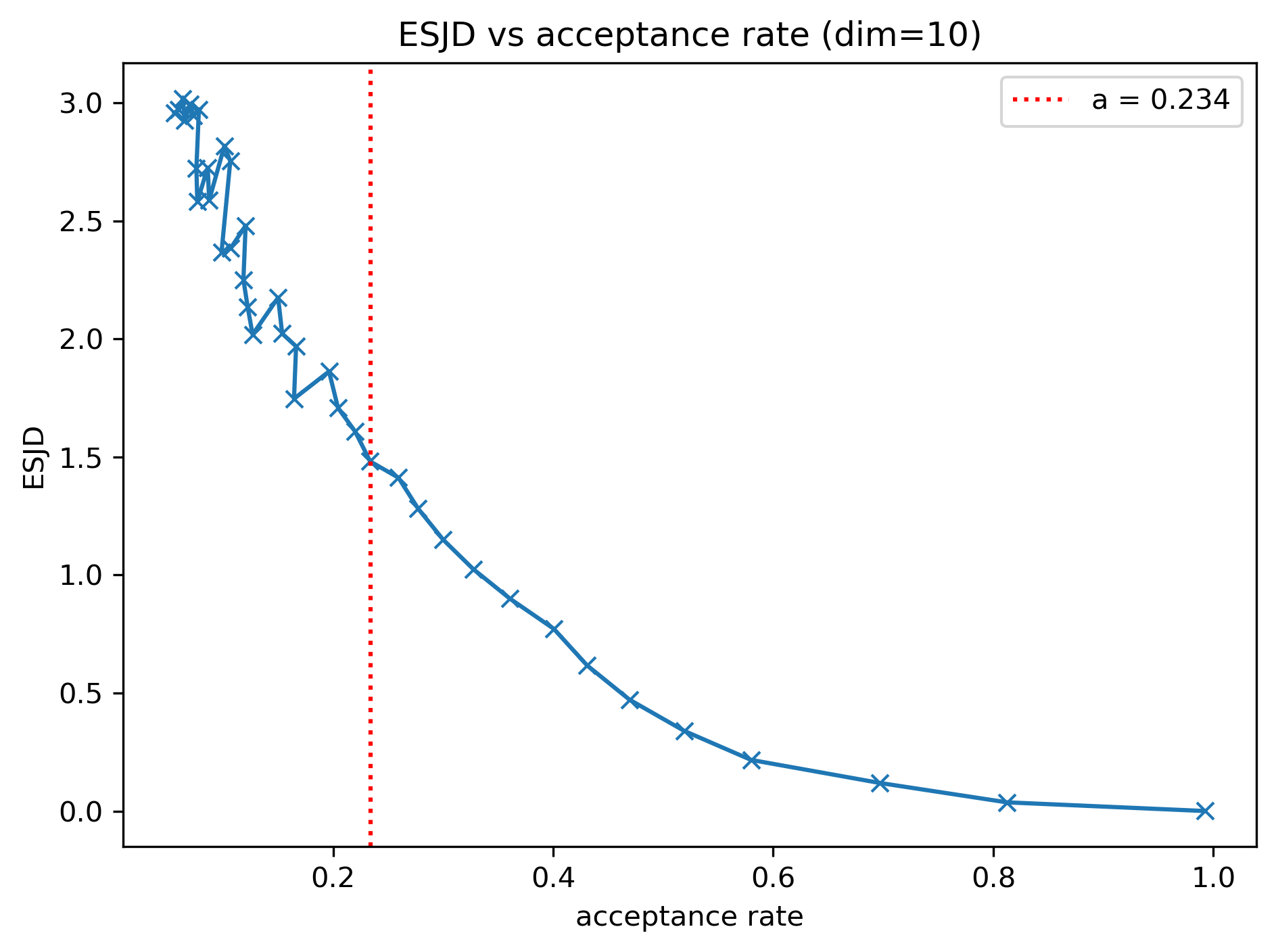}
\end{subfigure}

\vspace{0.5em}

\begin{subfigure}[t]{0.4\textwidth}
    \includegraphics[width=\linewidth]{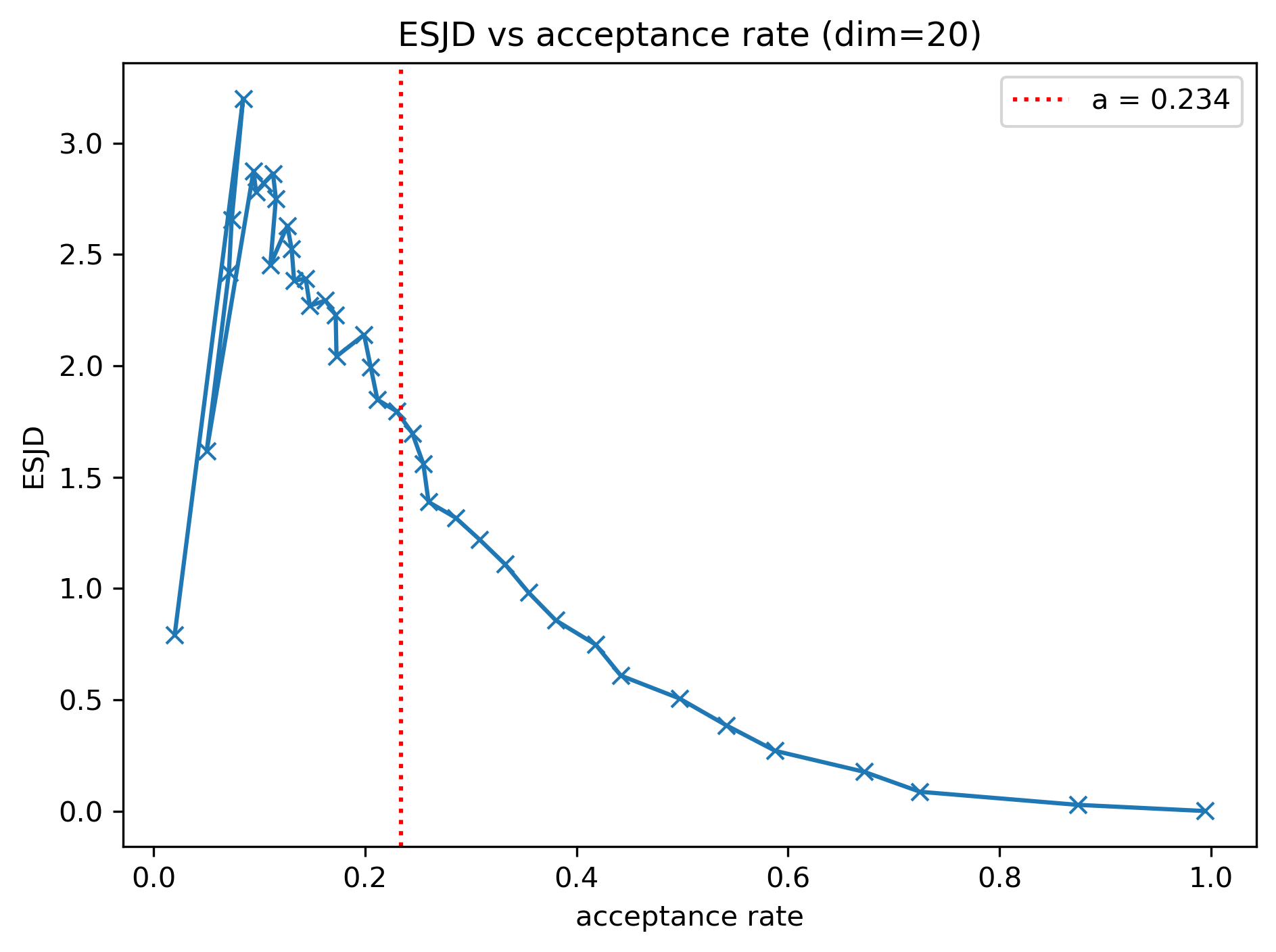}
\end{subfigure}
\begin{subfigure}[t]{0.4\textwidth}
    \includegraphics[width=\linewidth]{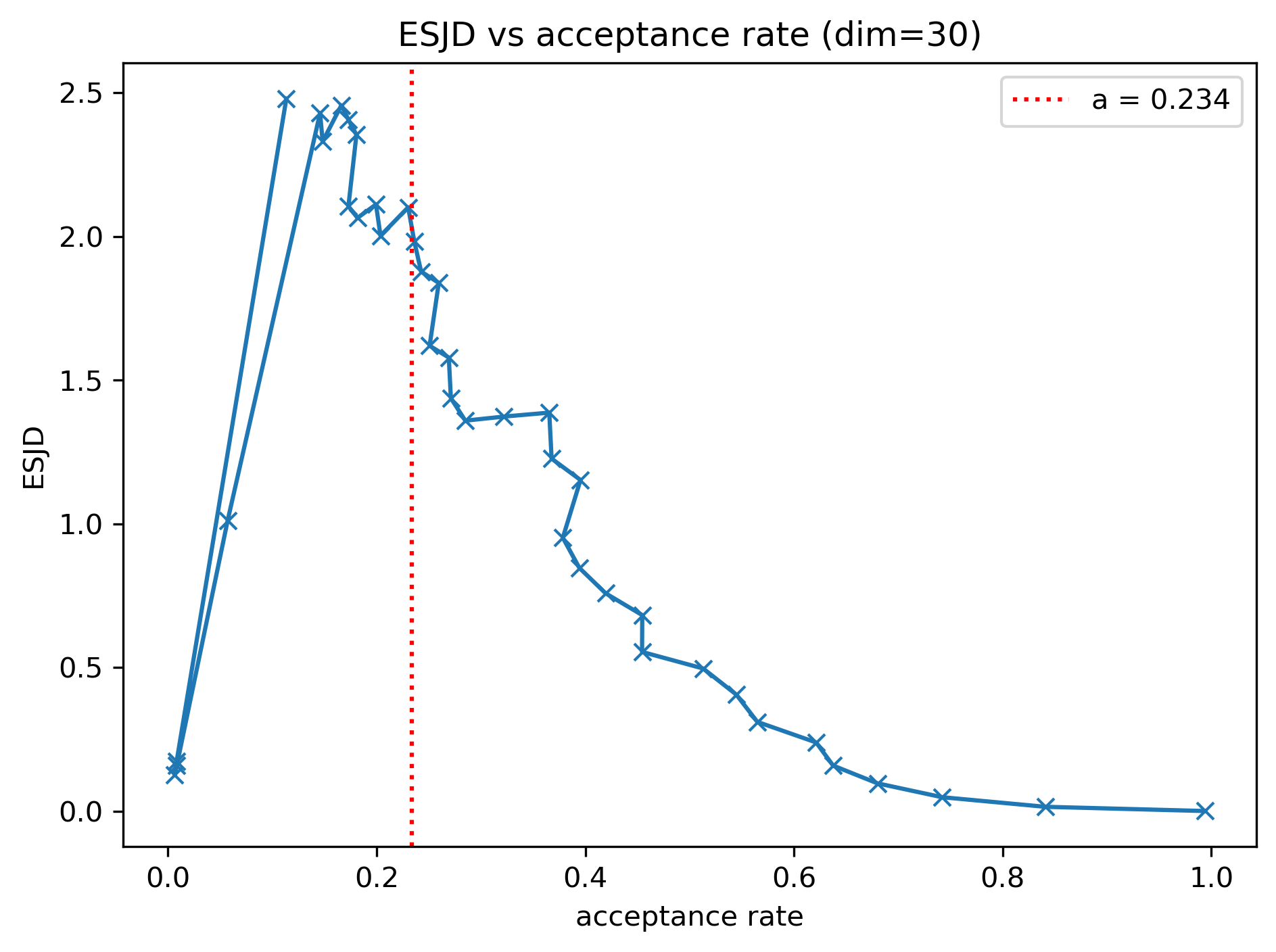}
\end{subfigure}

\caption{ESJD vs.\ acceptance rate for the Neal's funnel target $\pi_{12}$ under RWM with a Gaussian proposal in dimensions $d \in \{5, 10, 20, 30\}$ from top-left to bottom-right. Red dotted line indicates an acceptance rate of 0.234.}
\label{fig:esjd_neal_funnel}
\end{figure}

Similarly to the 2-dimensional Rosenbrock, we find that the ESJD increases monotonically as the acceptance rate reduces in lower dimensions. In higher dimensions ($d=20, 30$), we observe a similar trend, but at the extreme ends of the curve, we observe a complete failure of the RWM algorithm to accept any reasonable number of steps as the average acceptance rate decreases towards zero, leading to sudden random drops in ESJD. Notice that the curvature of the graph gets steeper with increasing dimension: it is nearly a straight line at $d=5$ and is much more curved by $d=30$.

Overall, this behaviour indicates an extreme sensitivity to the proposal scaling in higher dimensions. The practical implication is that even minor mistuning of the proposal variance will cause the sampler's efficiency to collapse to zero. This is a direct consequence of the target's geometry violating the conditions of Assumption~\ref{assump:sherlock2013}; the target has extreme eccentricity, and the Hessian is not locally stable.
Consequently, the 0.234 acceptance rate is not a relevant or useful heuristic for this target. The more important conclusion, which \citet{neal_slice_2003} himself meant to show with this example, is that for targets with strong non-linear dependencies, no single proposal scale can be efficient in both the narrow neck and the wide mouth of the funnel, leading to a sampler that inevitably fails as dimensionality grows. This experiment effectively demonstrates a hard boundary for the class of problems on which the 0.234 RWM heuristic can be successfully applied.

\section{Parallel Tempering Background}
\label{sec:pt_background}

This section offers a brief introduction to the parallel tempering method and its own optimal scaling framework that we conduct experiments on in Section \ref{sec:pt-simulations}. Many of the terms in Section \ref{sec:rwm_background} are re-used but have slightly different definitions for the parallel tempering context.

\subsection{Parallel Tempering Method}
The parallel tempering method \citep{Geyer1991MCMCMCdef}, also known as the Metropolis-coupled Markov chain Monte Carlo algorithm or the replica exchange method, is a version of the Metropolis algorithm that specialises in sampling from multimodal distributions. In this case, we sample from a target distribution with density $\pi(\bx)$ on a state space $\mathcal{X}$.
We define a sequence of tempered target densities $\pi^{\beta_j}(\bx)$ where $0 \leq \beta_n < \beta_{n-1} < \dots < \beta_1 < \beta_0 = 1$ are the inverse temperature values. We require that $\pi^{\beta_0}(\bx) = \pi(\bx)$ (that is, that $\beta_0 = 1$). 
Typically, the tempered target densities are simply powers of the original density, and we assume this for all parallel tempering experiments.
Then, we run a Markov chain for each of the $n+1$ values of $\beta$, each with its own tempered target density. Eventually, the $\beta_0 = 1$ chain, also called the ``cold'' chain, with its stationary density $\pi^{\beta_0}(\bx_0)$ should correspond to the original target density of interest $\pi(\bx)$, and the idea is that using information from other chains with higher temperatures (lower $\beta$) will speed up the mixing in the cold chain.

In each iteration of the algorithm, we alternate between \textit{within temperature} and \textit{temperature swap} moves. The former indicates that each chain takes a standard RWM update within its respective $\pi^{\beta_j}$. The latter indicates that for two chains with $\beta_j, \beta_k$ inverse temperature values that are adjacent in the sequence of $\beta$ values, switch the chain values of $\bx_j$ and $\bx_k$ with probability 

\begin{equation}
    \label{eq:swap-prob}
    \min \left\{1, \dfrac{\pi^{\beta_j}(\bx_k) \pi^{\beta_k}(\bx_j)}{\pi^{\beta_j}(\bx_j)\pi^{\beta_k}(\bx_k)}\right\}.
\end{equation}

\subsection{Optimal Scaling for Parallel Tempering} \label{subsec:pt_optimal_scaling_background}

Since the hot chains can explore the state space more quickly, we would like to maximise how frequently we can swap values from the hottest chain to the coldest chain so that the cold chain can mix faster and escape local modes. 
To do this, we maximise the effective speed with which the chain values move along in the inverse temperature domain.
The spacings of the inverse temperatures $\beta_i$'s are crucial to this efficiency.
If $\beta_j$ and $\beta_k$ are too far apart, we usually reject these swaps described by \eqref{eq:swap-prob}, but if they are too close, the swaps will not improve mixing. 
We would like to swap just the right amount to maximise mixing in the temperature domain.

The \textit{expected squared jumping distance} (ESJD) for parallel tempering thus refers to the expected squared jump in inverse temperatures.
Formalising this, when we attempt to swap the chain values between the inverse temperatures $\beta$ and $\gamma := \beta + \epsilon$ where $\beta, \epsilon > 0 $ and $\beta, \gamma \leq 1$, the swap is either accepted, in which case the values move a squared distance of $(\gamma - \beta)^2 = \epsilon^2$, or the swap is rejected, in which case the distance moved is 0. This leads to a very natural definition for ESJD, which is
\begin{equation}
    ESJD = \E[(\gamma - \beta)^2] = \epsilon^2 \times \E[\Pr(\text{swap accepted})].
\end{equation}
Maximising the asymptotic ESJD effectively maximises the efficiency of the attempted swap moves in providing mixing in the temperature space, or in other words, maximises the speed with which the chain values move in the inverse temperature space.

There is also a 0.234 swap acceptance rate theoretical result for parallel tempering. The required assumptions for the result are the following:

\begin{assumption}
\label{assump:atchade2011}
(\citet{atchade_towards_2011})
Let $\bx$ be a random variable from the target density $\pi$.
\begin{enumerate}
    \item The target density is an i.i.d.\ product $\pi(\bx) = \prod_{i=1}^d f(x_i)$ where $f$ is a single-dimensional component density. There are other smoothness and regularity restrictions on the target density taken from the main results of \citet{gelman_gilks_roberts_1997, roberts_rosenthal_2001}. 
    \item The tempered distributions are powers of the original density, so $\pi^{(\beta)}(\bx) = \prod_{i=1}^d f^{(\beta)}(x_i) = \prod_{i=1}^d (f(x_i))^{\beta}$.
    \item The inverse-temperature spread $\epsilon = d^{-1/2}\ell$ for some positive constant $\ell$. 
\end{enumerate}
\end{assumption}

Under these assumptions, choosing $\ell$ such that the inverse temperature spacing yields a swap acceptance probability of approximately 0.234 is also optimal in the sense that it maximises the ESJD of a parallel tempering algorithm.

\section{Parallel Tempering Simulations} \label{sec:pt-simulations}
In this section, we provide parallel tempering experiments that investigate how well the 0.234 optimal swap acceptance rate holds for the various multimodal target distributions in Section \ref{subsec:multimodal} in finite dimensions. The parallel tempering method is highly effective for exploring multimodal distributions \citep{atchade_towards_2011, hastings_monte_1970} and thus we aim our experiments at these distributions to learn more about their applicability in practice.

\subsection{Target Distributions}

\subsubsection{Multivariate Gaussian Distribution}
We first give an example of the standard multivariate Gaussian distribution to verify the correctness of our implementation. 
Note that the standard multivariate Gaussian as a target distribution satisfies the necessary conditions of \citet{atchade_towards_2011} that are stated in Assumption \ref{assump:atchade2011}. 

\subsubsection{Rough Carpet Distribution}

In this section, we use the same target distribution $\pi_4$ from (\ref{eq:target-rough-carpet}) as described in Section \ref{sec:rough-carpet}.
Here, we examine the simpler case with homogeneous scaling factors $f_d(\bx) = \prod_{i=1}^n f(x_i)$. In the individual component density $f(x)$ definition, we set $(\mu_1, \mu_2, \mu_3) = (-15,0,15)$.
We provide an example traceplot of the cold chain from one of the experiment simulations in Appendix \ref{app:plots} Figure \ref{fig:pt-trace-rough} to demonstrate what mixing looks like in the first three components (all components are i.i.d.). 
Unlike the standard random-walk Metropolis which cannot escape from the central mode in this setting of the target distribution, the values in the cold chain swap between modes considerably often due to the swapping of values between chains.

\subsubsection{Three Mixture Distribution}

In this section, we use the same target distribution as described in Section \ref{sec:3mode} \eqref{eq:3mode}. We set weights $w_1=w_2=w_3=\frac{1}{3}$, means $\boldsymbol{\mu}_1 = (-15, 0, 0, \dots, 0)$, $\boldsymbol{\mu}_2 = (0, 0, \dots, 0)$, $\boldsymbol{\mu}_3 = (15, 0, 0, \dots, 0)$ and covariances $\Sigma_1 = \Sigma_2 = \Sigma_3 = I_d$.
We provide an example traceplot of the cold chain in Appendix \ref{app:plots} Figure \ref{fig:pt-trace-3mix} to demonstrate what mixing looks like in the first three components.

\subsection{Methodology} \label{subsec:pt-methods}

We conduct experiments in dimension $d \in \{10, 20, 30\}$ that examine the trend of ESJD with swap acceptance rate on these distributions. 
For each distribution, we run many simulations where each simulation is a parallel tempering algorithm with 30 different average swap acceptance rate values. 
We explain how we set up these algorithms to match these swap rates (through constructing a specific inverse temperature ladder) in Section \ref{subsec:constructing-ladder}. 
Each parallel tempering algorithm runs for 500,000 iterations, so every individual chain in the algorithm takes 500,000 steps. We attempt a temperature swap every 100 steps.
Furthermore, we ran each algorithm instance over 20 different seeds and average the results to reduce the effects of randomness caused by a particular seed.

For each experiment, we set the tempered target distribution to simply be the original distribution raised to the power of $\beta$. We set the modes to be far away enough such that the other modes were not reached by a standard random-walk Metropolis algorithm after 100,000 iterations, but still reasonably close so that the parallel tempering method would show the cold chain values would swap between modes somewhat frequently.
The within-chain proposal distribution for each chain is given by the multivariate Gaussian \eqref{eq:gaussian_proposal} with $\sigma^2 = 2.38^2/d$, which is  suggested by \citet{gelman_gilks_roberts_1997} to be optimal for high dimensions. We calculate the optimal swap acceptance rate by finding each simulation's specified swap acceptance rate used to construct the iterative temperature ladder that maximises the ESJD, and take the average of the maximising acceptance rates over 20 seeds.

\subsubsection{Constructing an Inverse Temperature Ladder Iteratively}\label{subsec:constructing-ladder}

The most common method to construct an inverse temperature ladder selects the inverse temperatures using a geometric series spacing. 
However, since we are examining the 0.234 swap acceptance rule, we need to have a way of constructing the inverse temperature spacings such that the probability of a swap between adjacent chains is approximately 0.234 (or any other value). 
To construct an inverse temperature ladder with a desired swap acceptance rate $s$, we use an iterative procedure adapted from \cite[Section~2.2]{atchade_towards_2011}. 
Starting with $\beta_0 = 1$ and a minimum value $\beta_{min} = 0.01$, we iteratively add $\beta$'s to the ladder. Let $\beta_{curr}$ denote the most recent $\beta$ in the ladder. 
We initialize $\rho_n = 0.5$ ($n=1$ initially) and set $\beta^* = \beta_{curr} (1 + e^{\rho_n})^{-1}$. 
To determine if $\beta^*$ should be added to the ladder, we draw $N$ samples (our GPU implementation uses $N=\,$1,000,000) from the target distribution tempered by $\beta^*$ and $\beta_{curr}$, and calculate the average swap probability $a$. 
If $a$ is within $s \pm 0.0001$, we add $\beta^*$ to the ladder and set $\beta_{curr} = \beta^*$. Otherwise, we update $\rho_{n+1} = \rho_{n} + n^{-0.25} (a - s)$ and recalculate $\beta^*$. 
This process continues until $\beta^* \leq \beta_{min}$, at which point we add $\beta_{min}$ to the ladder and terminate.

\subsection{Results}

\begin{figure}[htbp]
\centering

\begin{subfigure}[t]{0.32\textwidth}
    \includegraphics[width=\linewidth]{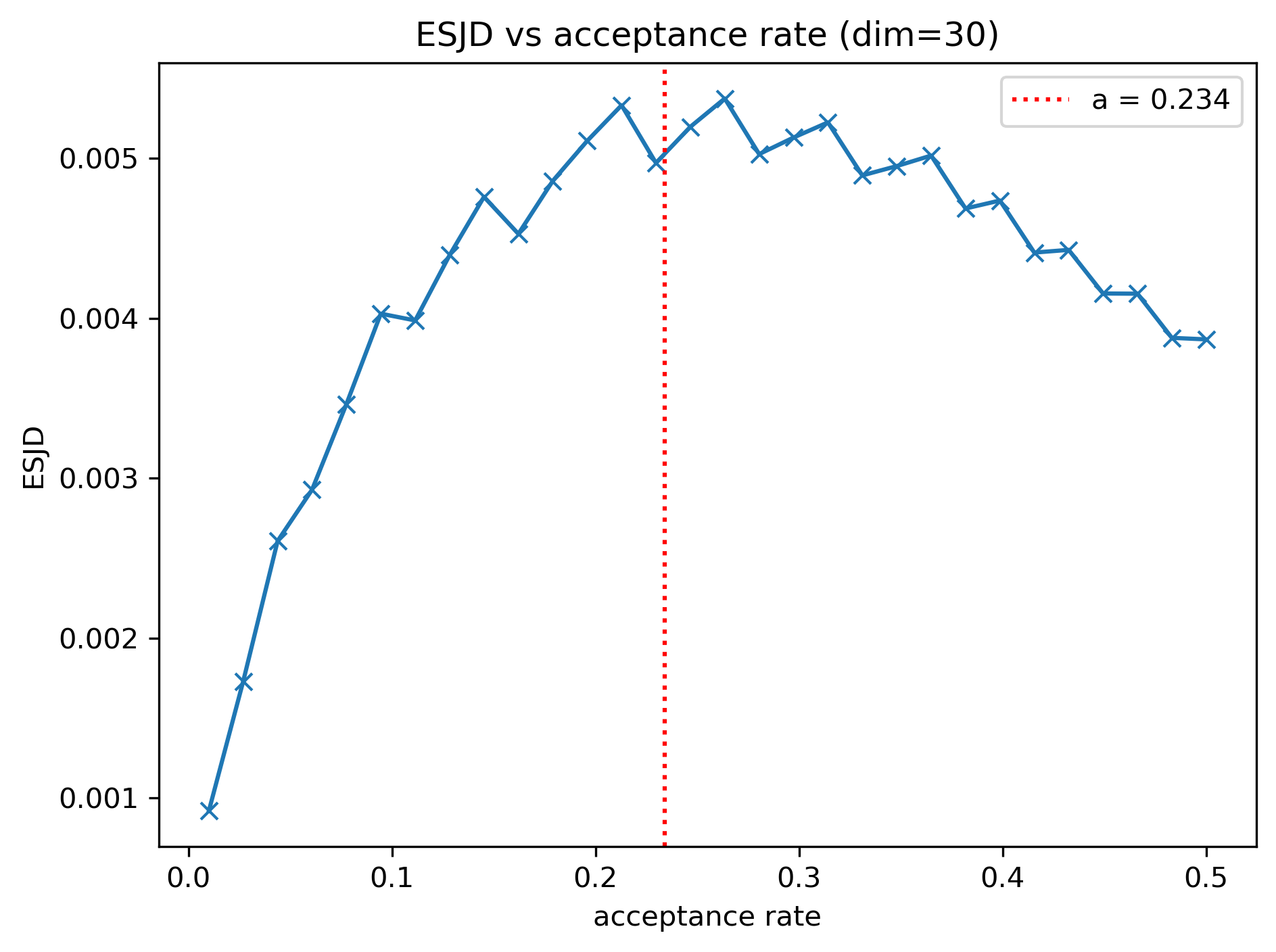}
    \caption{$\pi_2$: Gaussian}
\end{subfigure}
\hfill
\begin{subfigure}[t]{0.32\textwidth}
    \includegraphics[width=\linewidth]{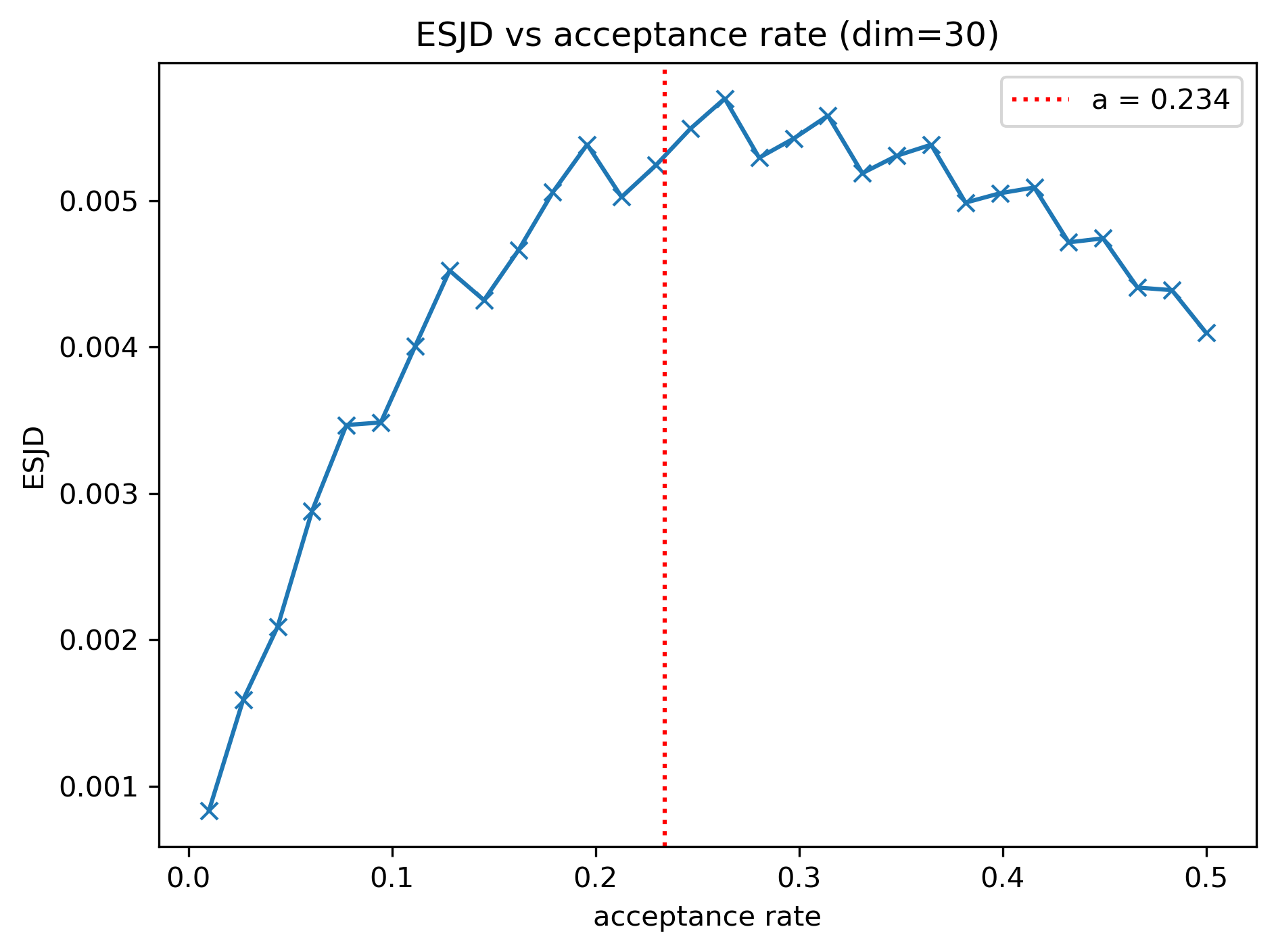}
    \caption{$\pi_4$: Rough Carpet}
\end{subfigure}
\hfill
\begin{subfigure}[t]{0.32\textwidth}
    \includegraphics[width=\linewidth]{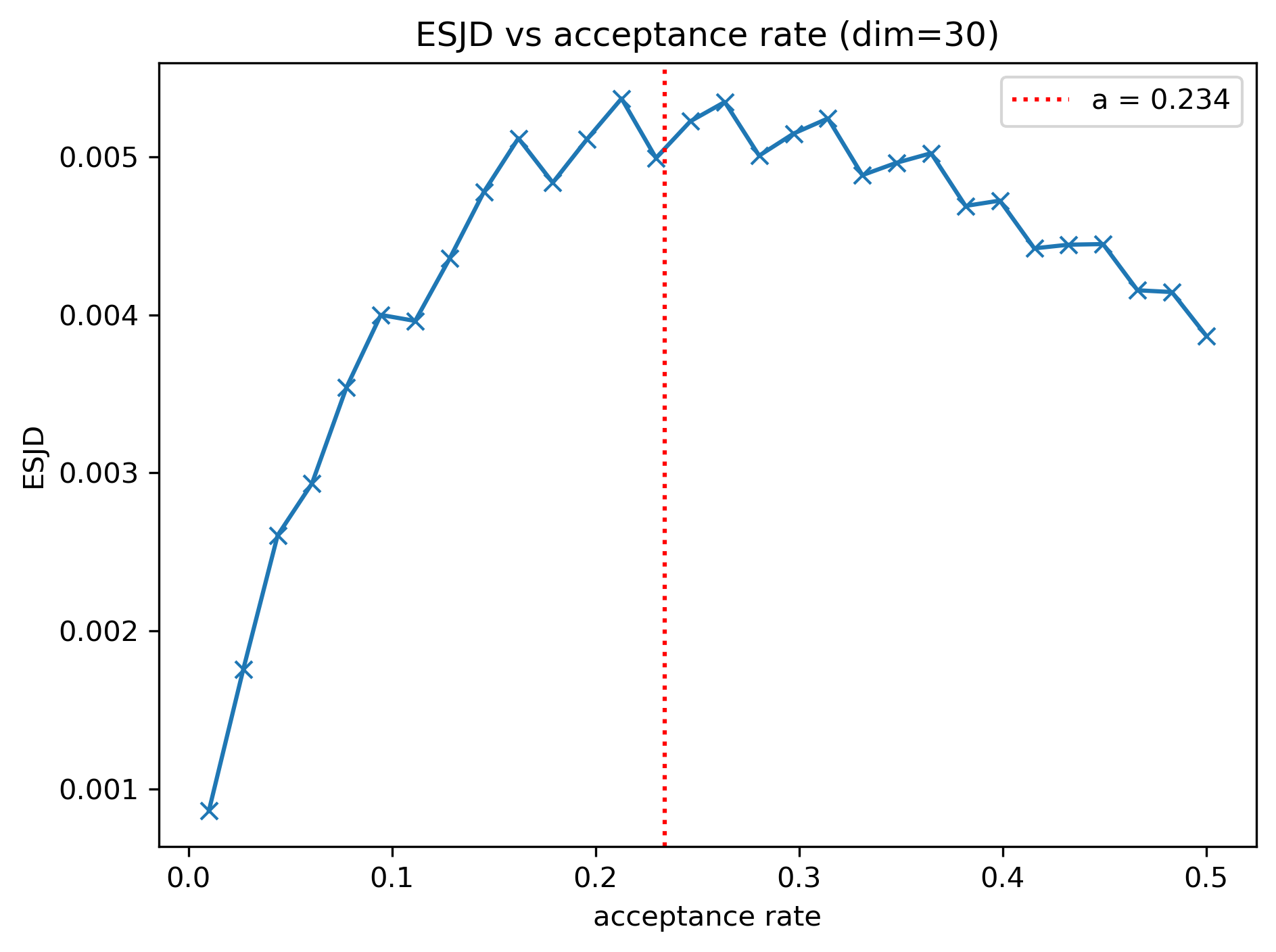}
    \caption{$\pi_6$: 3-Mixture}
\end{subfigure}

\caption{ESJD vs.\ swap acceptance rate for the parallel tempering algorithm with the $\pi_2, \pi_4, \pi_6$ target distributions in dimension $d=30$. Red dotted line indicates a swap acceptance rate of 0.234.}
\label{fig:pt_d30_esjd}
\end{figure}

The empirical optimal swap acceptance rates for the three target distributions are summarized in Table \ref{tab:pt_swap_rates}, and sample ESJD curves for $d=30$ shown in Figure \ref{fig:pt_d30_esjd}. The ESJD curves for PT are visibly more jagged than their RWM counterparts. This might be because the estimation of ESJD in this context is more susceptible to Monte Carlo error even with a large number of simulation steps, since the efficiency or acceptance rate of a swap move depends on the joint stationary distribution of two adjacent chains, and this is repeated for each temperature construction. Another possible reason might be the determinism in the way $\rho_n$ changes. Future work could investigate alternative algorithms to compute a fixed discrete temperature ladder that provides a certain swap rate prior to running the PT algorithm, perhaps by computing each change in $\rho_n$ from a distribution. Nonetheless, clear trends emerge from the results.
    \begin{table}[htbp]
\centering
\caption{Empirical swap acceptance rates that maximize ESJD for various PT target densities. Based on Fig \ref{fig:pt_d30_esjd}, the jagged pattern of the ESJD curves suggests that the true optimal acceptance rates are in a neighbourhood of these values. See Section \ref{subsec:pt-methods} for calculation of optimal swap acceptance rates.}
\label{tab:pt_swap_rates}
\begin{tabular}{lccc}
\toprule
\textbf{Target} & \(d=10\) & \(d=20\) & \(d=30\) \\
\midrule
$\pi_2$: Standard Multivariate Gaussian       & 0.2618 & 0.2685 & 0.2474 \\
$\pi_4$: Rough Carpet                & 0.2660 & 0.2803 & 0.2761 \\
$\pi_6$: 3-Mixture                   & 0.2609 & 0.2660 & 0.2372 \\
\bottomrule
\end{tabular}
\end{table}

For both the standard Gaussian ($\pi_2$) and the non-product Three-Mixture ($\pi_6$) targets, the optimal swap acceptance rate is consistently in the neighbourhood of 0.234. For the Three-Mixture target, the rate converges cleanly towards the theoretical value, from 0.261 in $d=10$ to 0.237 in $d=30$. The Gaussian target shows a similar pattern, with an optimal rate of 0.247 at d=30.

The standard Gaussian result provides a successful validation of our experimental setup against the theory of \citet{atchade_towards_2011}, as this target satisfies the i.i.d. product form assumption. More importantly, the result for the Three-Mixture target suggests that the 0.234 heuristic for PT swaps is robust to violations of the i.i.d. product assumption, much like its RWM counterpart. Although the target is a sum of densities, its underlying geometric regularity (a shell composed of three isotropic modes) seems sufficient for the optimal scaling 0.234 result to hold. For a practitioner, this suggests that constructing a temperature ladder with a swap acceptance probability of 0.234 is a viable and likely near-optimal strategy, even for complex, non-product multimodal targets, provided they are not pathologically structured.

Notably, the Rough Carpet target ($\pi_4$) shows a deviation. The optimal swap acceptance rate is consistently higher than 0.234 and increases away from the theoretical value with increasing dimension. This is surprising because the Rough Carpet target does satisfy the i.i.d. product form and other conditions of Assumption \ref{assump:atchade2011} by \citet{atchade_towards_2011}.

There are two plausible explanations that we favour here. The failure of the 0.234 rule here could be because the extreme multimodality of the target could reduce the practical utility of the "hot" chains. The target possesses $3^d$ modes, and its probability mass is spread thinly across this vast landscape. For a hot chain where the inverse temperature $\beta$ is close to 0, the tempered density becomes extremely flat and loses the slight peak that a unimodal target would have. Consequently, it fails to efficiently sample from the specific modal regions that are of interest to the colder chains; the information flow from hot to cold chains is broken. To swap successfully more often, the temperature steps must be made smaller, which forces the swap acceptance rate to be higher.

An alternative and more simple explanation is that, due to the jaggedness of the ESJD vs swap acceptance rate trend in Fig \ref{fig:pt_d30_esjd}, the discrete temperature ladder is only approximately optimal, and the true empirically optimal swap rate is somehow confounded by determinism in the ladder construction algorithm. As mentioned previously, alternative methods to construct temperature ladders yielding certain swap acceptance rates prior to running a PT algorithm could be an area for future work.

\section{Discussion}\label{sec:discussion}

We have presented an extensive empirical investigation into the generalizability of the 0.234 acceptance rate heuristic for both Random-Walk Metropolis (RWM) and Parallel Tempering (PT) algorithms, using a variety of target and proposal distributions.
Our results demonstrate that the 0.234 rule's utility extends beyond the classical i.i.d. product setting of \citet{gelman_gilks_roberts_1997}. The heuristic remains approximately optimal for: (i) non-Gaussian Laplace and Uniform proposal distributions that satisfy a ``shell'' condition with a less stringent symmetry requirement than spherical symmetry; (ii) multimodal targets such as the ``rough carpet'' i.i.d product with $3^d$ modes, and non-product targets constructed as a sum of isotropic components, such as our Three-Mixture model; and (iii) even for highly correlated, non-linear targets like the Full and Hybrid Rosenbrock in sufficiently high dimensions ($d \ge 20$). These findings provide empirical validation for the generalized theory of \citet{sherlock_optimal_2013}, which predicts that as long as the target's geometry is sufficiently regular and isotropic in the high-dimensional limit, the 0.234 rule should hold. Our results in Section \ref{subsec:non-gaussian-proposal} even suggest that weaker conditions on the symmetry of the proposal than those required by \citet{sherlock_optimal_2013} could be sufficient.

That being said, there are cases where 0.234 does not seem as relevant. The main case is in very low dimensions; in all our experiments with varying target densities, the optimal acceptance rate strays from 0.234 below a certain threshold which varies for each target density. 
In particular, targets composed of independent, low-dimensional anisotropic blocks (Even Rosenbrock) or those that are globally anisotropic in low dimensions (2D Rosenbrock), have an optimal acceptance rate significantly below 0.234. In these cases, we hypothesize the high-dimensional averaging effect that stabilizes the target's geometry has not occurred, and the sampler's performance is dictated by the challenging local geometry. As the dimension increases, the law of large numbers mitigates the effect of any single scaling factor or local correlation, so the target becomes effectively less eccentric, complying with the theory of \citet{sherlock_optimal_2013}, and the optimal acceptance rate recovers towards 0.234. On the other hand, the continuous hypercube and Neal's funnel completely fail the 0.234 heuristic. Although the hypercube is continuously differentiable everywhere except on its boundary, which has zero measure, an intuitive argument is that as $d$ increases, the probability that the proposal has all the values inside the hypercube decreases. As for the funnel, no single proposal scaling can be globally efficient, which is just a limitation of the RWM algorithm.

Next, we discuss the optimal swap acceptance rate figure for parallel tempering and compare it with the optimal acceptance rate for random-walk Metropolis. 
On one hand, the results for the multivariate Gaussian and three-mixture density examples show that the 0.234 swap acceptance rate figure may be optimal even in lower dimensions. 
Although costly, if the practitioner is willing and able to construct the inverse temperature ladder iteratively with a spacing given by a swap acceptance probability, the spacing dictated by a 0.234 swap acceptance probability may be optimal for the practitioner. 
However, the theoretical 0.234 figure may not always be applicable: 
despite satisfying the i.i.d. conditions of the PT optimal scaling theory \citep{atchade_towards_2011}, the extreme multimodality ($3^d$ modes) of the ``rough carpet'' product density appears to render the hot chains ineffective  in our experiments, with the caveat that the apparent failure of the 0.234 rule could partly be due to the jaggedness of our experimental ESJD vs swap acceptance rate curves. One hypothesis for this observed failure is that because the probability mass is spread so thinly, the hot chains fail to sample from meaningful modal regions, breaking the flow of information to the cold chain. This forces the algorithm to use smaller temperature steps (and thus higher swap acceptance rates) to achieve mixing, representing a practical failure of the 0.234 PT rule.

Not only that, constructing a temperature ladder iteratively with the “optimal spacing” is challenging and time-consuming. 
It requires the ability to draw samples from the target distribution, which is itself the main goal of the parallel tempering algorithm. 
Without a direct way to sample from the target, a practitioner could run an MCMC algorithm with the $\beta$ value being proposed for the new addition to the ladder, but this may be time-consuming with many attempts required until a suitable $\beta$ is found for each step of the ladder.

Since this optimal spacing may not yield the optimal ESJD in practice for finite, lower-dimensional targets and with a finite number of runs, an alternative approach is to adaptively adjust the spacings of the temperature ladder during the algorithm’s runs based on its recent ESJD performance over a recent window as guidance to practitioners. 
This is known as adaptive parallel tempering \citep{Miasojedow_2012_adaptivePTMCMC}.
Previous literature shows that adaptive parallel tempering algorithms can achieve substantial efficiency gains over a standard geometric temperature spacing \citep{vousden_dynamic_2015} while still converging to the target distribution \citep{Miasojedow_2012_adaptivePTMCMC} under the assumptions that the target distribution is sufficiently regular and has tails decaying faster than exponentially. 
Adaptive parallel tempering may suffice to be ``good enough'' for most practitioners whilst avoiding the costly temperature construction algorithm. 

Lastly, while we have established that the 0.234 acceptance rate for RWM is still fairly good even in lower dimensions and on some target distributions that may not have the i.i.d.\ product form, we provide some precautions to the practitioner. 
The 0.234 acceptance rate is a good heuristic to tune the algorithm to, yet there are still many cases where it is not optimal in terms of efficiency, such as the continuous hypercube example in Section \ref{subsec:hypercube}.
Even if the 0.234 acceptance rate is in fact optimal, the acceptance rate alone does not necessarily guarantee good samples, as illustrated by Figure \ref{fig:10m no escape}. 
What literature \citep{roberts_rosenthal_2001, sherlock_theoryandpractice_2010} indicates is that another very important thing in practice beyond just the acceptance rate is defining a good proposal covariance matrix $\Sigma_d$ that is approximately proportional to the target covariance matrix $\Sigma_\pi$; doing so can significantly increase the asymptotic relative efficiency. 
Realistically, it may be impossible to know $\Sigma_\pi$ in advance or even provide a good estimate for it. 
In fact, understanding how we can extend this result to correlated targets is an ongoing challenge \citep{jun_yang_2020} and immediately relevant to the practitioner since real-world data tends to have correlations between variables. 
Practitioners may consider using the adaptive Metropolis algorithm \citep{haario_heikki_saksman_2001} to dynamically update the algorithm's proposal covariance matrix. 
This adaptive algorithm is very useful and yields significant speed benefits in low and high dimensions \citep{craiu_2009_adaptiveMCMC, roberts_rosenthal_2009_adaptiveMCMC_examples}, but an important potential pitfall to note is that adaptive methods are not guaranteed to converge to the correct target density; in many cases, they may fail to converge or converge to something completely different \citep{roberts_rosenthal_2005_ergodicity, roberts_rosenthal_2009_adaptiveMCMC_examples}.

\section*{Acknowledgements}
Resources used in preparing this research were provided, in part, by the Province of Ontario, the Government of Canada through CIFAR, and companies sponsoring the Vector Institute. 
We also thank the anonymous reviewer(s) for their helpful comments on earlier versions of this paper.

\section*{Declarations}

The authors did not receive any funding for this submitted work and have no relevant financial or non-financial competing interests. 
The code implementing the experiments and generating the figures used in this work are publicly available on GitHub at 
\url{https://github.com/aidanmrli/rwm-pt-pytorch}.

During the preparation of revisions, the authors used generative AI large language models (Gemini 2.5 Pro Preview 05-06, Claude 4.0 Sonnet) to assist with coding and manuscript revision. Generative AI was utilized to help generate, debug, and optimize code for the simulations. The new generated code based on PyTorch allows for the use of Graphics Processing Units (GPUs) when running RWM and PT algorithms to significantly speed up simulations and run more simulations at once. For manuscript revision, Generative AI was used to improve phrasing and clarity on text written by the authors. The authors thoroughly reviewed and edited all AI-generated output and take full responsibility for the scientific integrity and final content of this publication.   

\bibliography{main.bib}
\appendix

\section{Plots for RWM simulations} \label{app:plots}

\begin{figure}[htbp]
\centering

\begin{subfigure}[t]{0.32\textwidth}
    \includegraphics[width=\linewidth]{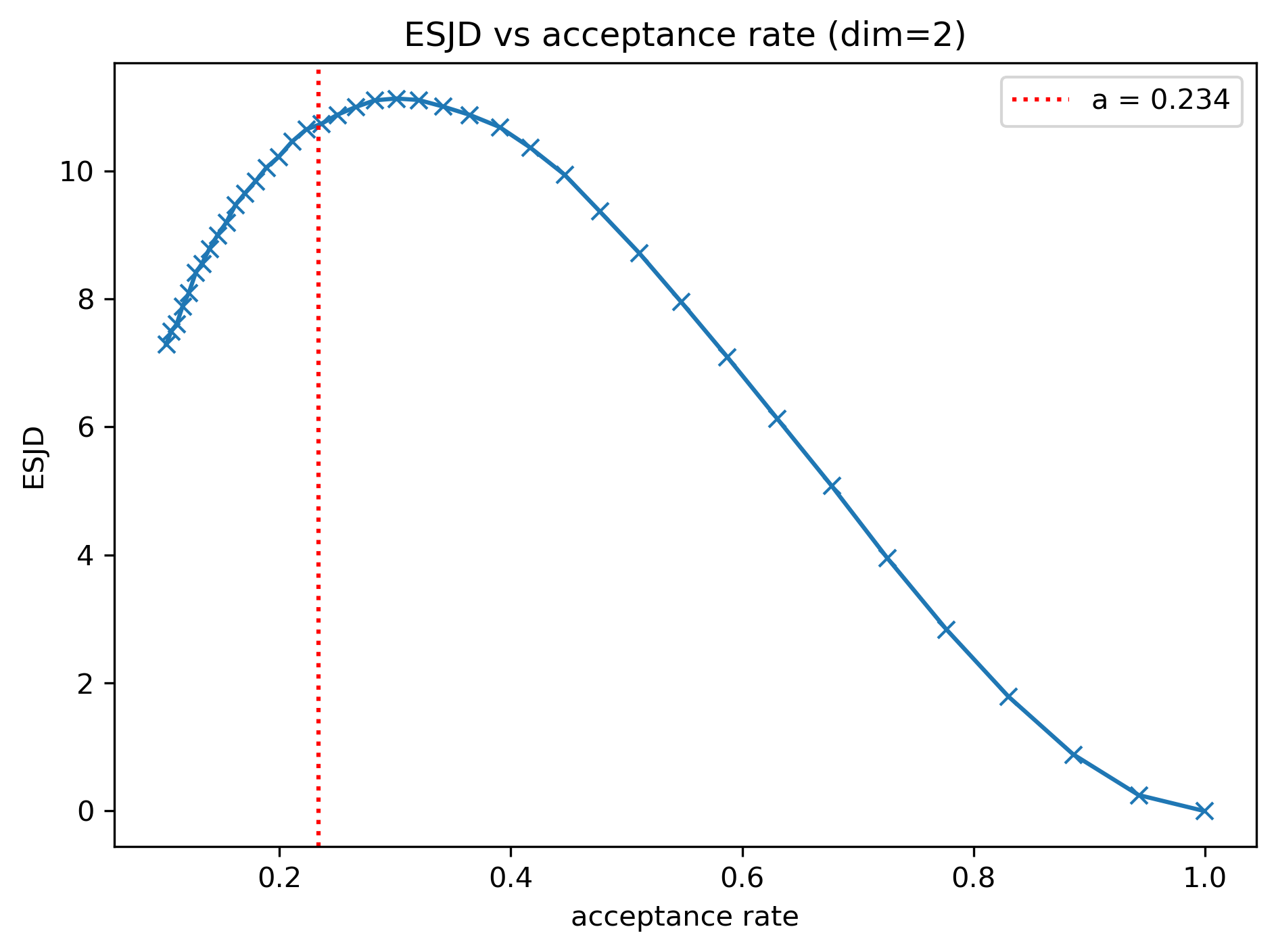}
\end{subfigure}
\hfill
\begin{subfigure}[t]{0.32\textwidth}
    \includegraphics[width=\linewidth]{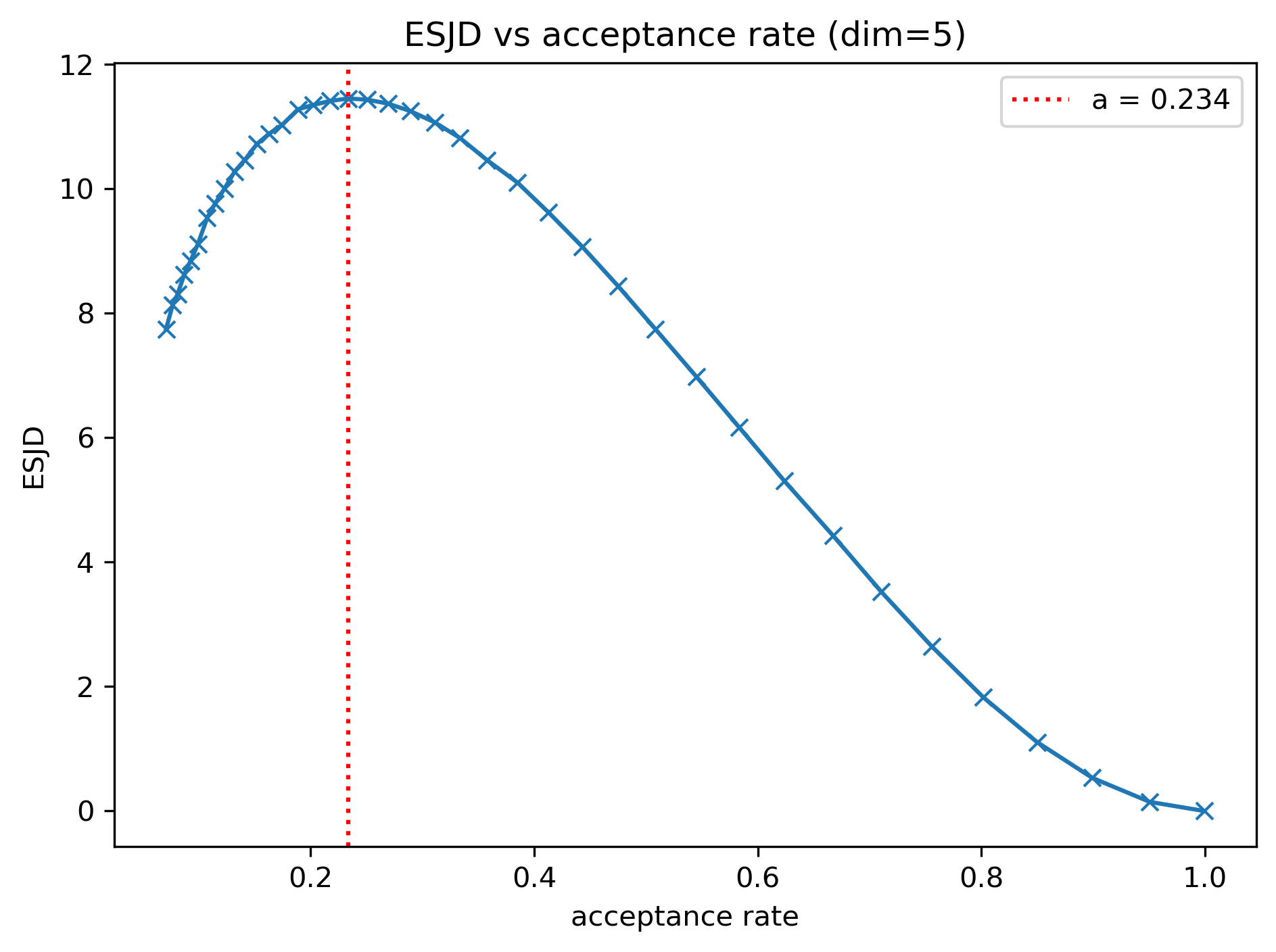}
\end{subfigure}
\hfill
\begin{subfigure}[t]{0.32\textwidth}
    \includegraphics[width=\linewidth]{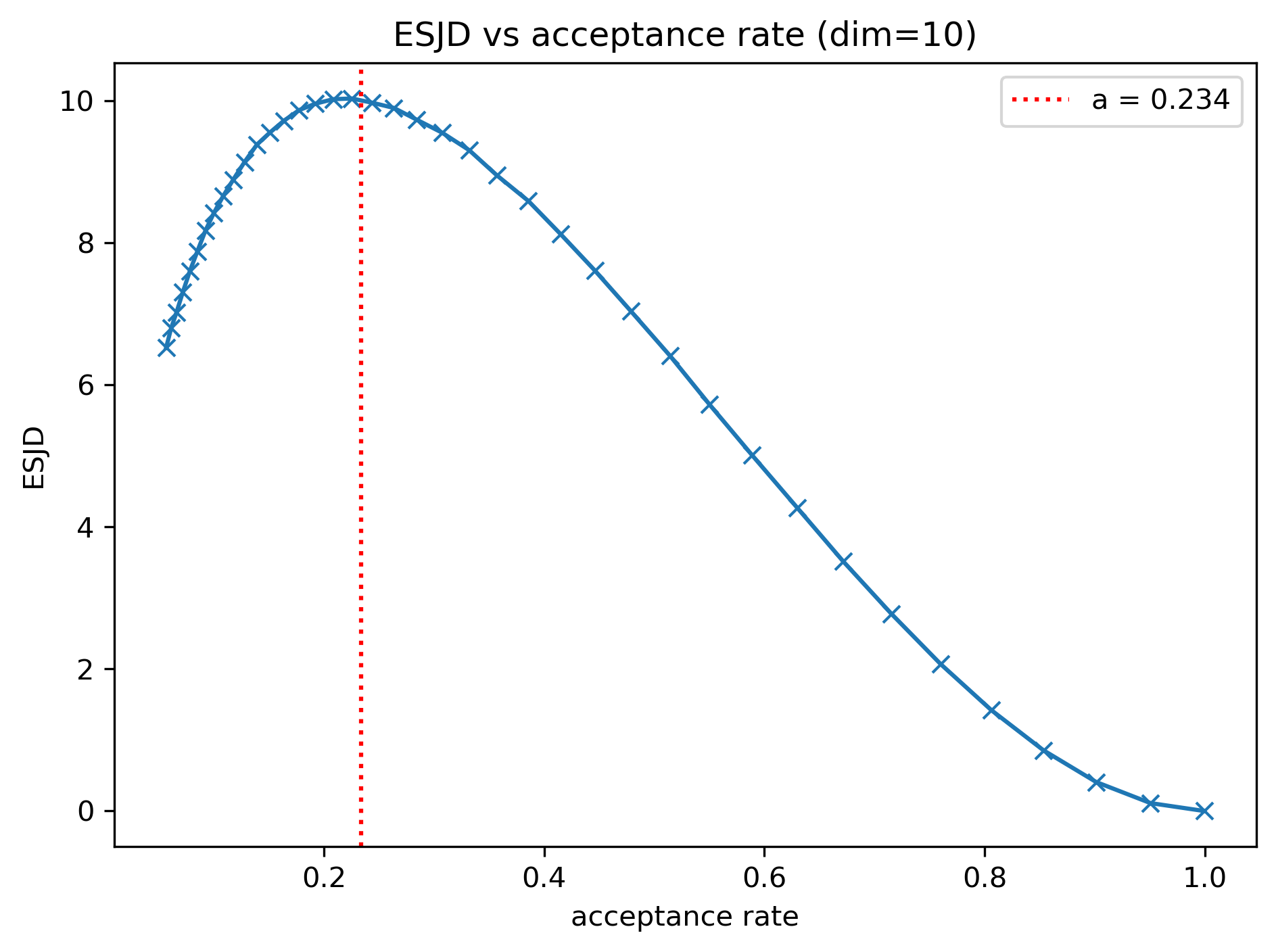}
\end{subfigure}

\vspace{0.5em}

\begin{subfigure}[t]{0.32\textwidth}
    \includegraphics[width=\linewidth]{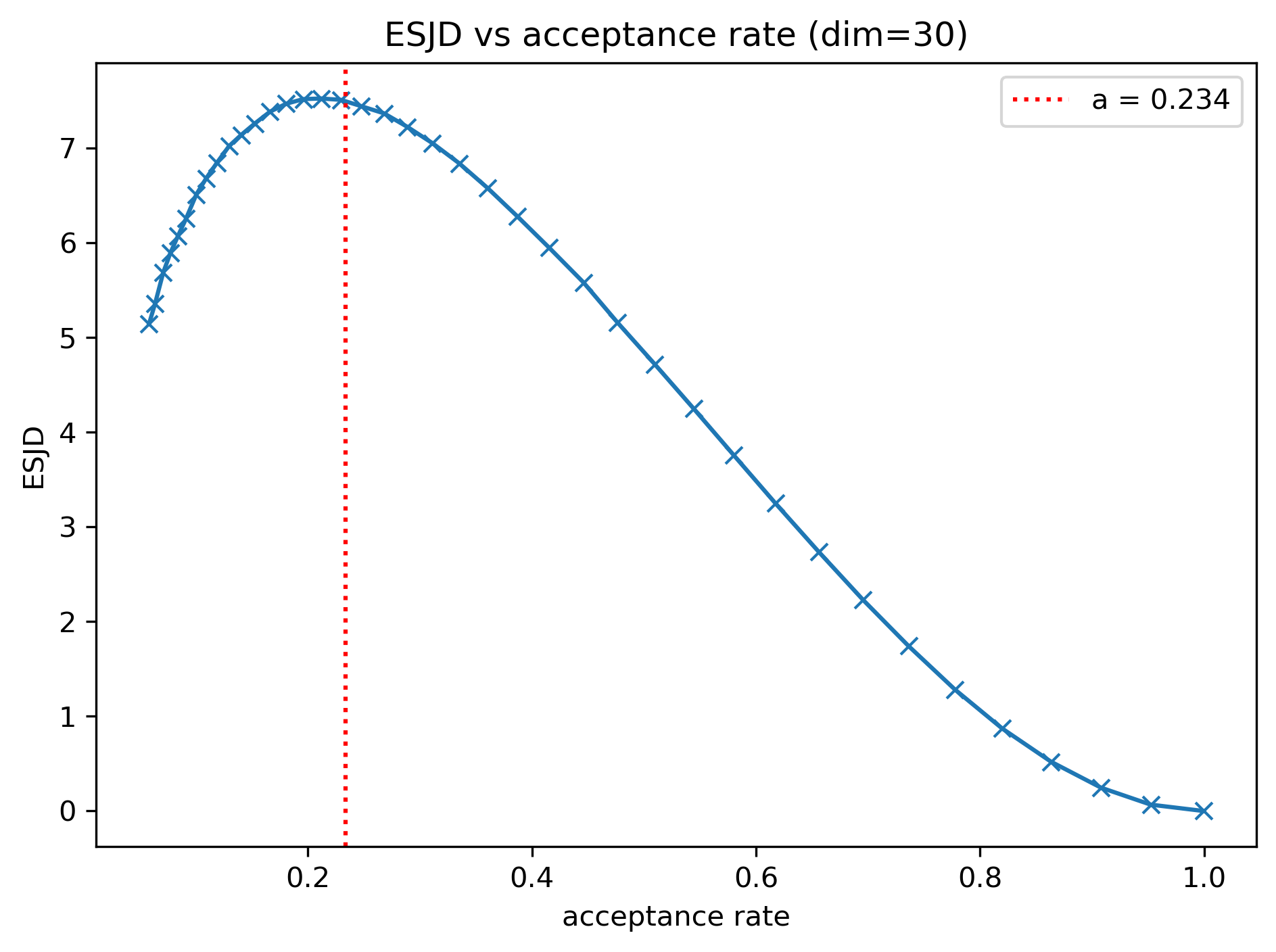}
\end{subfigure}
\hfill
\begin{subfigure}[t]{0.32\textwidth}
    \includegraphics[width=\linewidth]{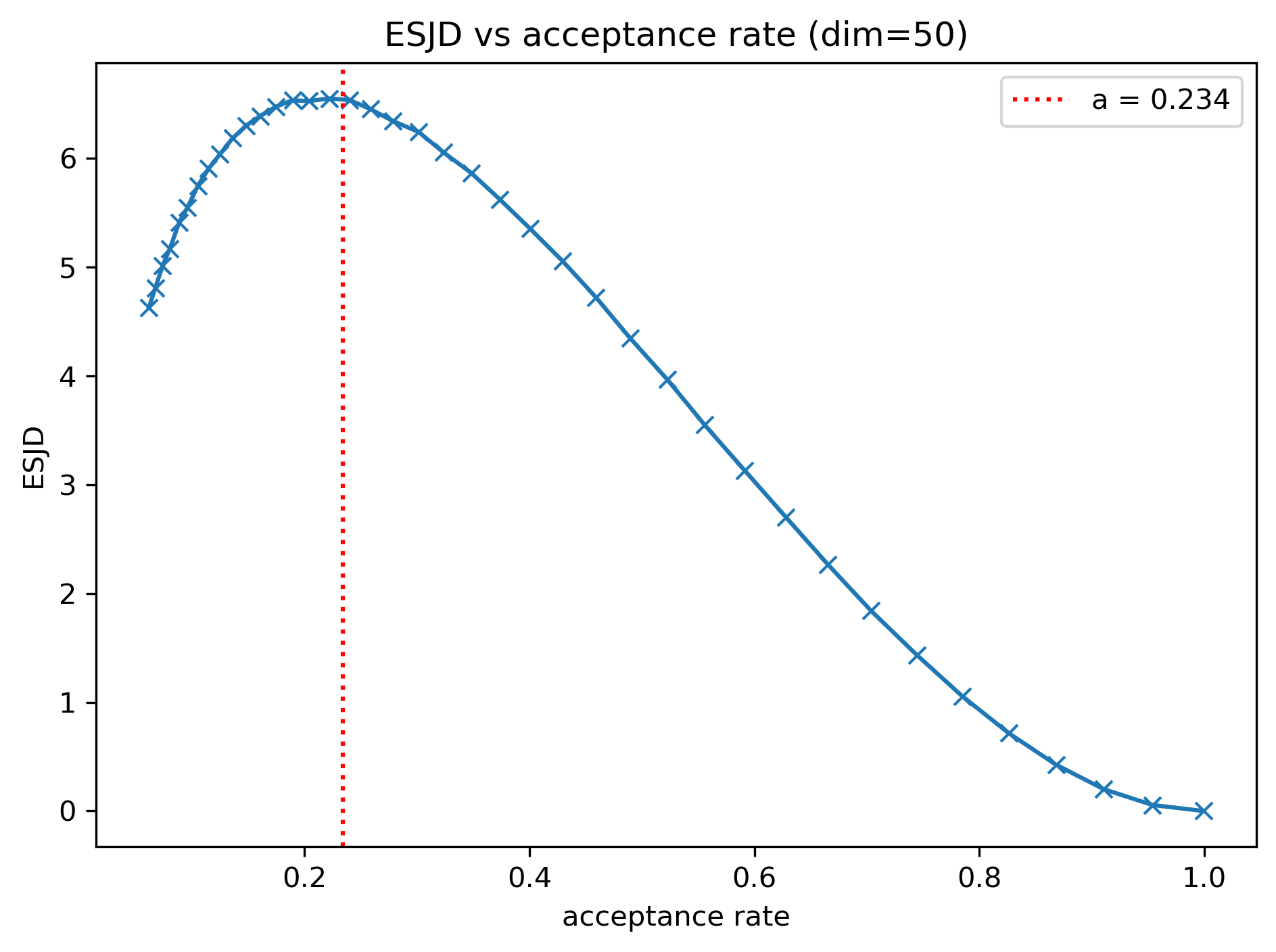}
\end{subfigure}
\hfill
\begin{subfigure}[t]{0.32\textwidth}
    \includegraphics[width=\linewidth]{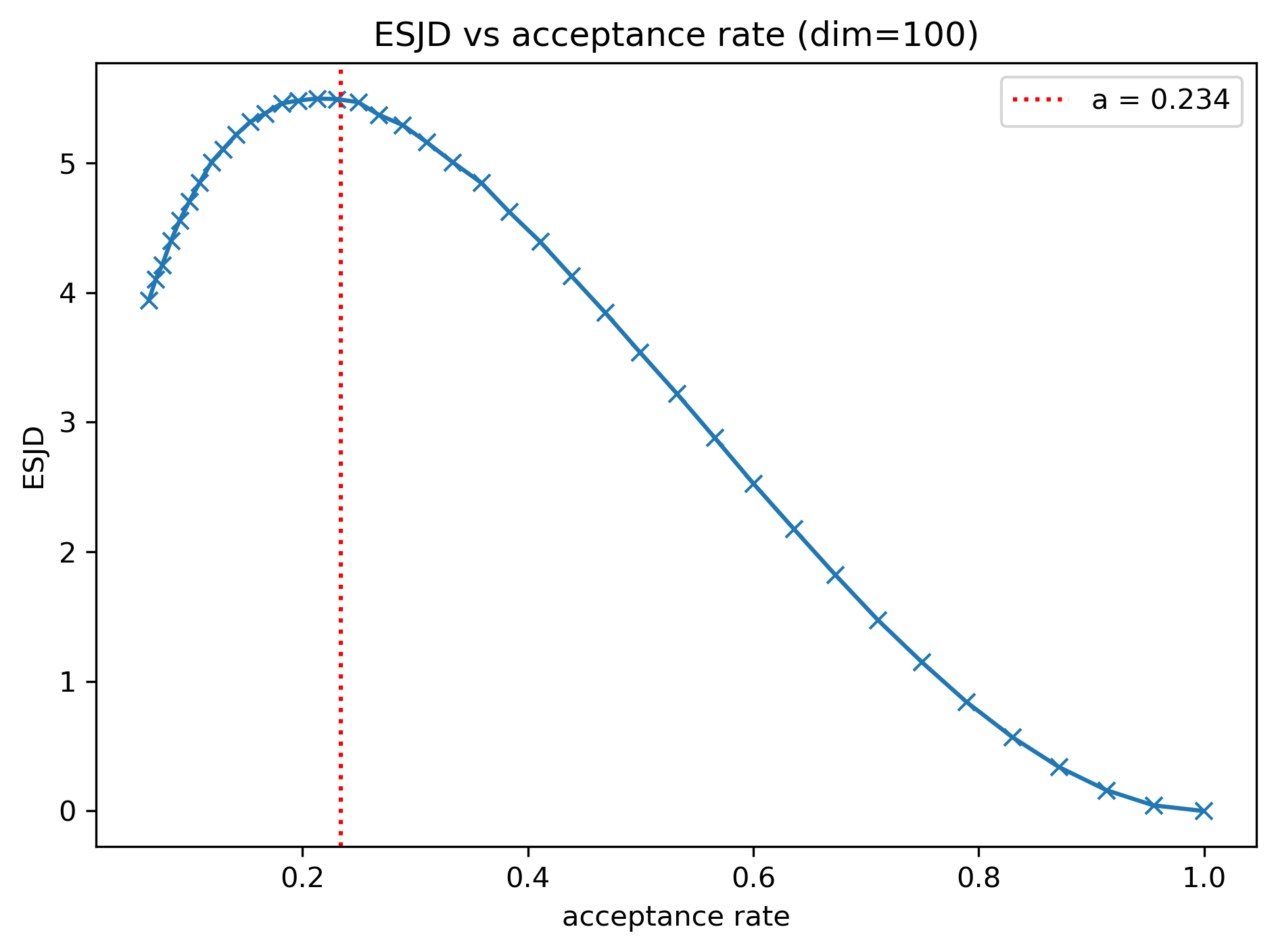}
\end{subfigure}

\caption{ESJD vs.\ acceptance rate for the i.i.d.\ Gamma(3, 2) target distribution $\pi_2$ (Eq \ref{eq:targets_gamma_beta}) under RWM with a Gaussian proposal in dimensions $d \in \{2, 5, 10, 30, 50, 100\}$ from top-left to bottom-right. Red dotted line indicates an acceptance rate of 0.234.}
\label{fig:esjd_gamma_iid}
\end{figure}


\begin{figure}[htbp]
\centering

\begin{subfigure}[t]{0.32\textwidth}
    \includegraphics[width=\linewidth]{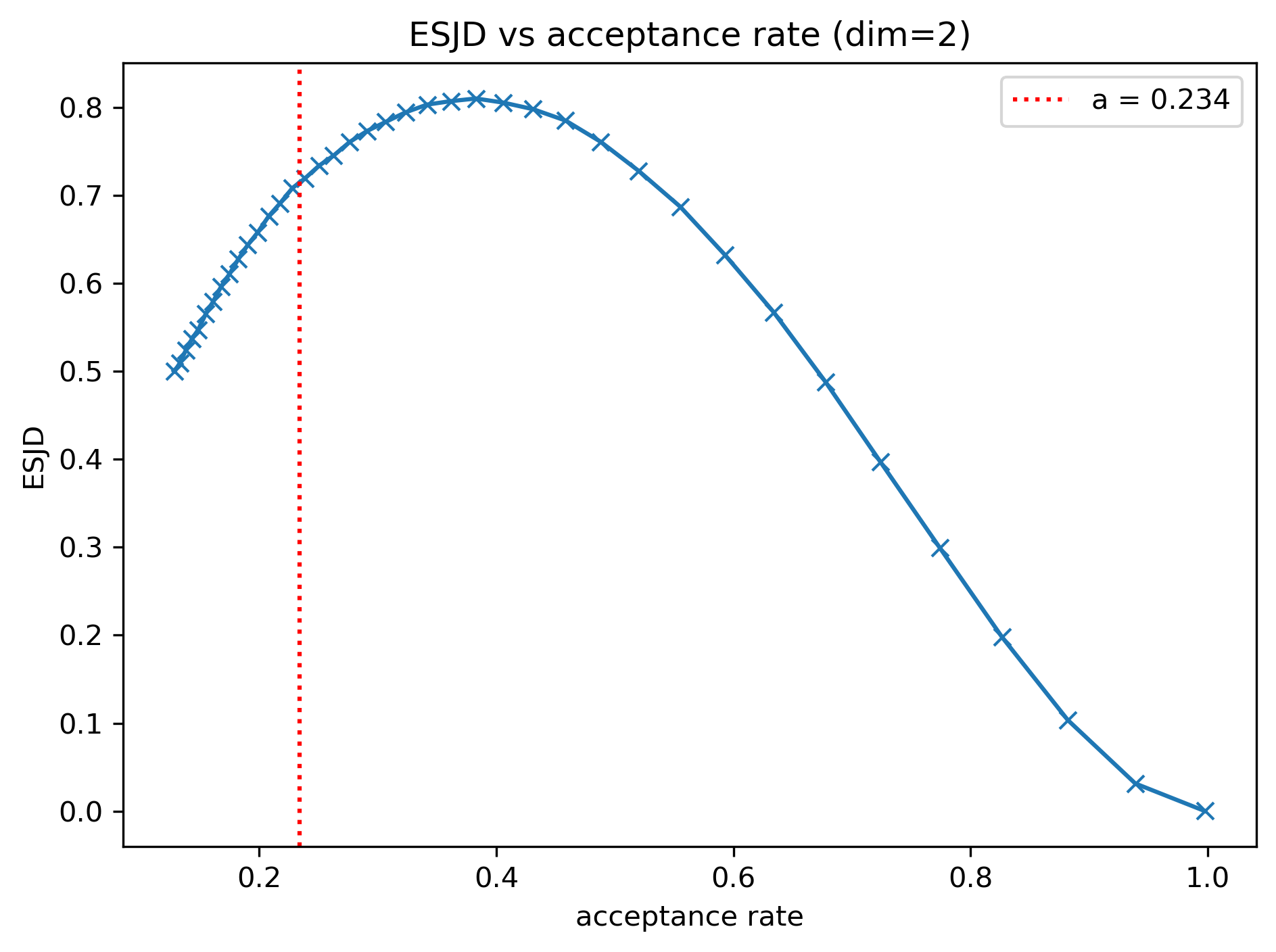}
\end{subfigure}
\hfill
\begin{subfigure}[t]{0.32\textwidth}
    \includegraphics[width=\linewidth]{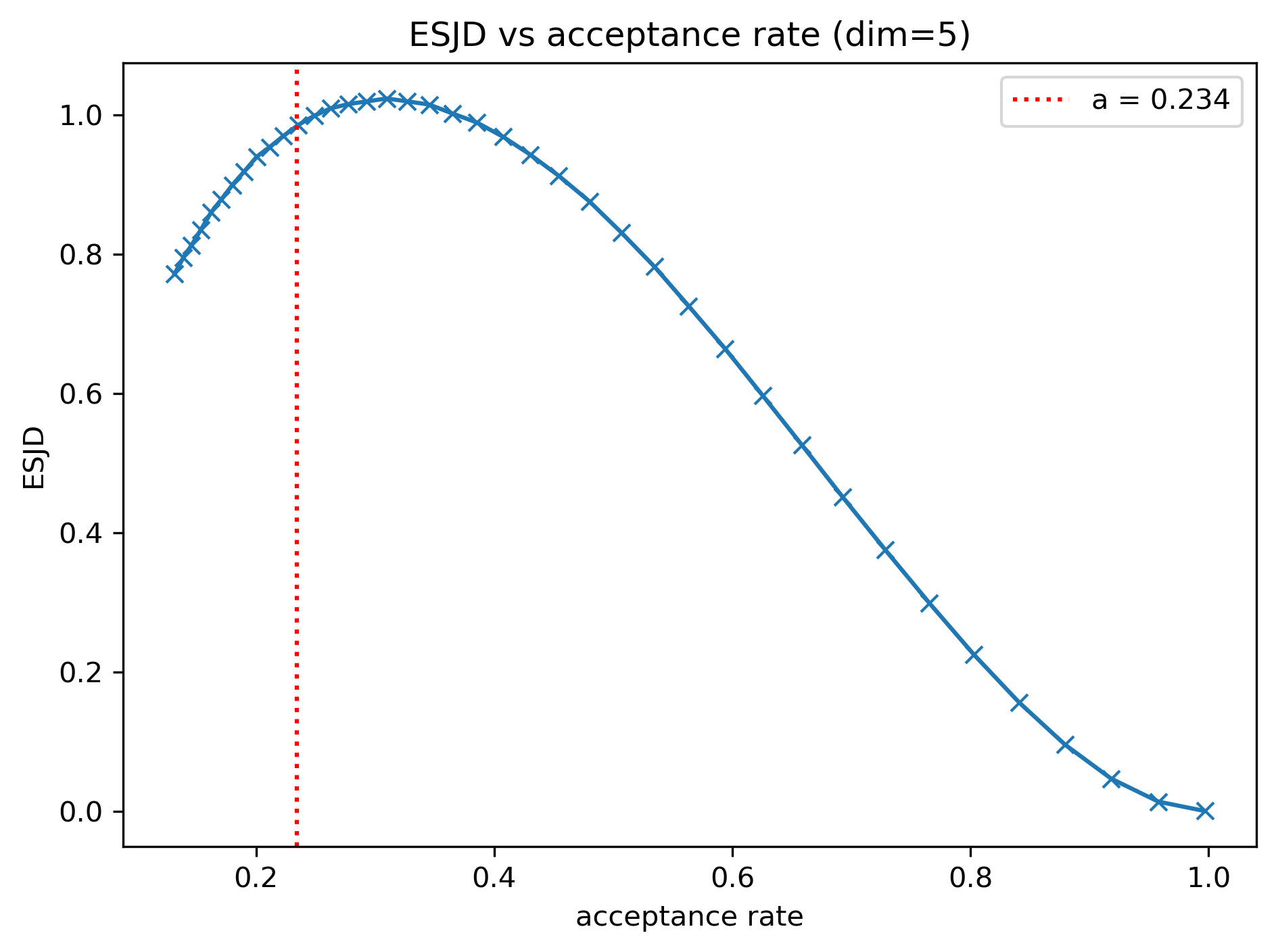}
\end{subfigure}
\hfill
\begin{subfigure}[t]{0.32\textwidth}
    \includegraphics[width=\linewidth]{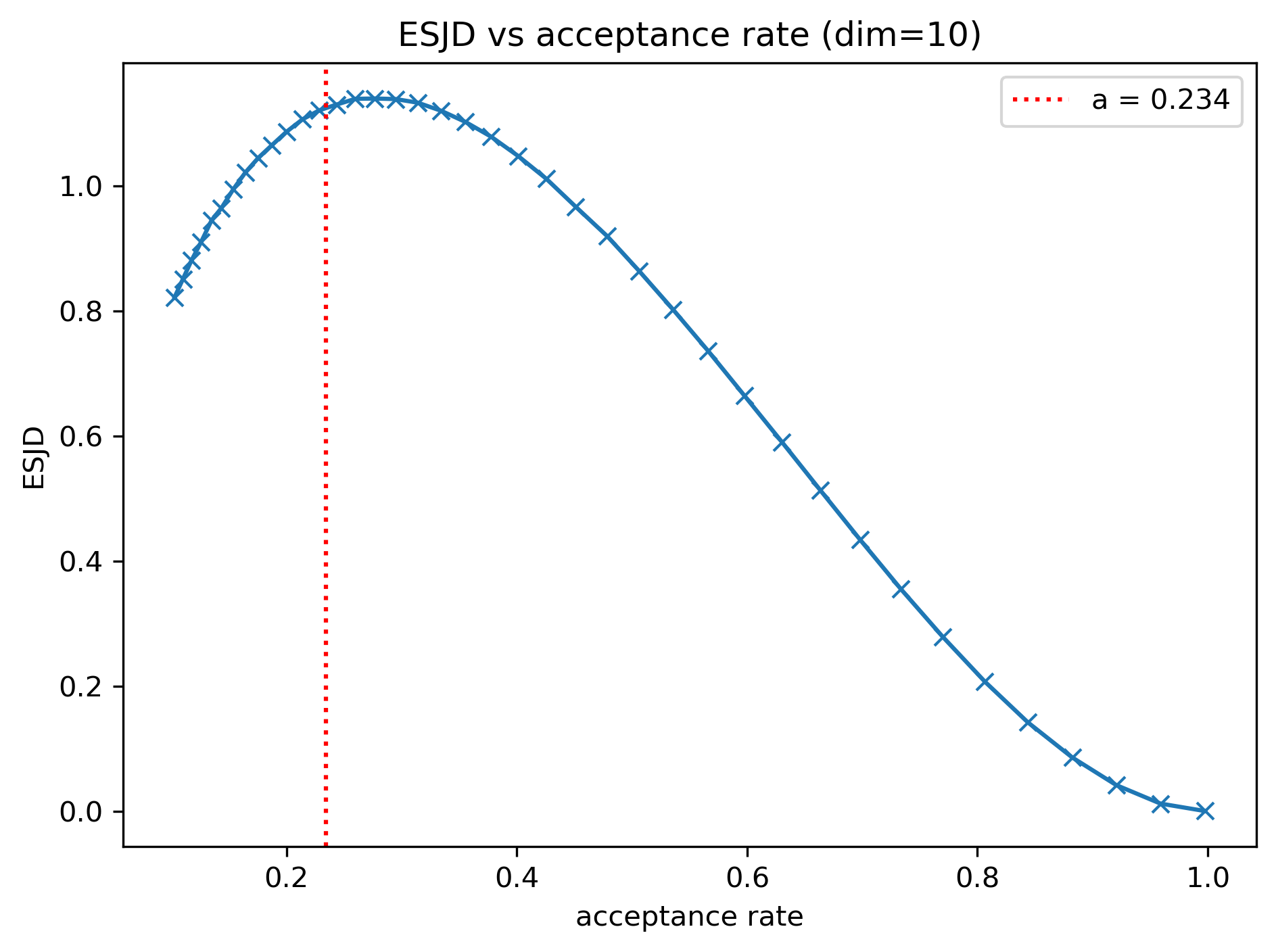}
\end{subfigure}

\vspace{0.5em}

\begin{subfigure}[t]{0.32\textwidth}
    \includegraphics[width=\linewidth]{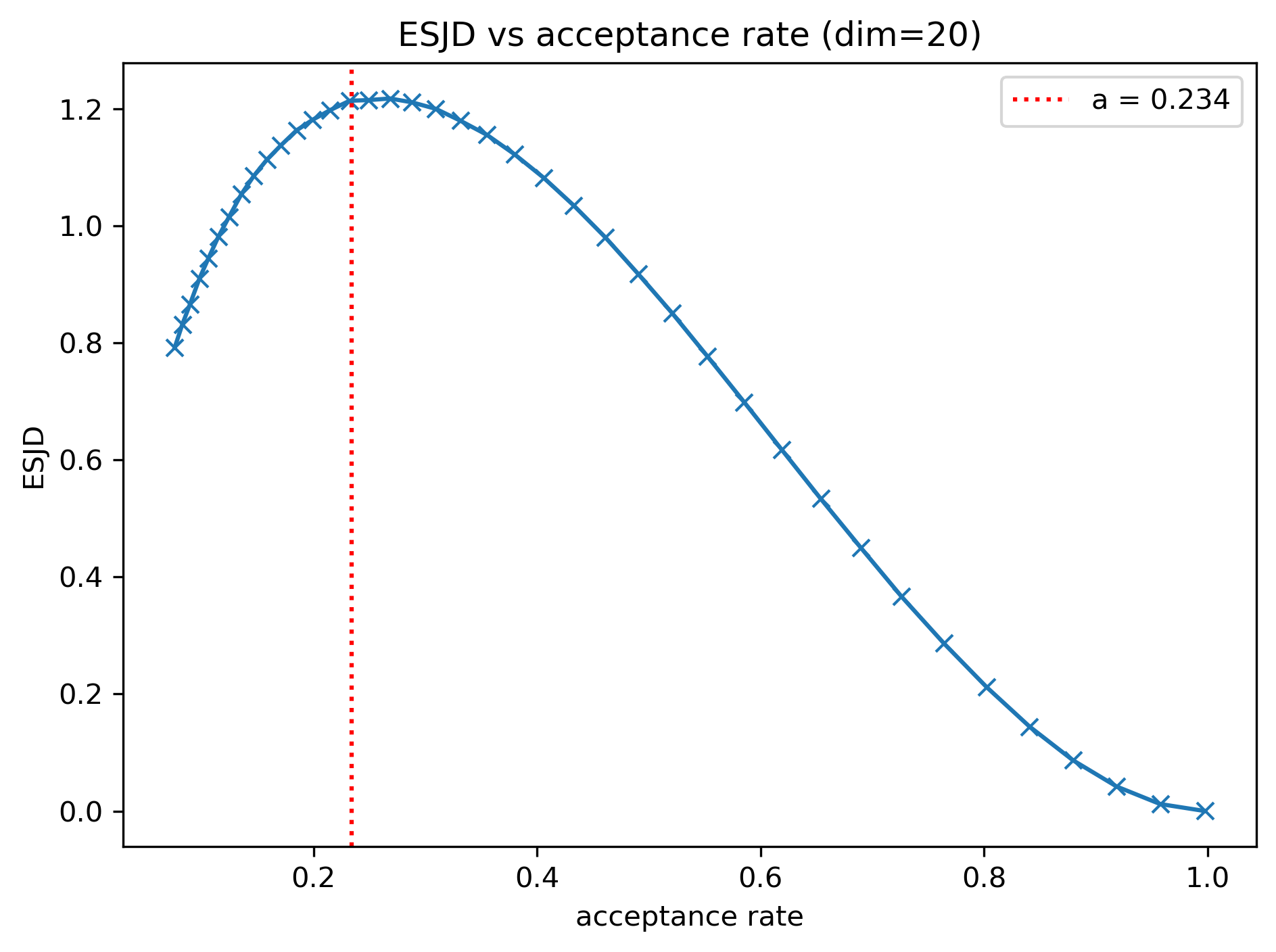}
\end{subfigure}
\hfill
\begin{subfigure}[t]{0.32\textwidth}
    \includegraphics[width=\linewidth]{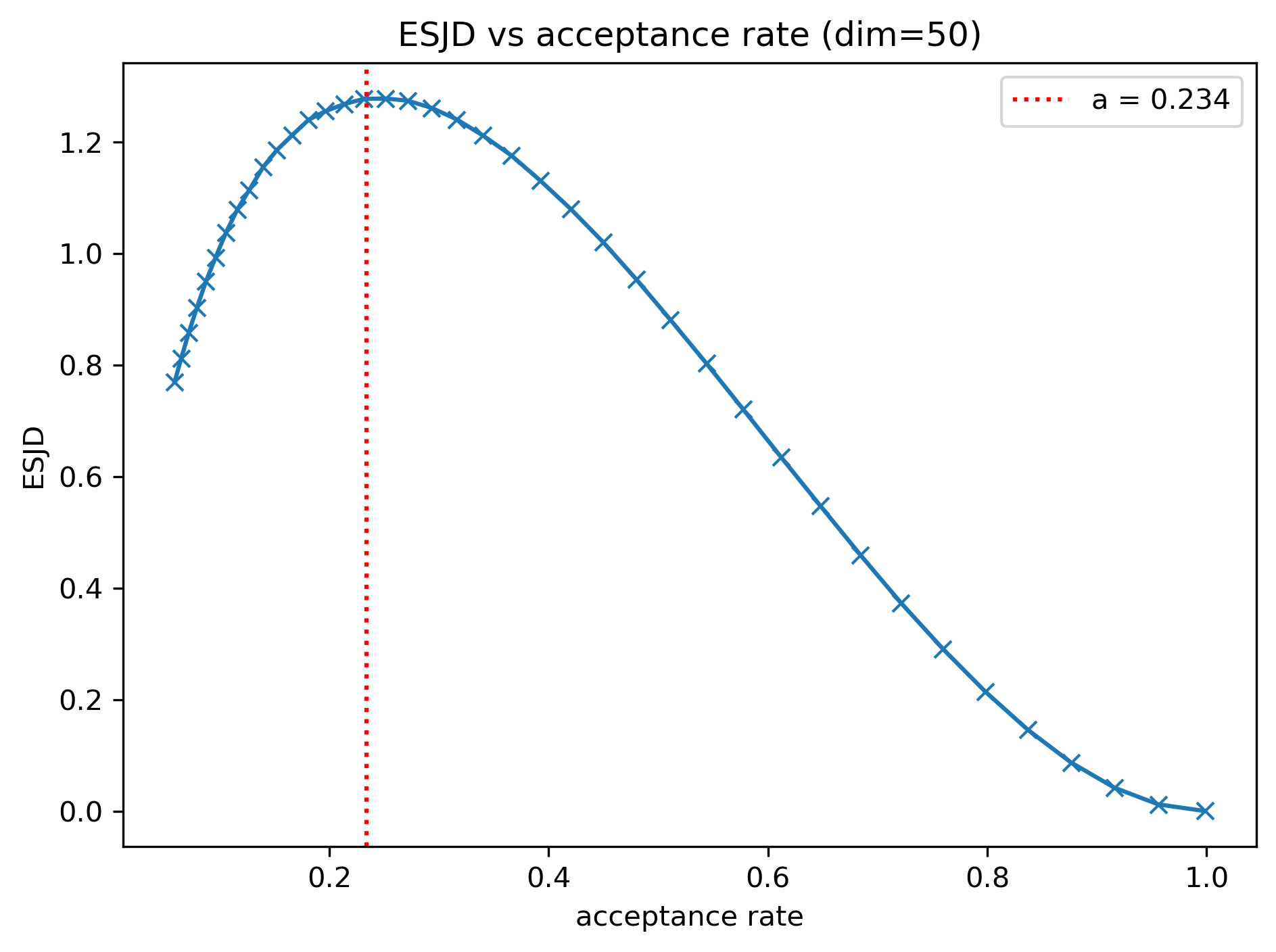}
\end{subfigure}
\hfill
\begin{subfigure}[t]{0.32\textwidth}
    \includegraphics[width=\linewidth]{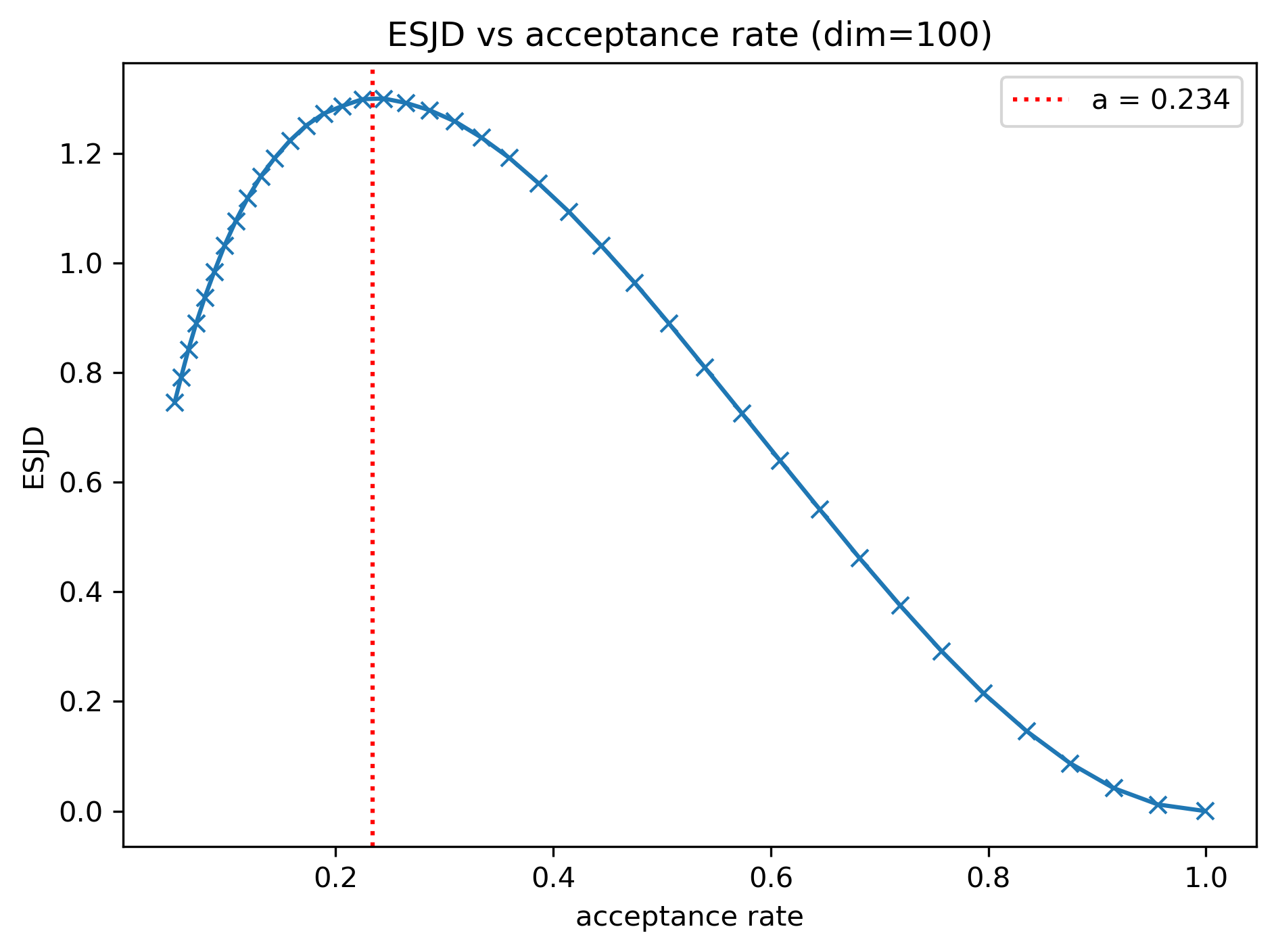}
\end{subfigure}

\caption{ESJD vs.\ acceptance rate for the standard multivariate Gaussian target distribution $\pi_3$ under RWM with a multivariate Laplace proposal $Q_L$ (Eq \ref{eq:gaussian_target_laplace_uniform_proposals}) in dimensions $d \in \{2, 5, 10, 30, 50, 100\}$ from top-left to bottom-right. Red dotted line indicates an acceptance rate of 0.234.}
\label{fig:esjd_prop_laplace}
\end{figure}


\begin{figure}[htbp]
\centering

\begin{subfigure}[t]{0.32\textwidth}
    \includegraphics[width=\linewidth]{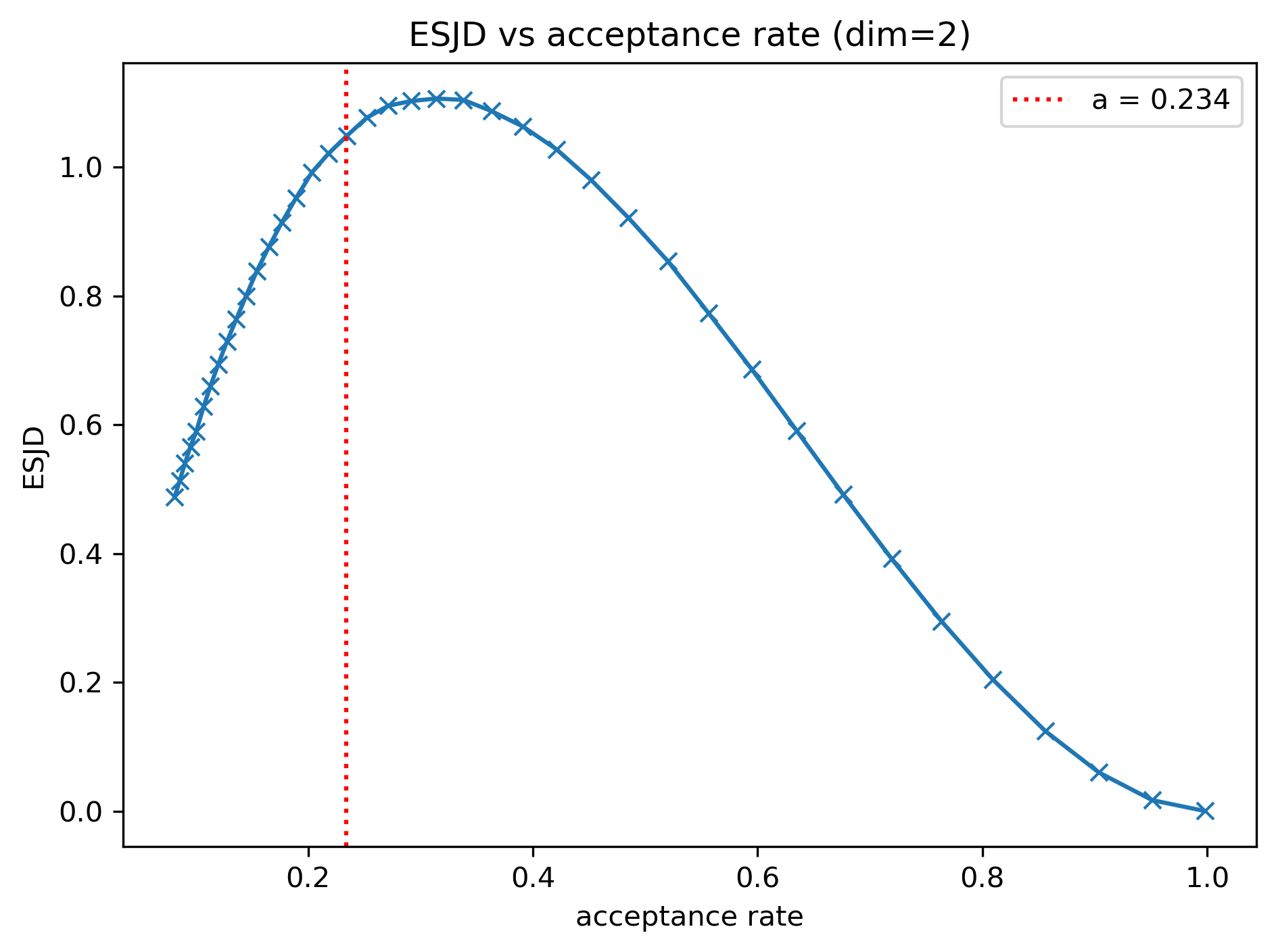}
\end{subfigure}
\hfill
\begin{subfigure}[t]{0.32\textwidth}
    \includegraphics[width=\linewidth]{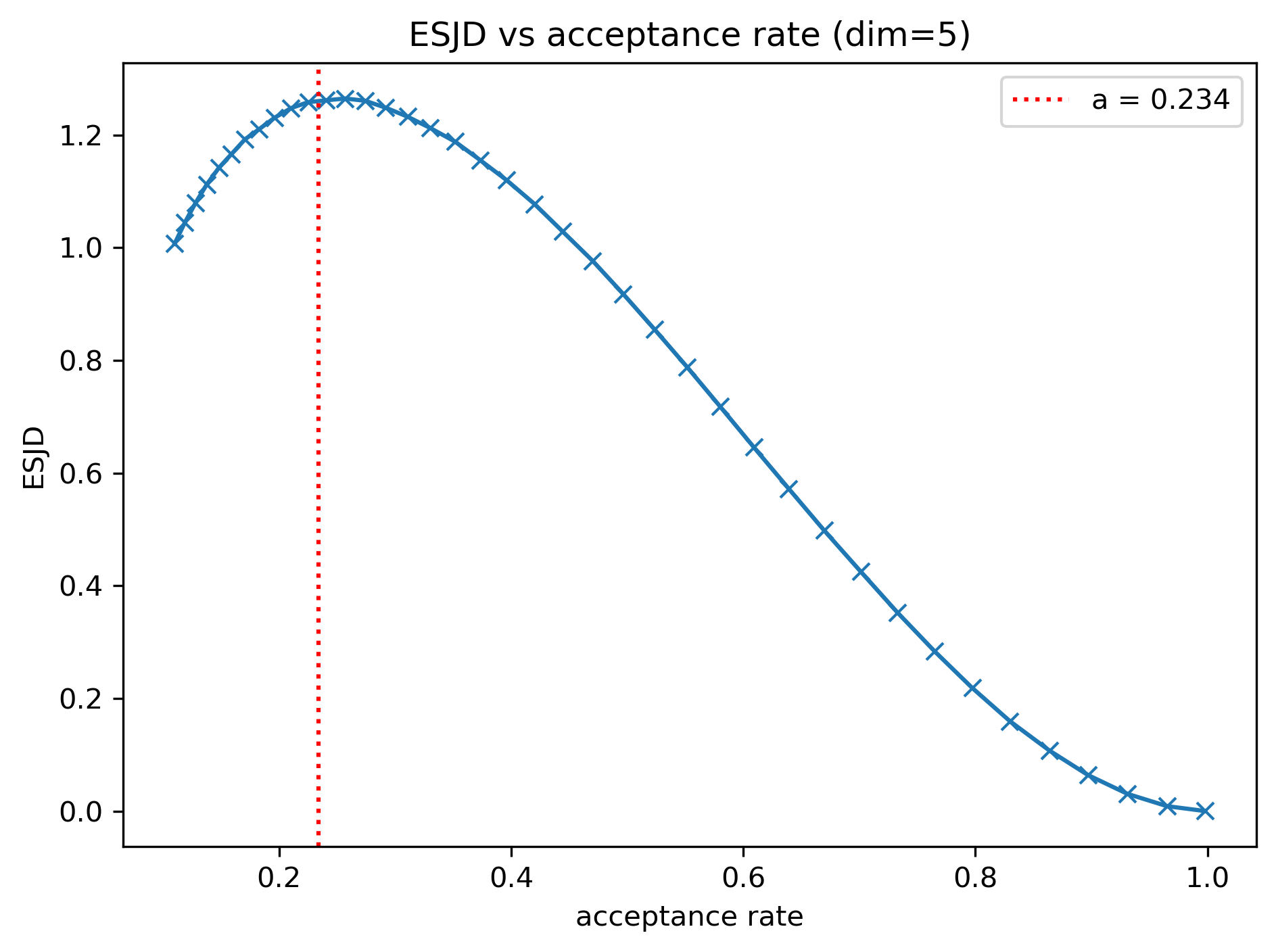}
\end{subfigure}
\hfill
\begin{subfigure}[t]{0.32\textwidth}
    \includegraphics[width=\linewidth]{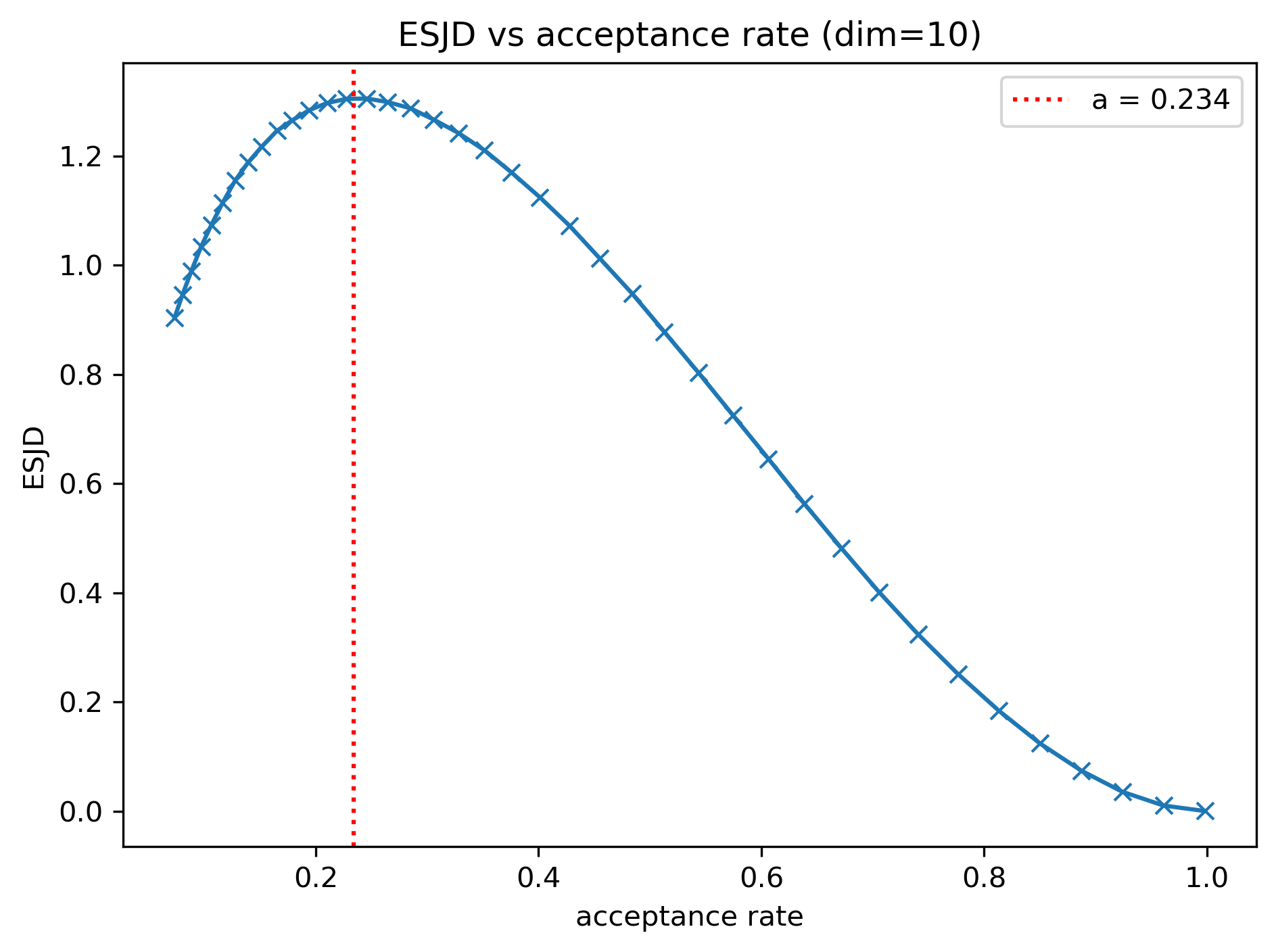}
\end{subfigure}

\vspace{0.5em}

\begin{subfigure}[t]{0.32\textwidth}
    \includegraphics[width=\linewidth]{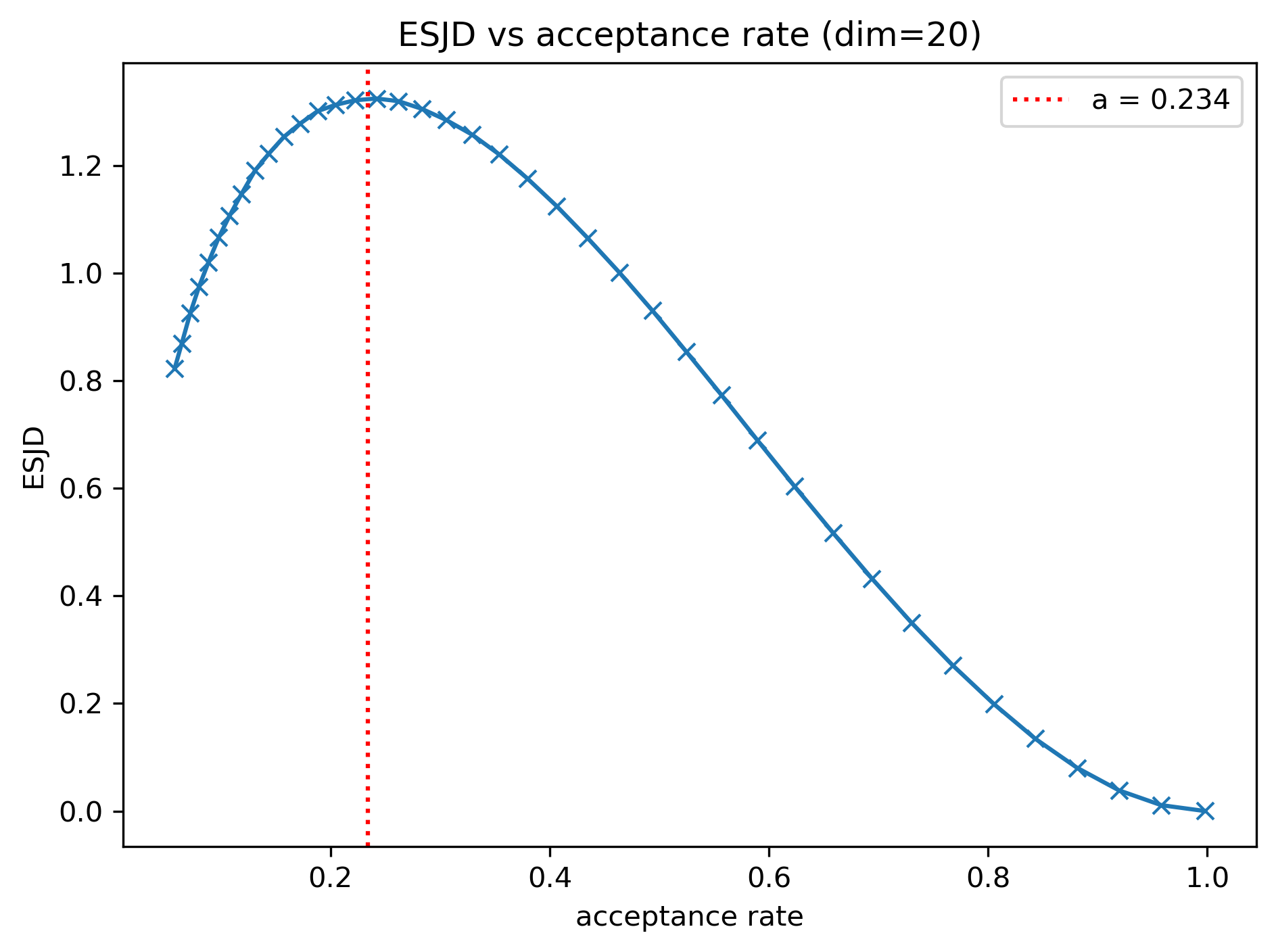}
\end{subfigure}
\hfill
\begin{subfigure}[t]{0.32\textwidth}
    \includegraphics[width=\linewidth]{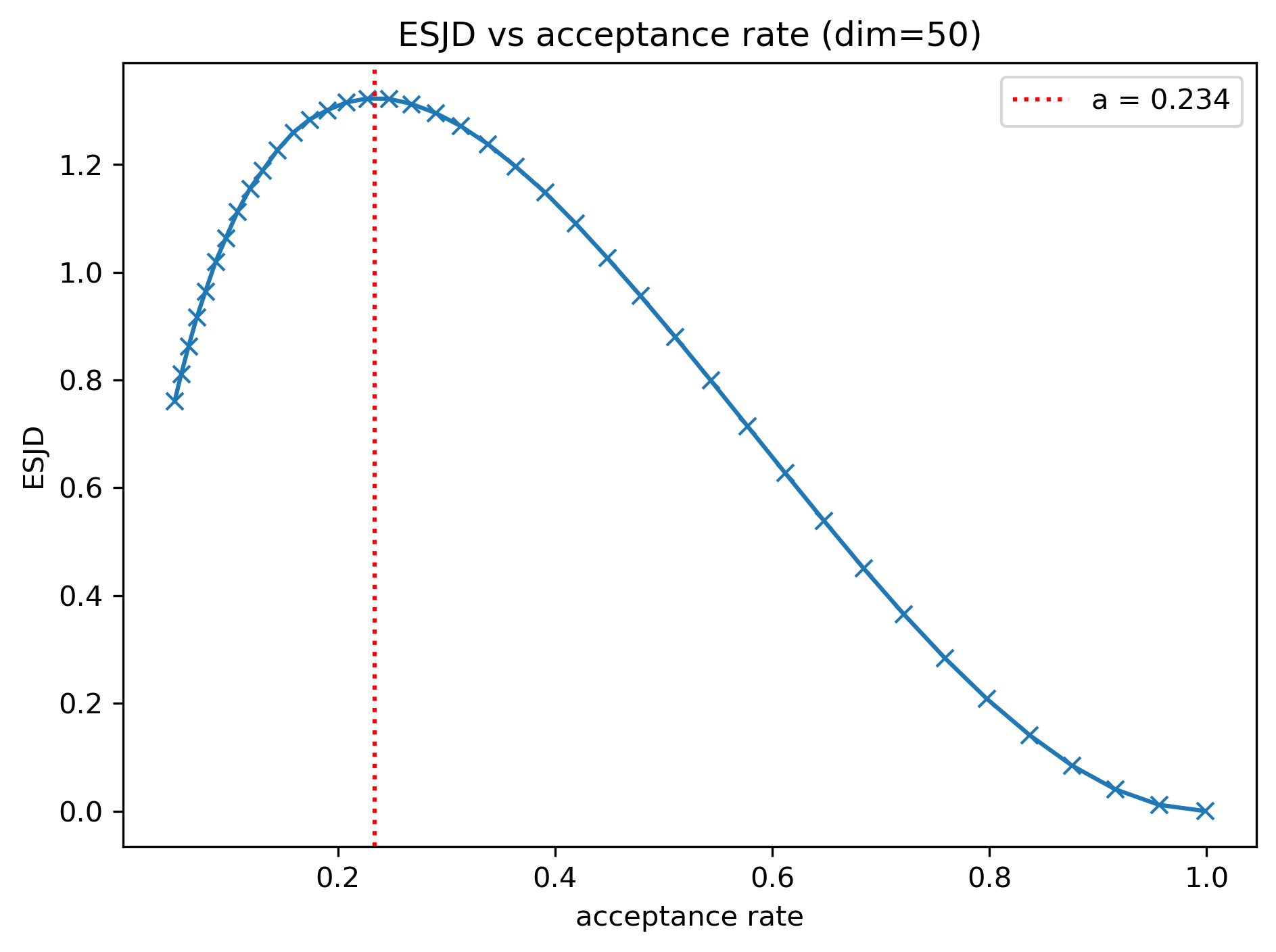}
\end{subfigure}
\hfill
\begin{subfigure}[t]{0.32\textwidth}
    \includegraphics[width=\linewidth]{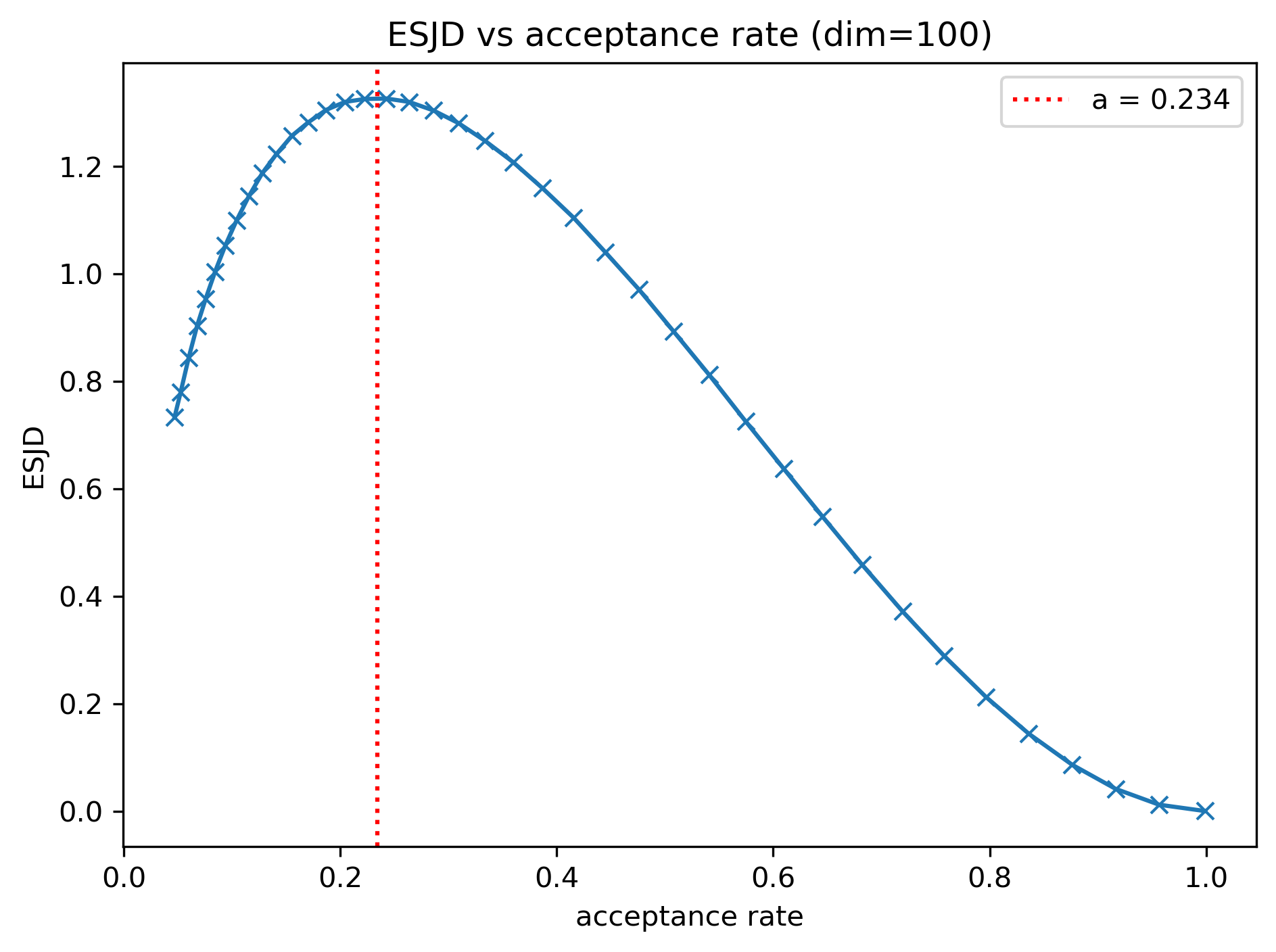}
\end{subfigure}

\caption{ESJD vs.\ acceptance rate for the standard multivariate Gaussian target distribution $\pi_3$ under RWM with a multivariate Uniform proposal $Q_U$ (Eq \ref{eq:gaussian_target_laplace_uniform_proposals}) in dimensions $d \in \{2, 5, 10, 30, 50, 100\}$ from top-left to bottom-right. Red dotted line indicates an acceptance rate of 0.234.}
\label{fig:esjd_prop_uniform}
\end{figure}


\begin{figure}[ht]
    \centering
    \begin{subfigure}{0.32\textwidth}
        \includegraphics[width=\linewidth, valign=t]{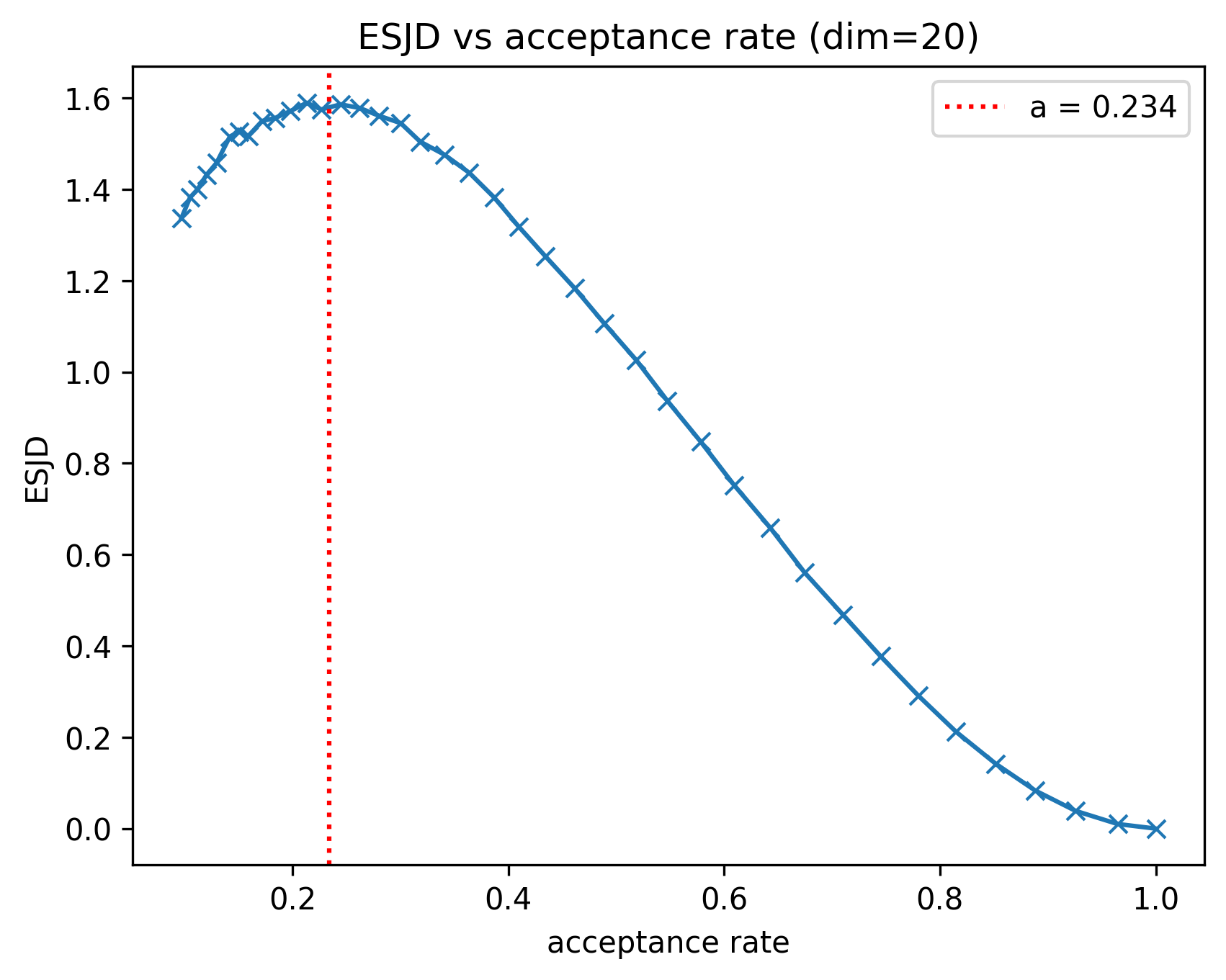}
    \end{subfigure}
    \begin{subfigure}{0.32\textwidth}
        \includegraphics[width=\linewidth, valign=t]{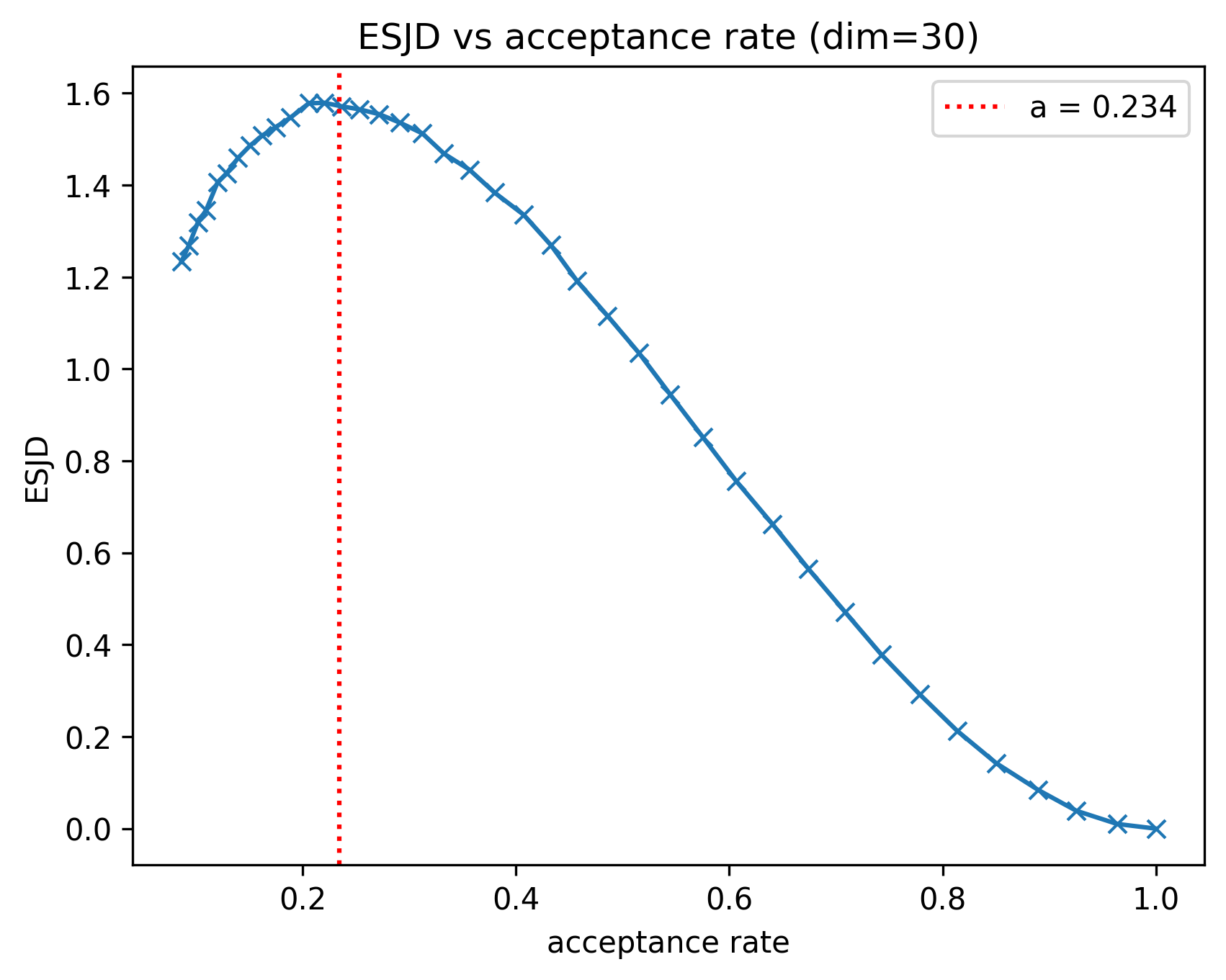}
    \end{subfigure}
    \begin{subfigure}{0.32\textwidth}
        \includegraphics[width=\linewidth, valign=t]{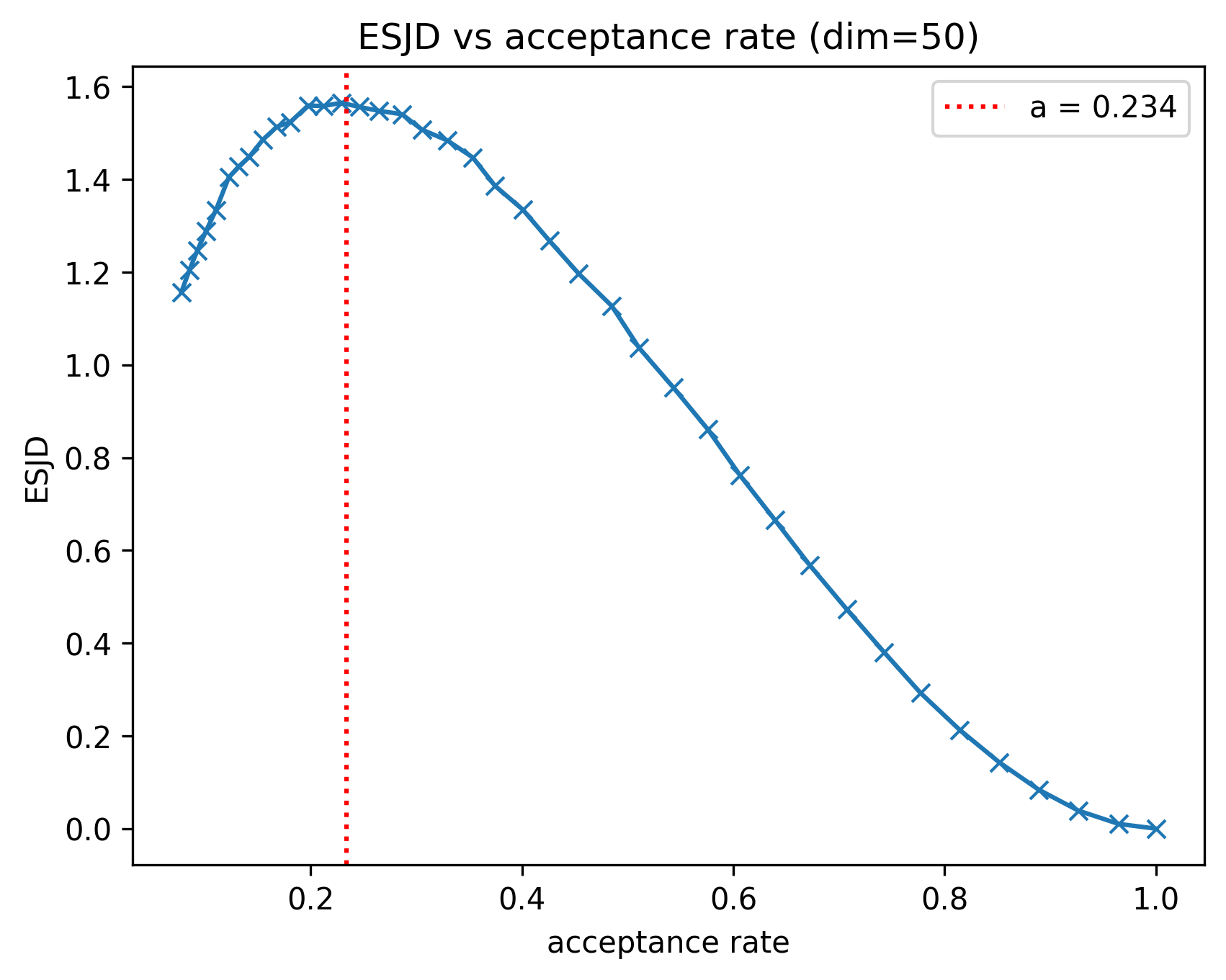}
    \end{subfigure}
    \caption{ESJD vs.\ acceptance rate for the i.i.d.\ rough carpet target distribution $\pi_4$ (Eq \ref{eq:target-rough-carpet}) under RWM with a Gaussian proposal in dimensions $d \in \{20,30,50\}$ from left to right. Red dotted line indicates an acceptance rate of 0.234.}
    \label{fig:esjd_rough_carpet}
\end{figure}

\begin{figure}[ht]
    \centering
    \begin{subfigure}{0.32\textwidth}
        \includegraphics[width=\linewidth, valign=t]{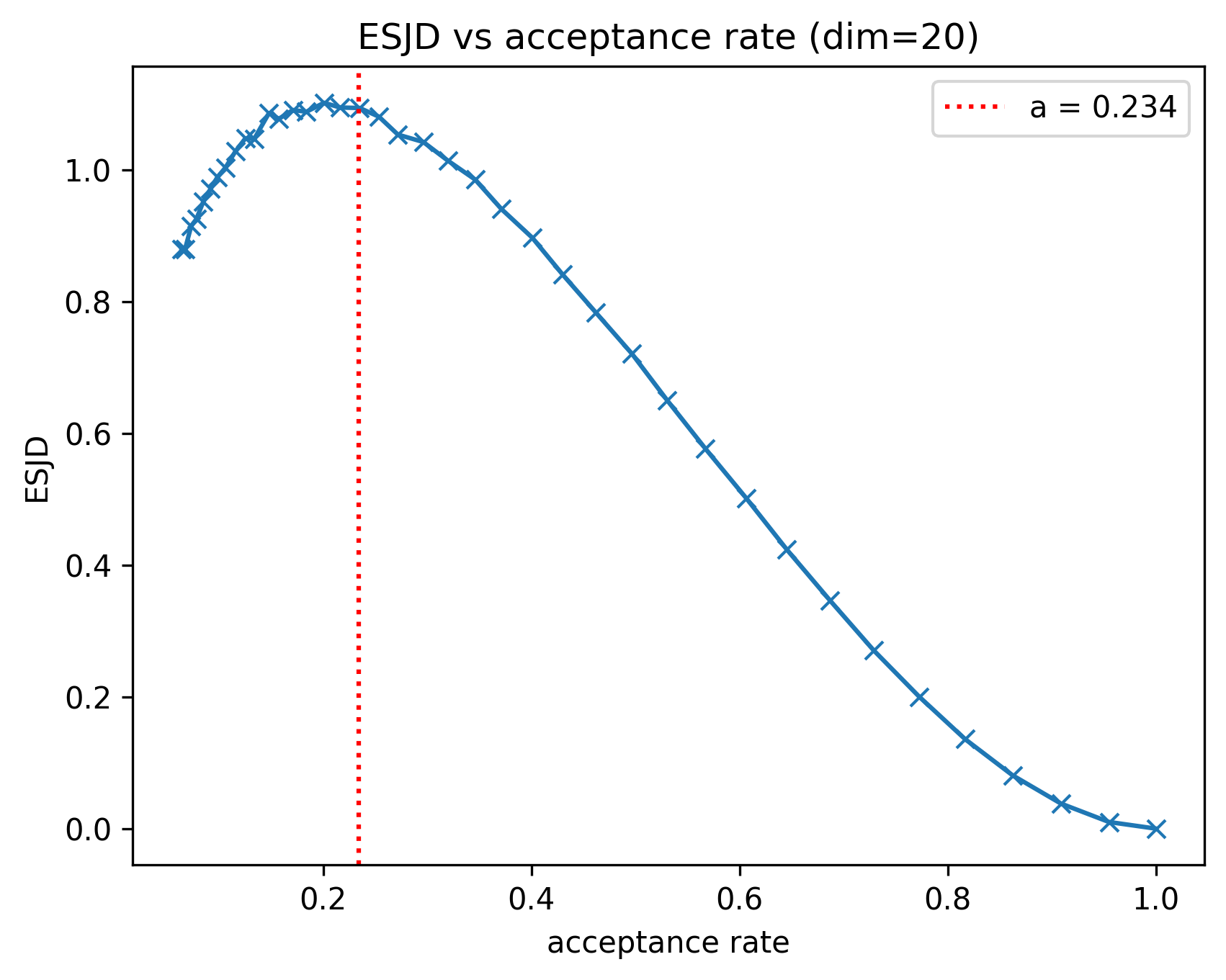}
    \end{subfigure}
    \begin{subfigure}{0.32\textwidth}
        \includegraphics[width=\linewidth, valign=t]{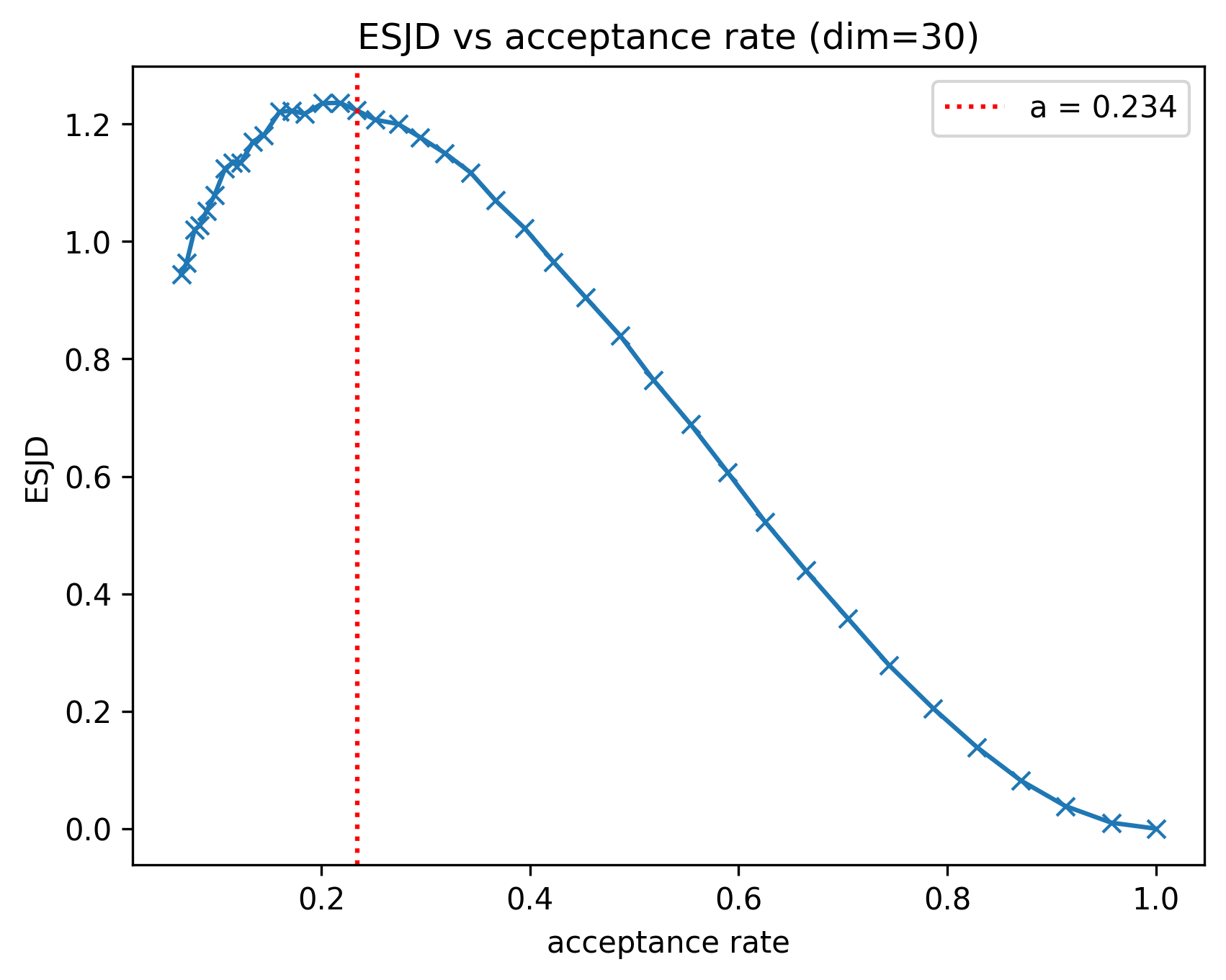}
    \end{subfigure}
    \begin{subfigure}{0.32\textwidth}
        \includegraphics[width=\linewidth, valign=t]{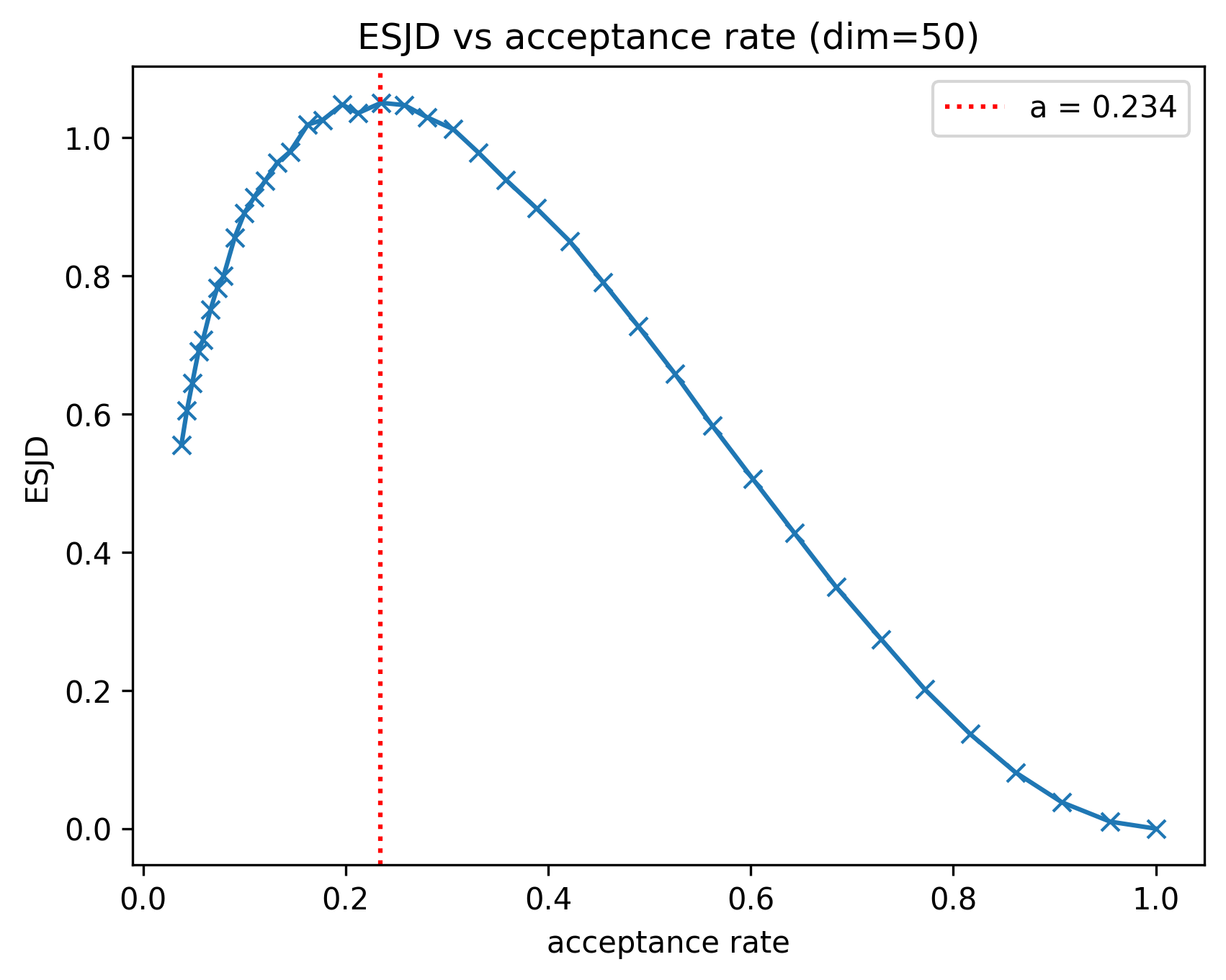}
    \end{subfigure}
    \caption{ESJD vs.\ acceptance rate for the inhomogeneously scaled i.i.d.\ rough carpet target distribution $\pi_5$ (Eq \ref{eq:target-rough-carpet-scaled}) under RWM with a Gaussian proposal in dimensions $d \in \{20,30,50\}$ from left to right. Red dotted line indicates an acceptance rate of 0.234.}
    \label{fig:esjd_rough_carpet_scaled}
\end{figure}


\begin{figure}[ht]
    \centering
    \begin{subfigure}{0.32\textwidth}
        \includegraphics[width=\linewidth, valign=t]{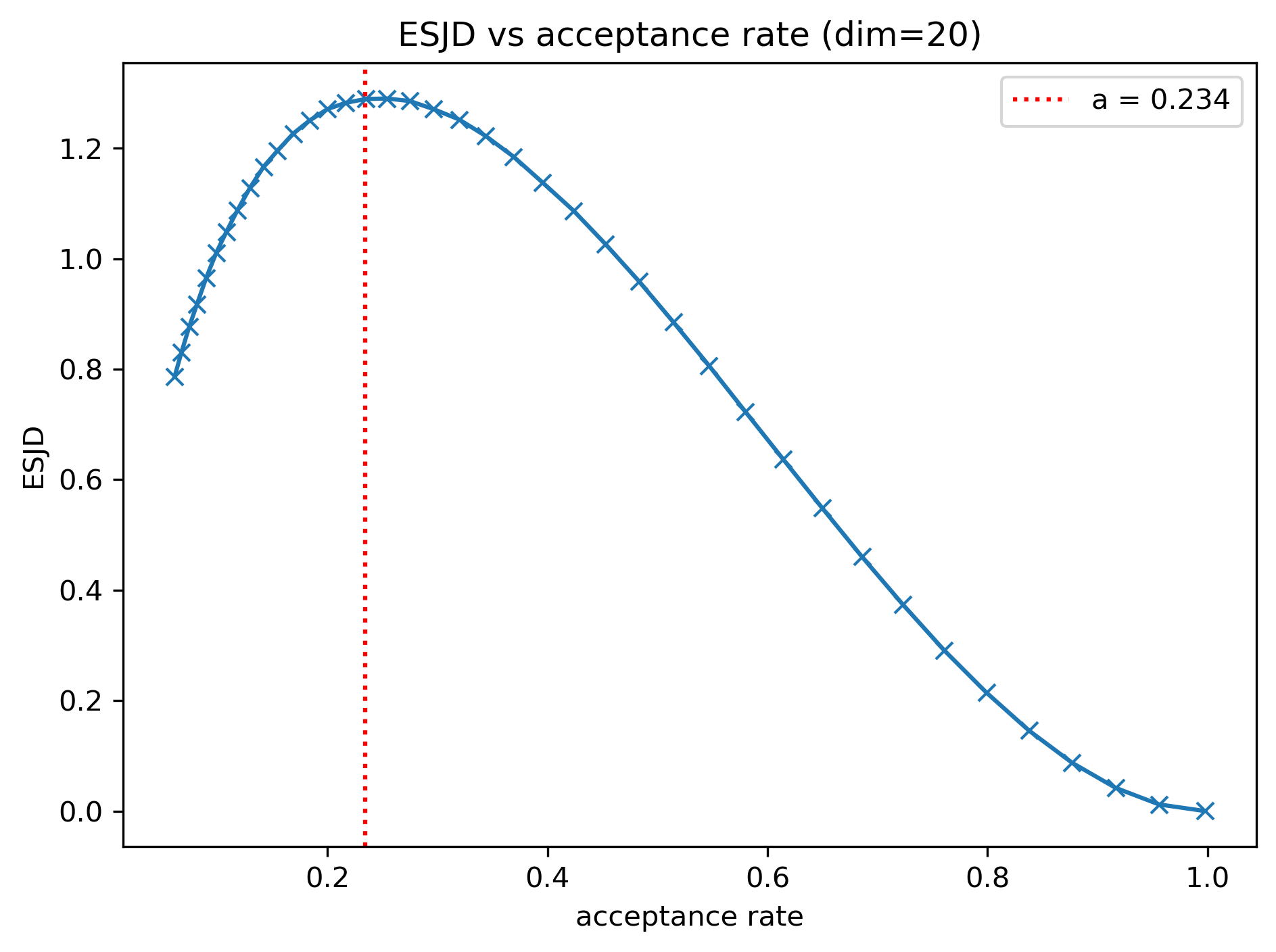}
    \end{subfigure}
    \begin{subfigure}{0.32\textwidth}
        \includegraphics[width=\linewidth, valign=t]{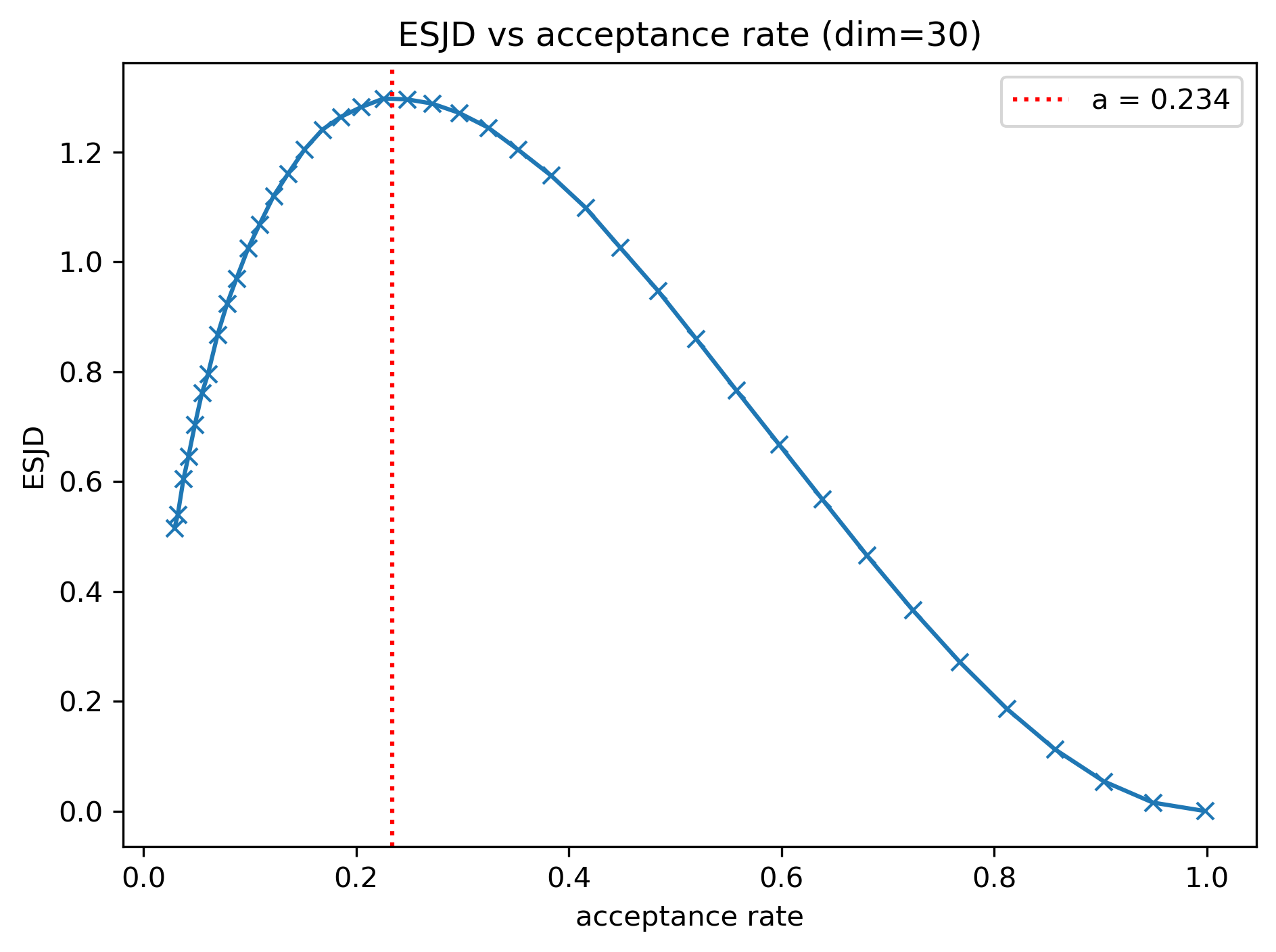}
    \end{subfigure}
    \begin{subfigure}{0.32\textwidth}
        \includegraphics[width=\linewidth, valign=t]{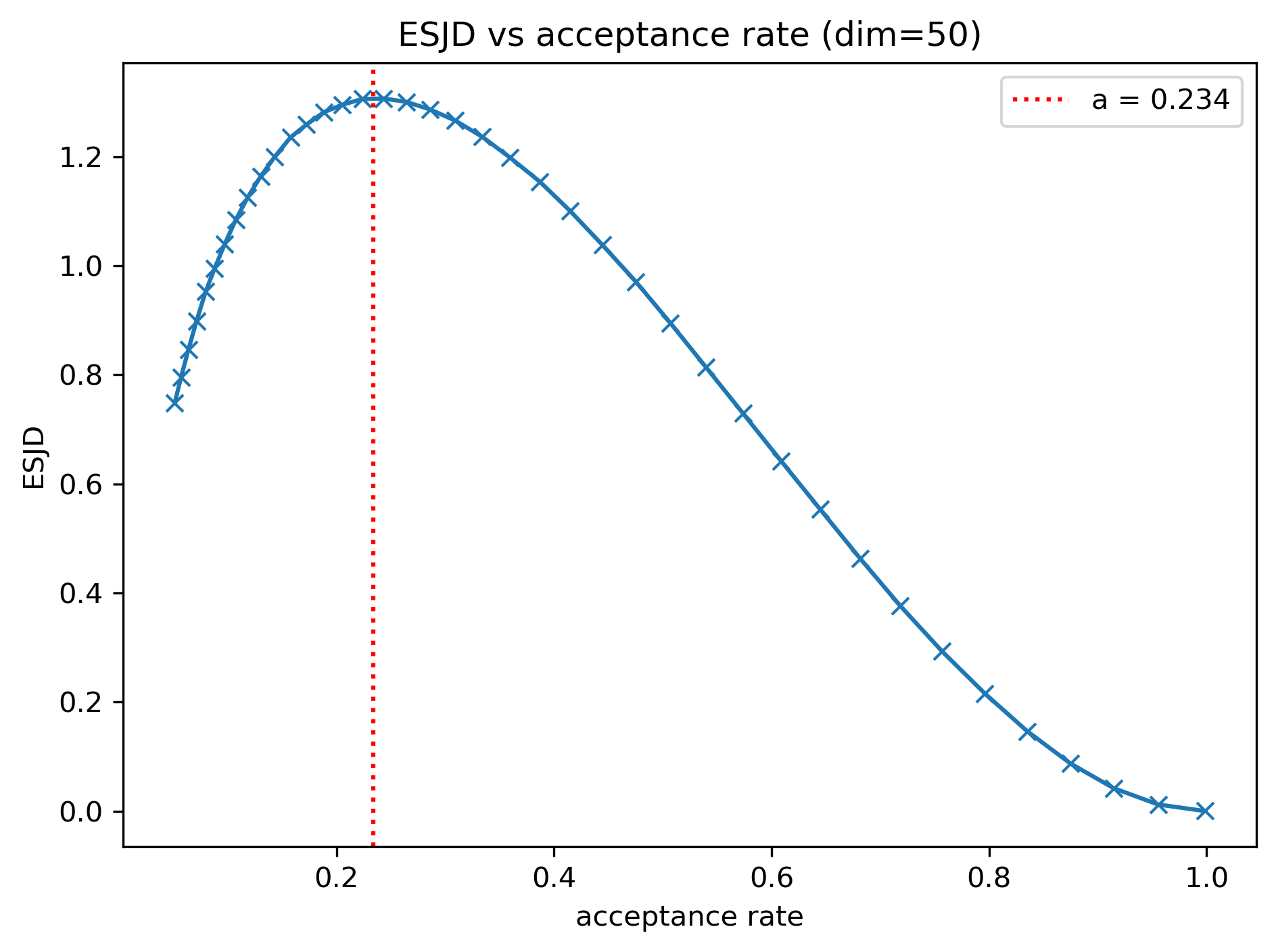}
    \end{subfigure}
    \caption{ESJD vs.\ acceptance rate for the three mixture target distribution $\pi_6$ (Eq \ref{eq:3mode}) under RWM with a Gaussian proposal in dimensions $d \in \{20,30,50\}$ from left to right. Red dotted line indicates an acceptance rate of 0.234.}
    \label{fig:esjd_3mix}
\end{figure}

\begin{figure}[ht]
    \centering
    \begin{subfigure}{0.32\textwidth}
        \includegraphics[width=\linewidth, valign=t]{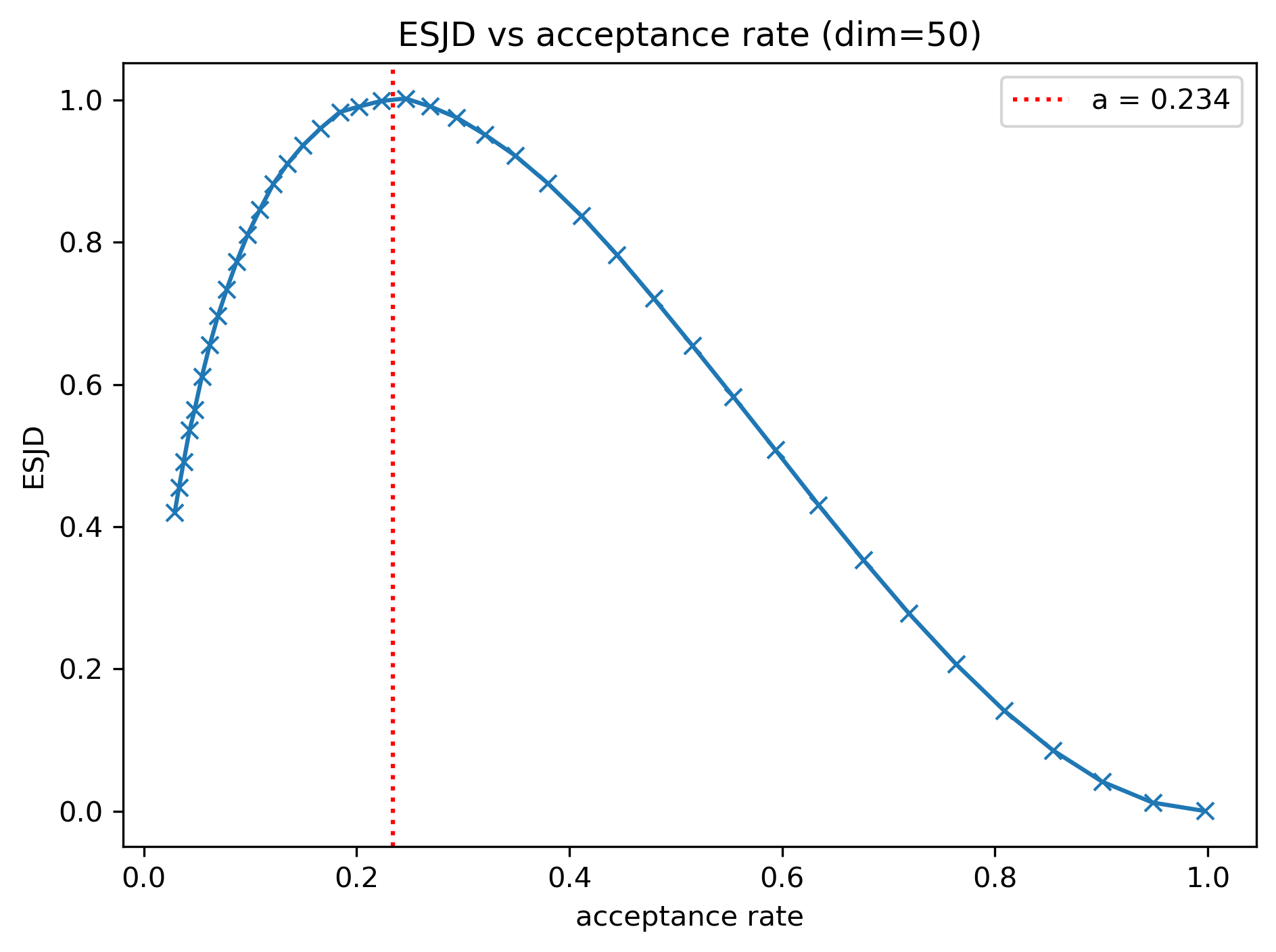}
    \end{subfigure}
    \begin{subfigure}{0.32\textwidth}
        \includegraphics[width=\linewidth, valign=t]{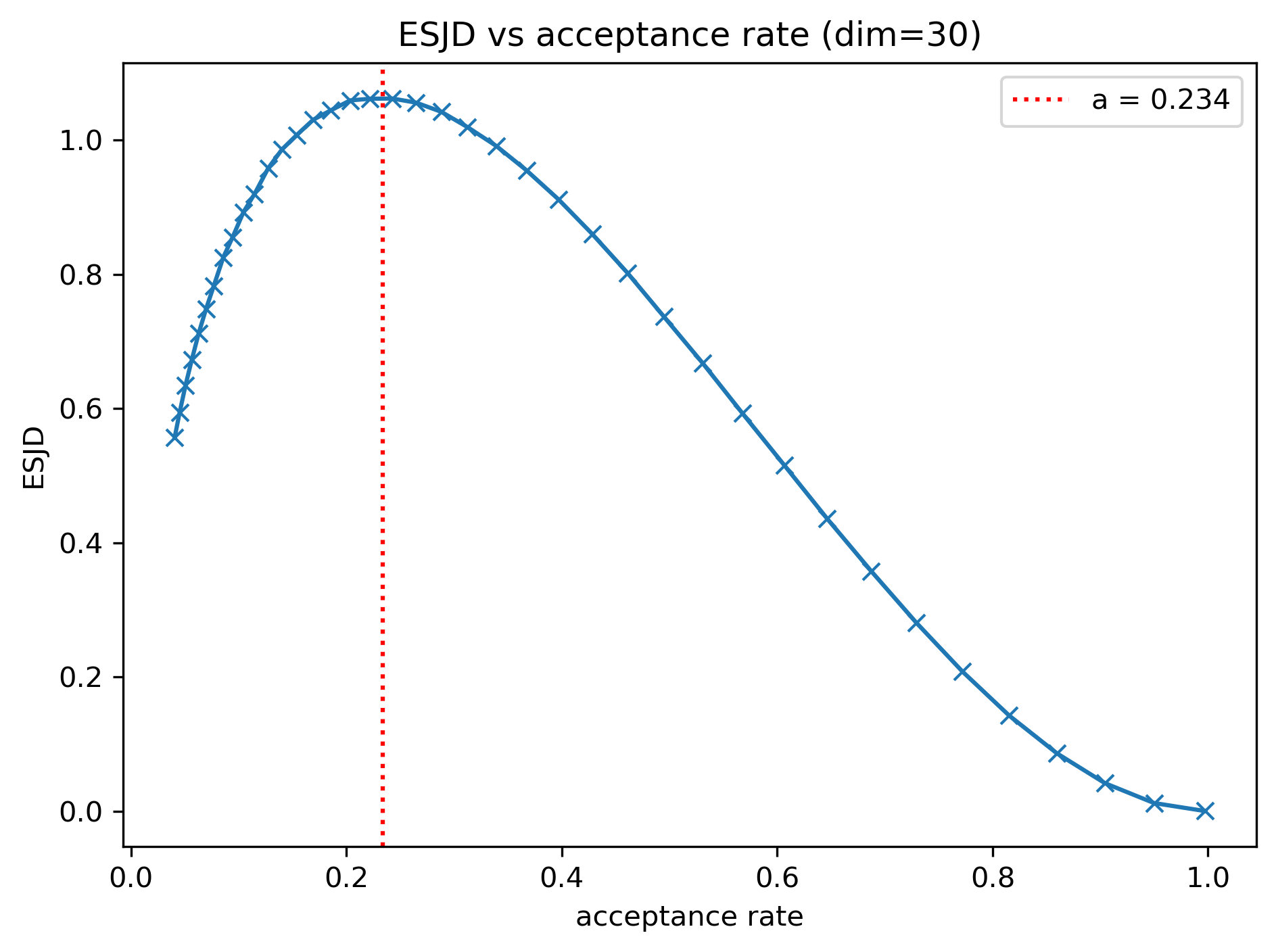}
    \end{subfigure}
    \begin{subfigure}{0.32\textwidth}
        \includegraphics[width=\linewidth, valign=t]{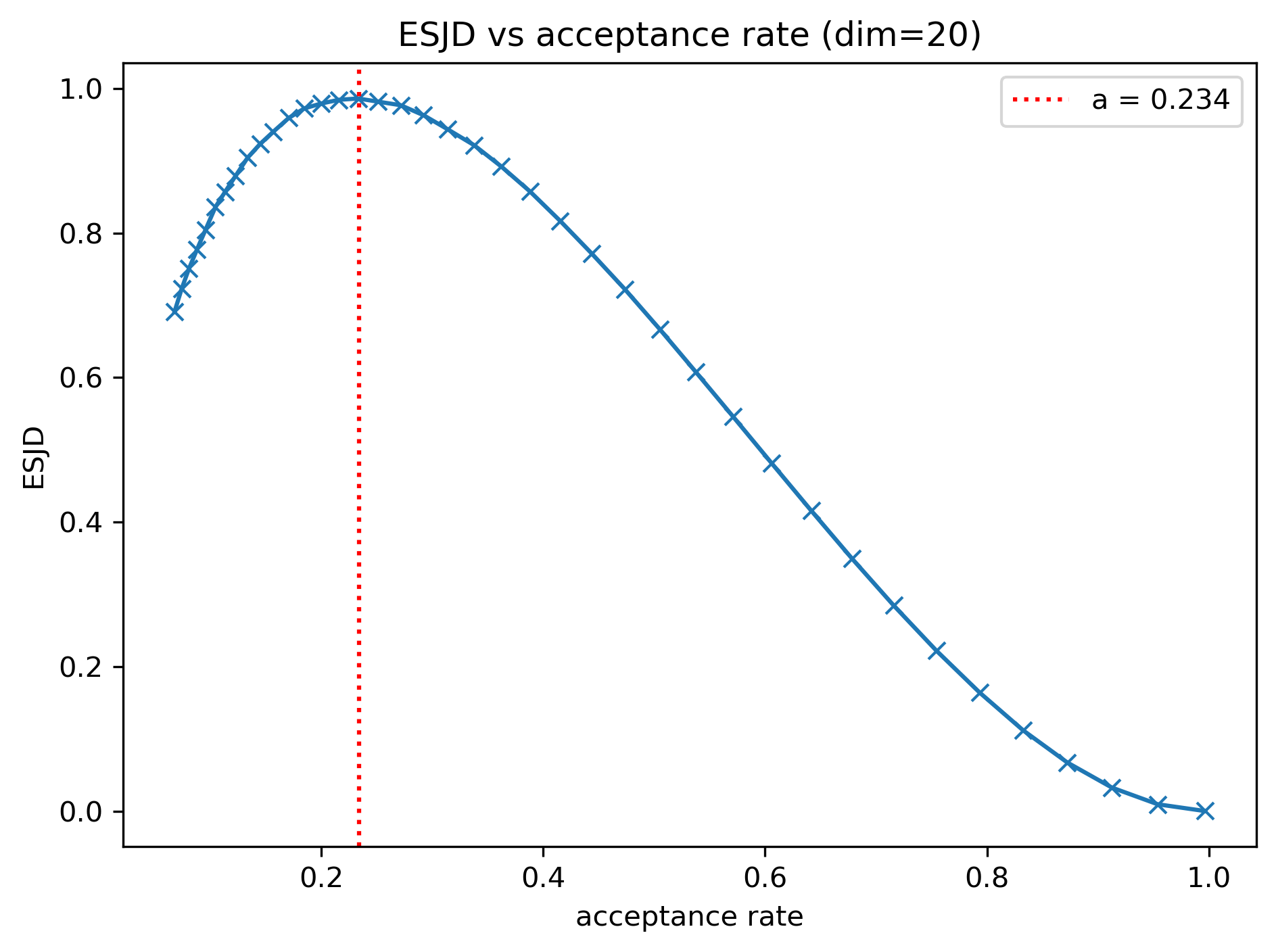}
    \end{subfigure}
    \caption{ESJD vs.\ acceptance rate for the inhomogeneously scaled three mixture target distribution $\pi_7$ (Eq \ref{eq:3mode}) under RWM with a Gaussian proposal in dimensions $d \in \{20,30,50\}$ from left to right. Red dotted line indicates an acceptance rate of 0.234.}
    \label{fig:esjd_3mix_scaled}
\end{figure}


\begin{figure}[htbp]
\centering

\begin{subfigure}[t]{0.4\textwidth}
    \includegraphics[width=\linewidth]{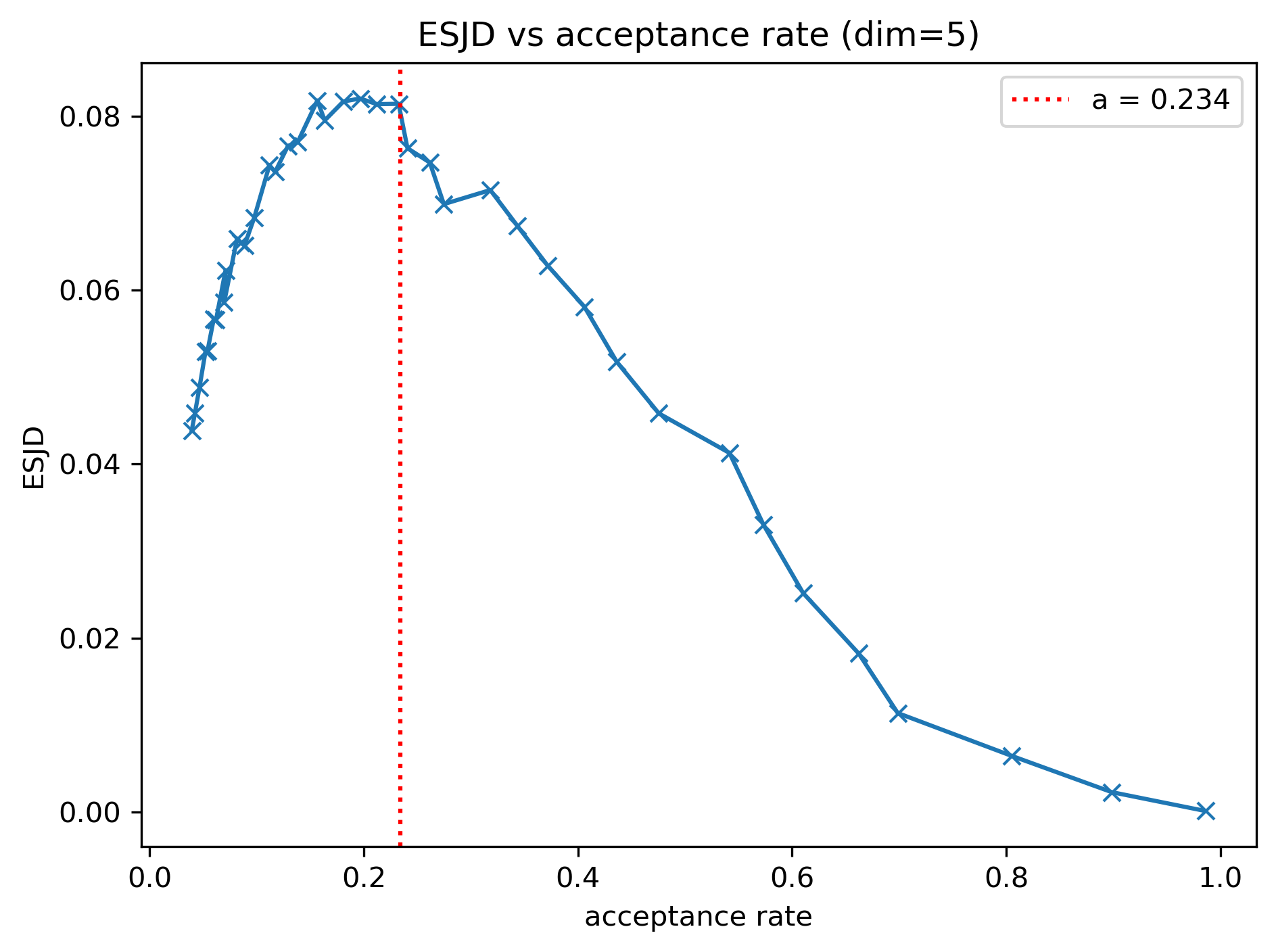}
\end{subfigure}
\begin{subfigure}[t]{0.4\textwidth}
    \includegraphics[width=\linewidth]{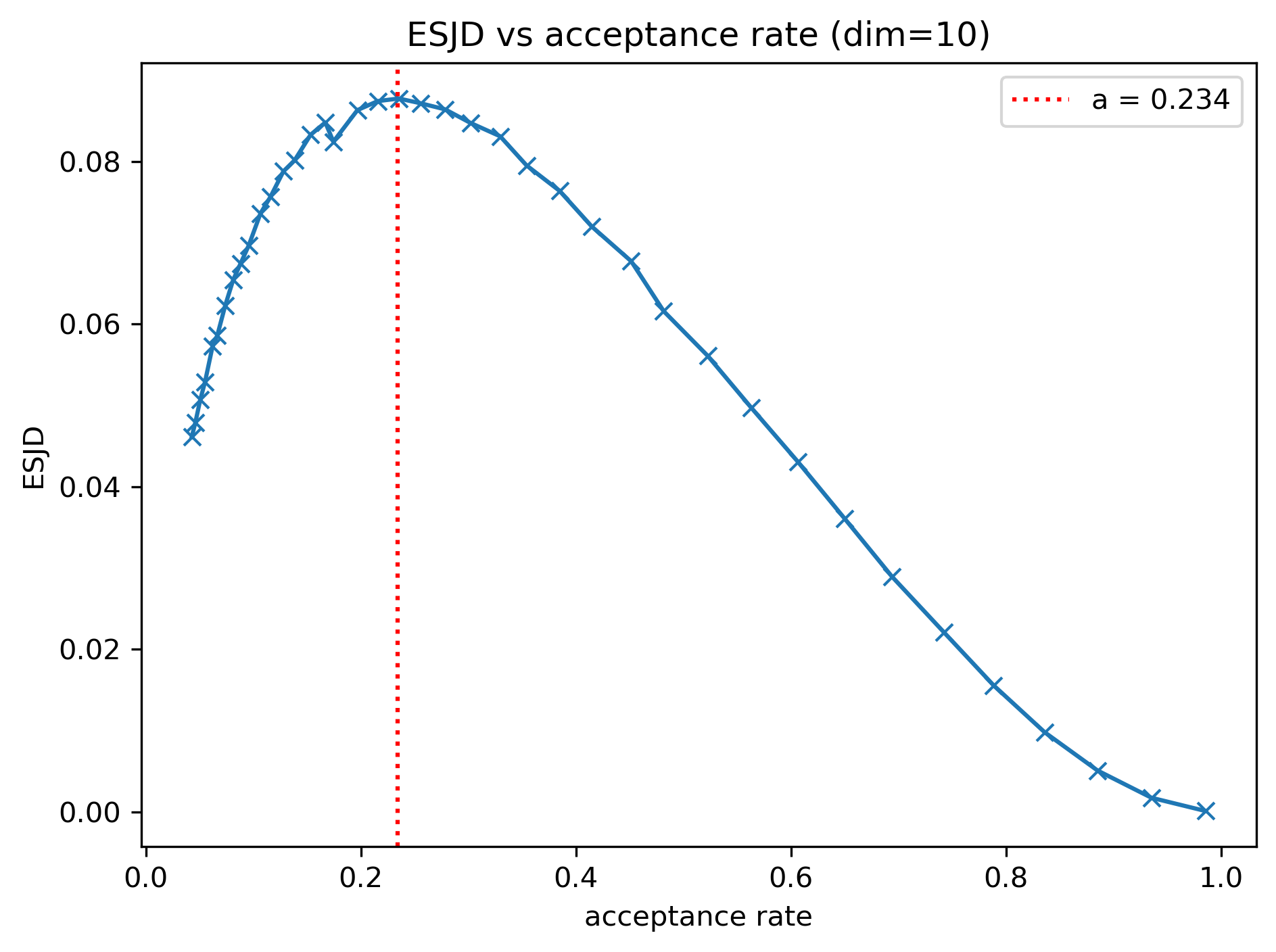}
\end{subfigure}

\vspace{0.5em}

\begin{subfigure}[t]{0.4\textwidth}
    \includegraphics[width=\linewidth]{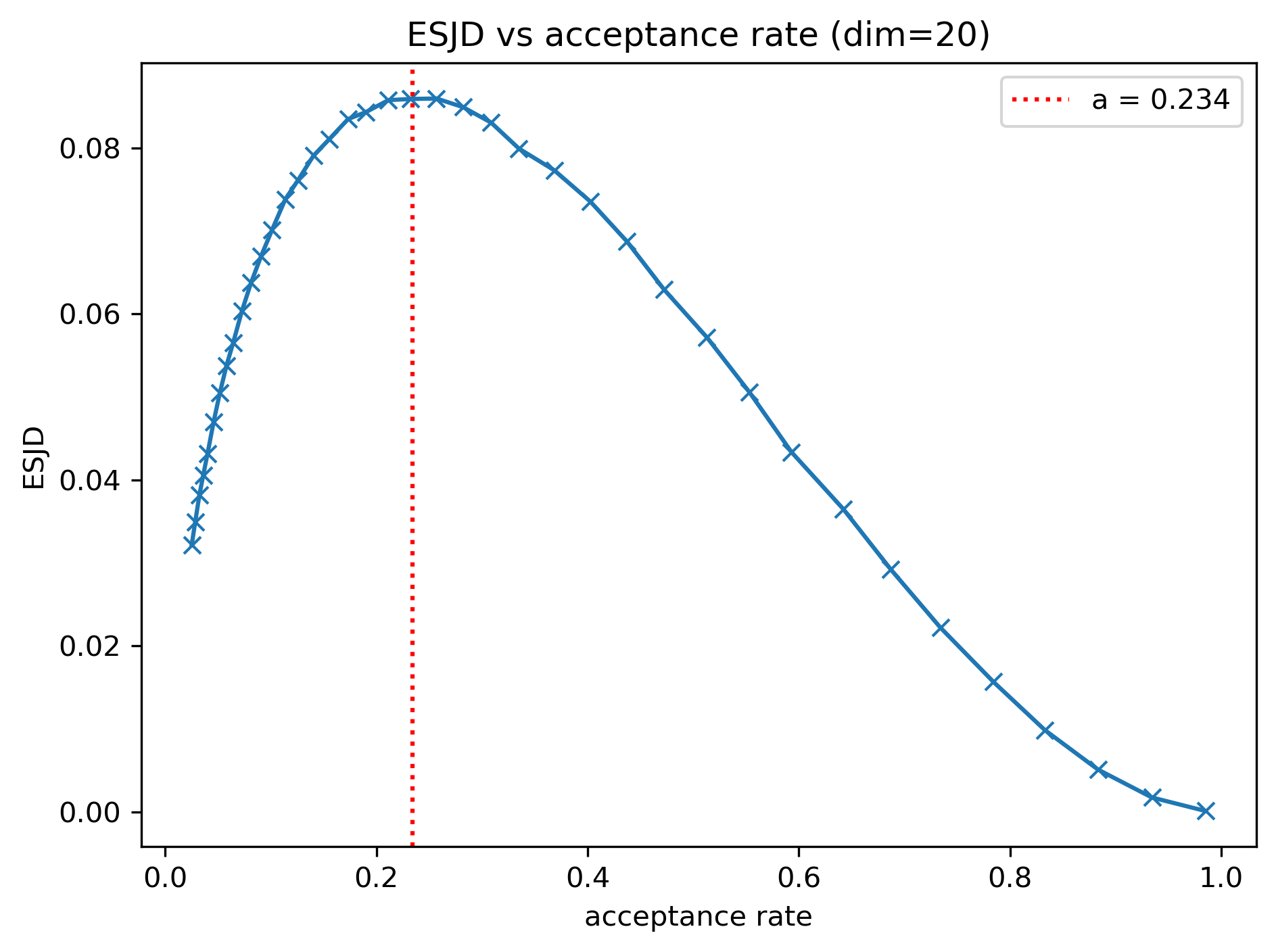}
\end{subfigure}
\begin{subfigure}[t]{0.4\textwidth}
    \includegraphics[width=\linewidth]{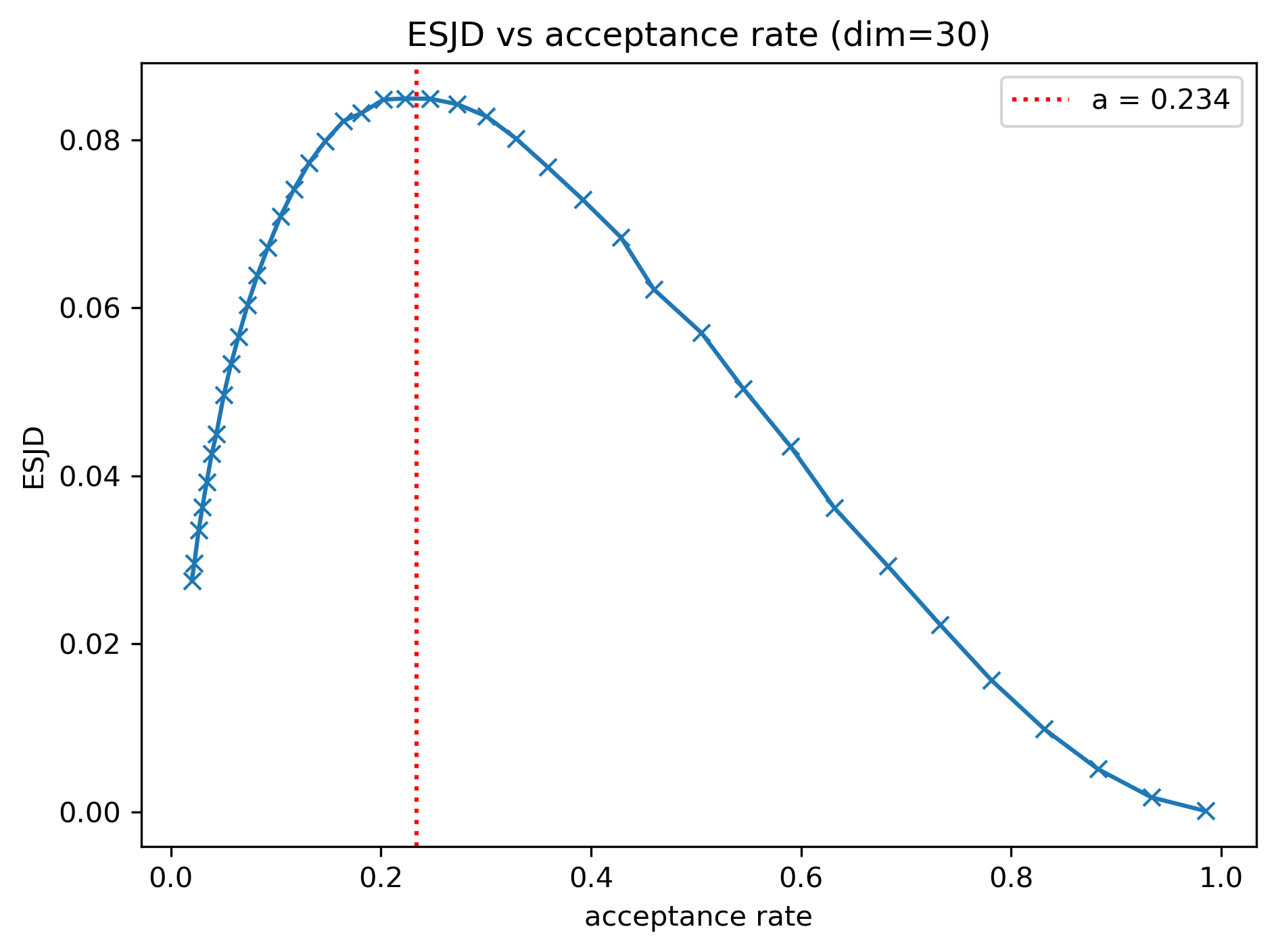}
\end{subfigure}

\caption{ESJD vs.\ acceptance rate for the Full Rosenbrock target $\pi_{8}$ (Eq \ref{eq:full_rosenbrock}) under RWM with a Gaussian proposal in dimensions $d \in \{5, 10, 20, 30\}$ from top-left to bottom-right. Red dotted line indicates an acceptance rate of 0.234.}
\label{fig:esjd_rosenbrock_full}
\end{figure}


\begin{figure}[htbp]
\centering

\begin{subfigure}[t]{0.4\textwidth}
    \includegraphics[width=\linewidth]{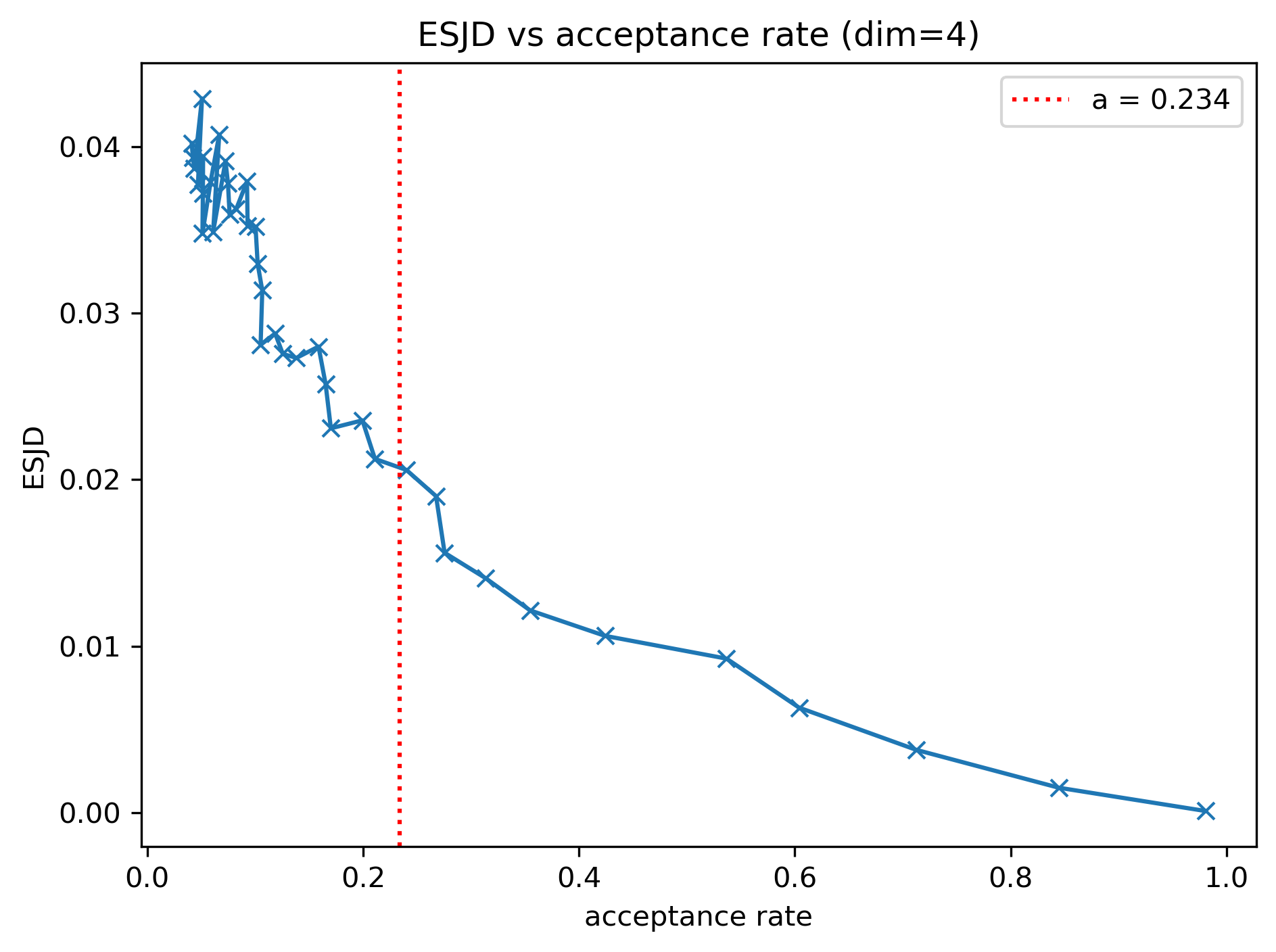}
\end{subfigure}
\begin{subfigure}[t]{0.4\textwidth}
    \includegraphics[width=\linewidth]{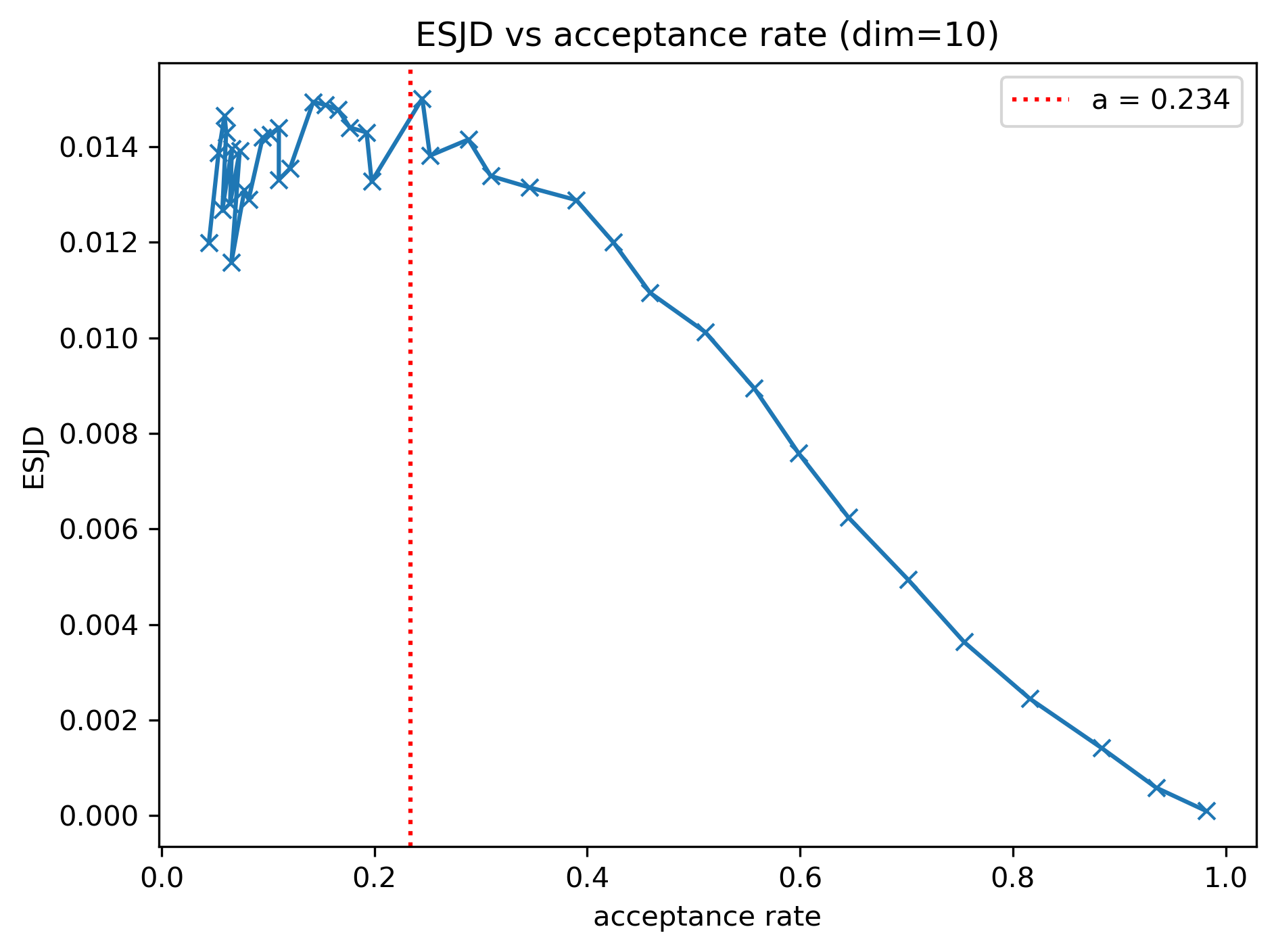}
\end{subfigure}

\vspace{0.5em}

\begin{subfigure}[t]{0.4\textwidth}
    \includegraphics[width=\linewidth]{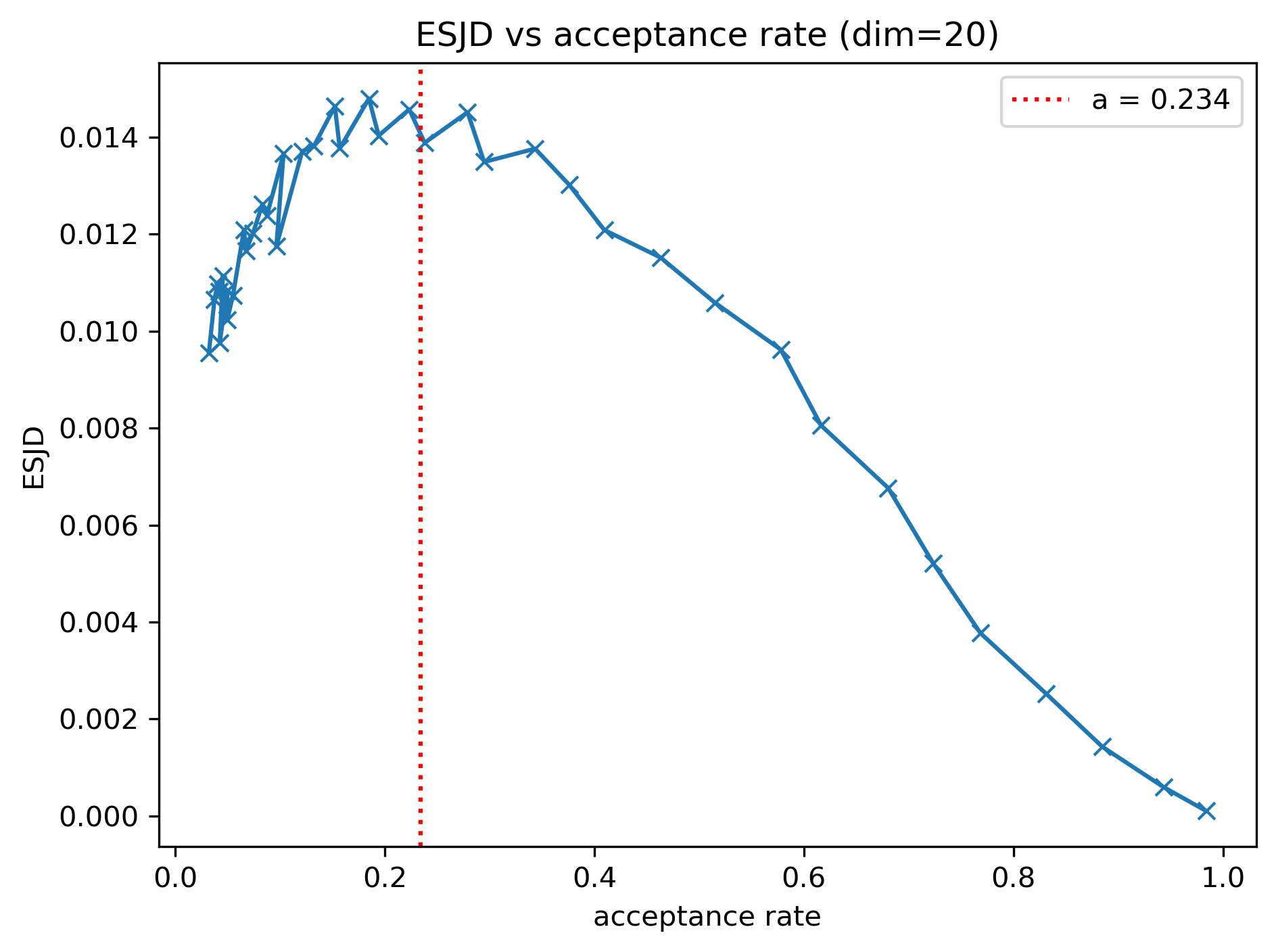}
\end{subfigure}
\begin{subfigure}[t]{0.4\textwidth}
    \includegraphics[width=\linewidth]{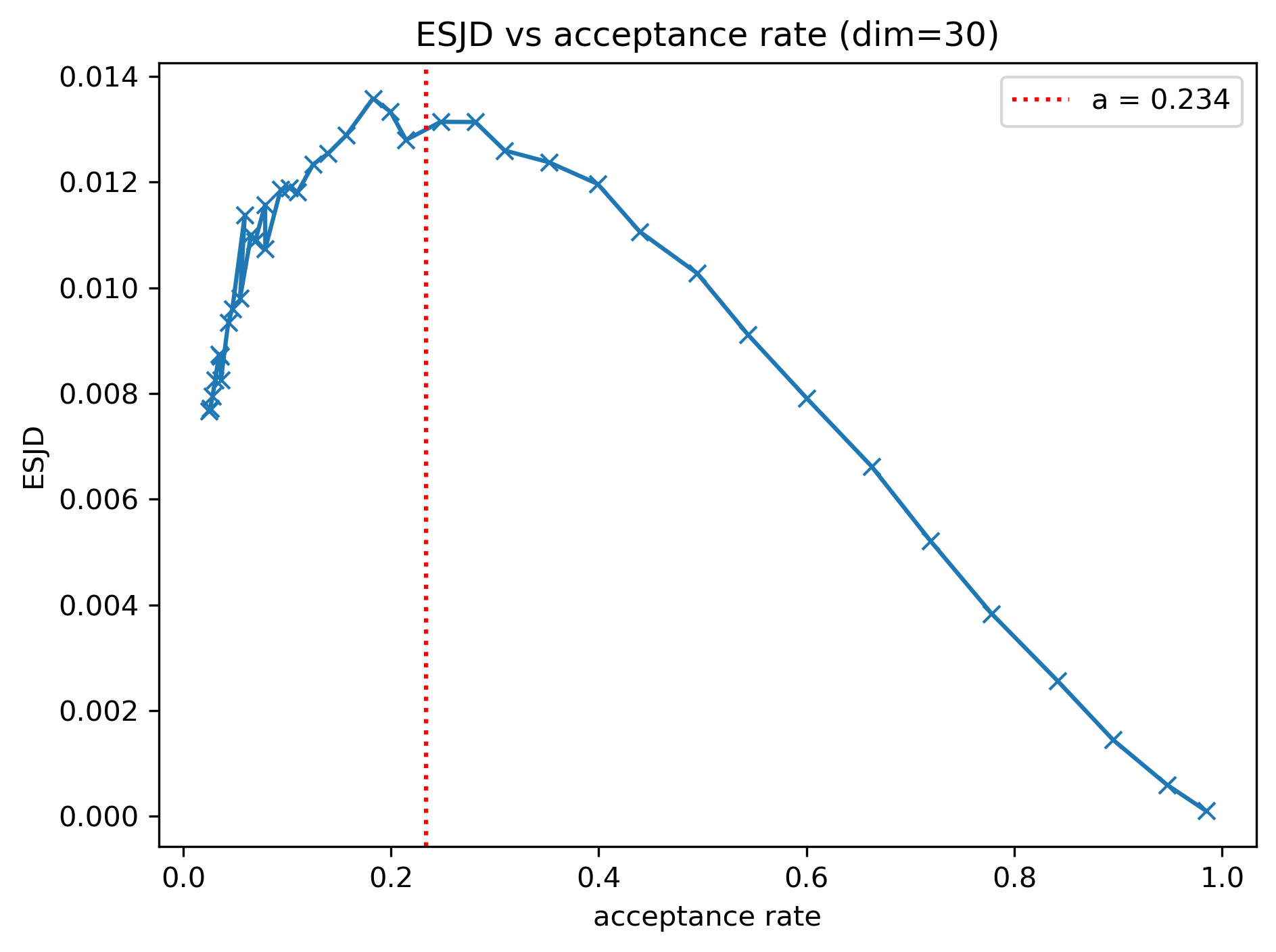}
\end{subfigure}

\caption{ESJD vs.\ acceptance rate for the Even Rosenbrock target $\pi_{9}$ (Eq \ref{eq:even_rosenbrock}) under RWM with a Gaussian proposal in dimensions $d \in \{5, 10, 20, 30\}$ from top-left to bottom-right. Red dotted line indicates an acceptance rate of 0.234.}
\label{fig:esjd_rosenbrock_even}
\end{figure}


\begin{figure}[htbp]
\centering
\begin{subfigure}[t]{0.4\textwidth}
    \includegraphics[width=\linewidth]{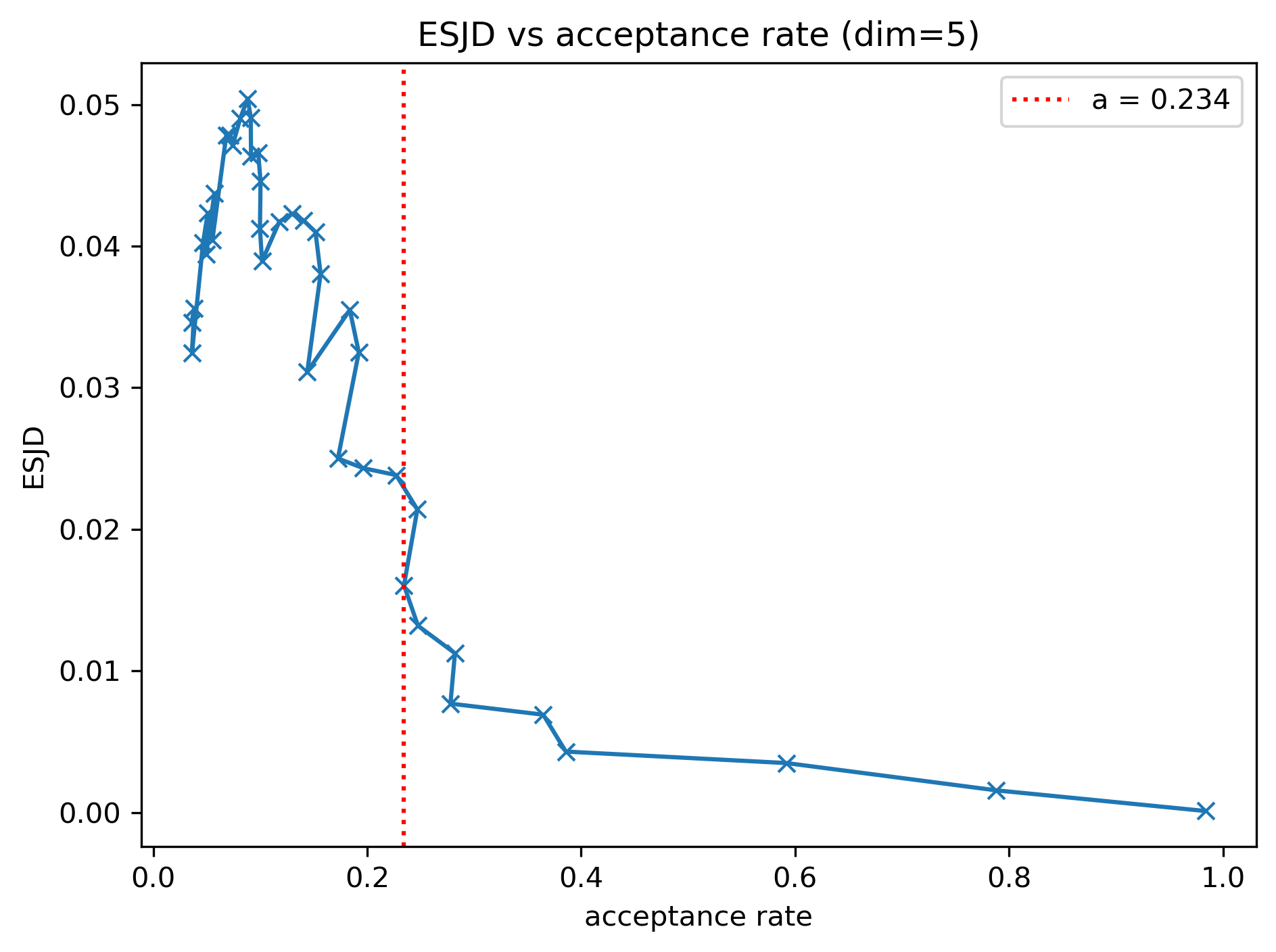}
\end{subfigure}
\begin{subfigure}[t]{0.4\textwidth}
    \includegraphics[width=\linewidth]{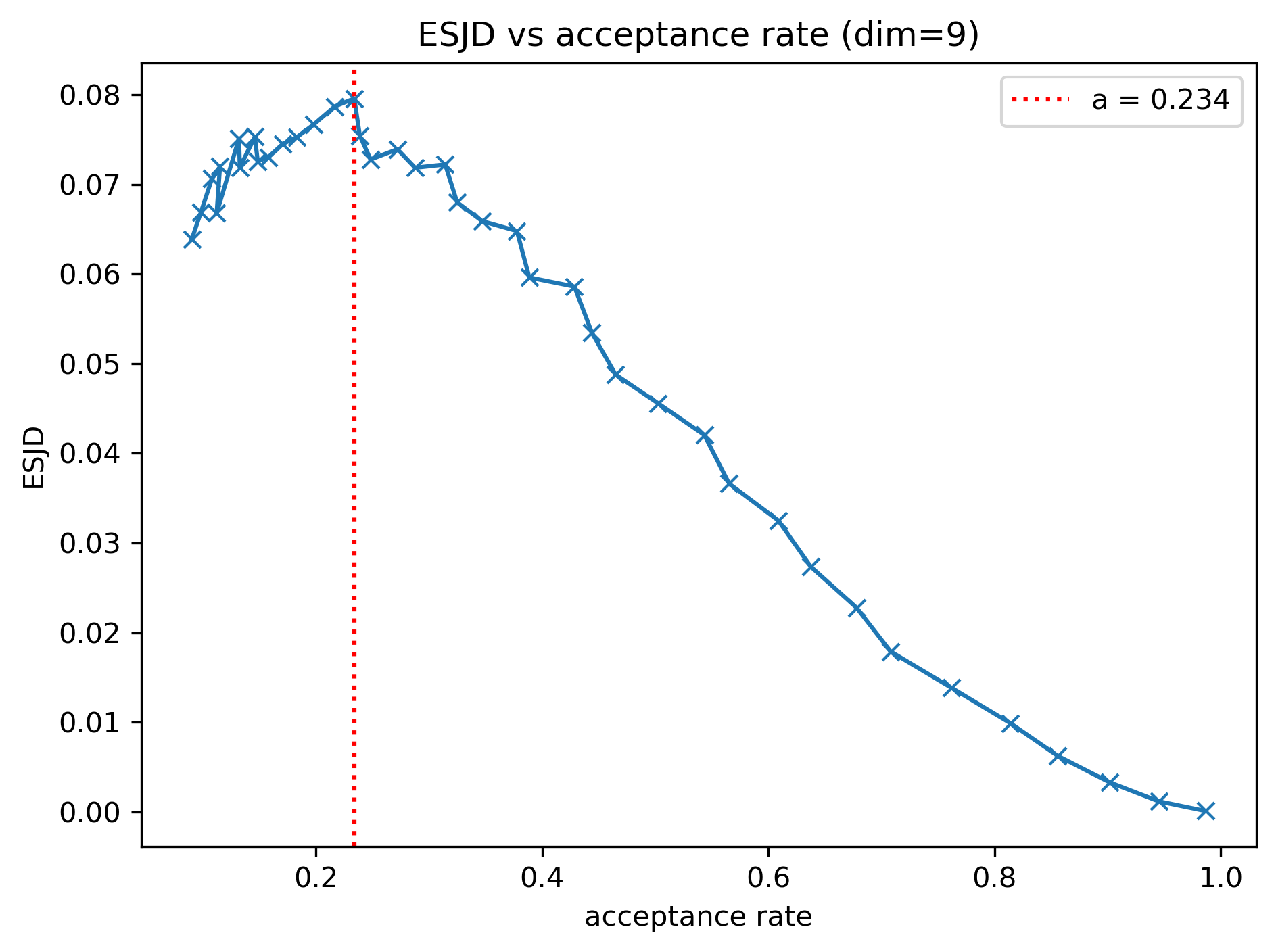}
\end{subfigure}

\vspace{0.5em}

\begin{subfigure}[t]{0.4\textwidth}
    \includegraphics[width=\linewidth]{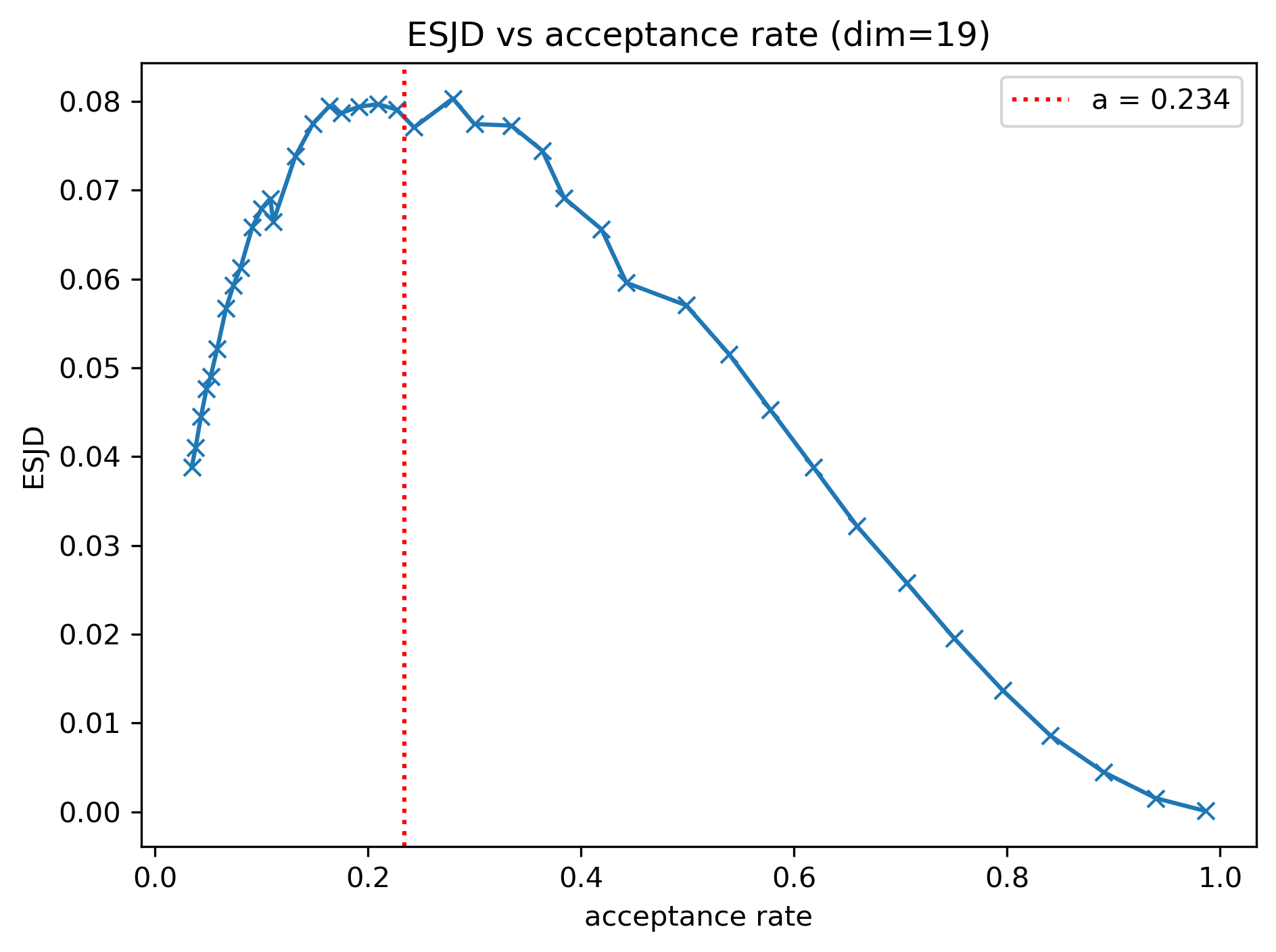}
\end{subfigure}
\begin{subfigure}[t]{0.4\textwidth}
    \includegraphics[width=\linewidth]{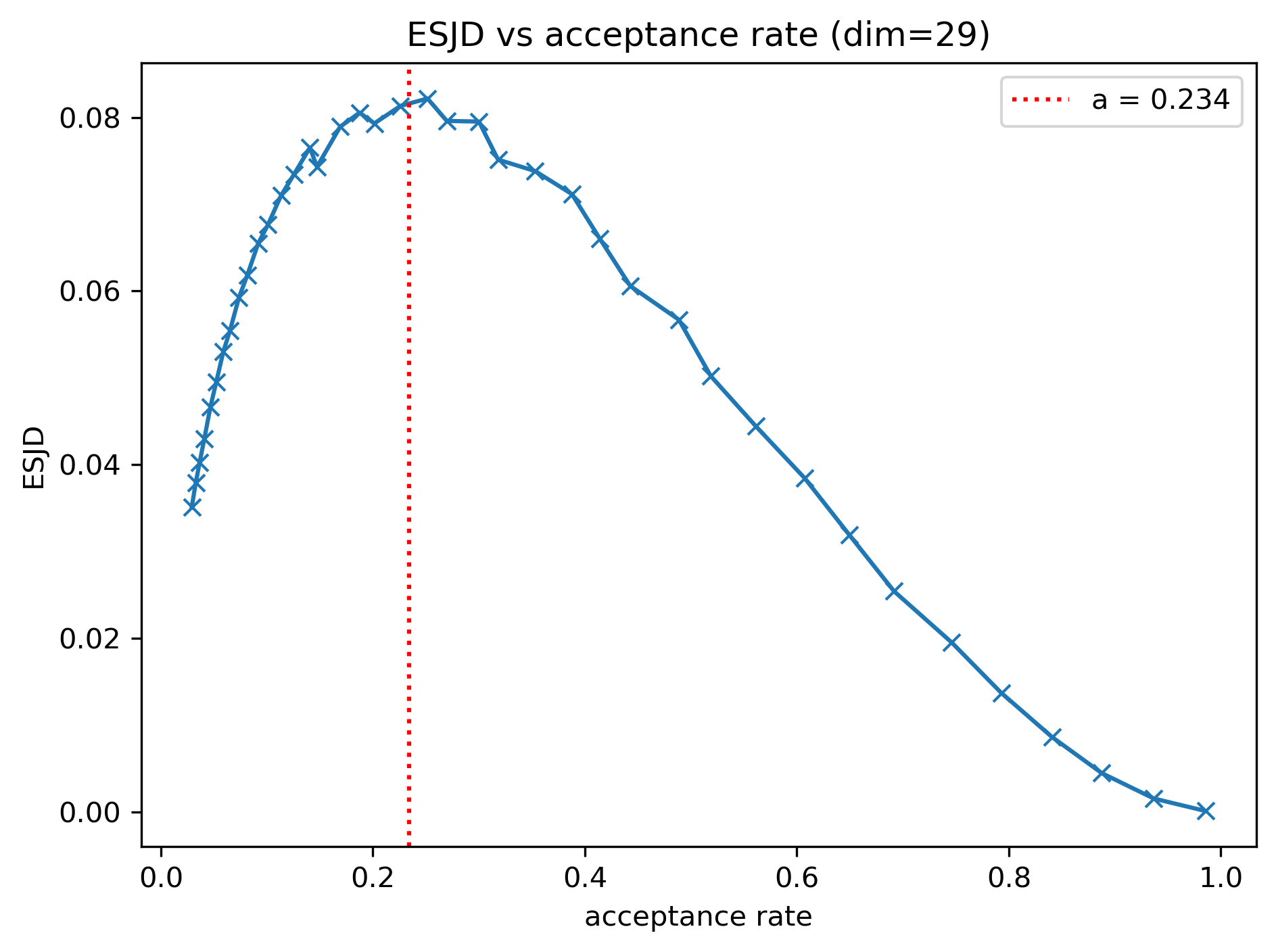}
\end{subfigure}

\caption{ESJD vs.\ acceptance rate for the Hybrid Rosenbrock target $\pi_{10}$ (Eq \ref{eq:hybrid_rosenbrock}) under RWM with a Gaussian proposal in dimensions $d \in \{5, 9, 19, 29\}$ from top-left to bottom-right. Red dotted line indicates an acceptance rate of 0.234.}
\label{fig:esjd_rosenbrock_hybrid}
\end{figure}


\begin{figure}[h]
    \centering
    \includegraphics[width=0.4\textwidth]{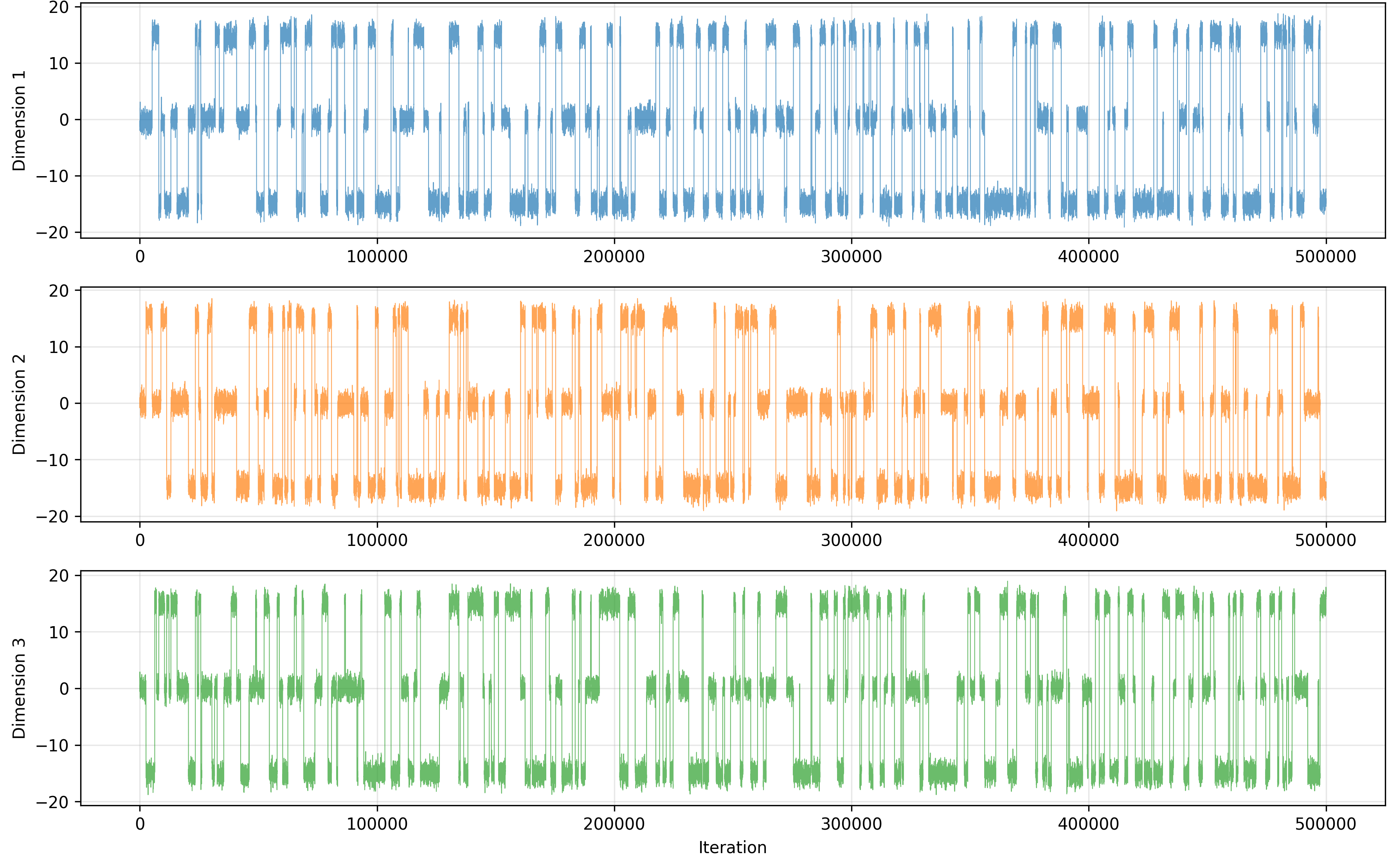}
    \caption{Traceplot of the first three dimensions of the cold chain from one of the parallel tempering simulations on the rough carpet target distribution $\pi_4$ (Eq \ref{eq:target-rough-carpet}).}
    \label{fig:pt-trace-rough}
\end{figure}

\begin{figure}[h]
    \centering
    \includegraphics[width=0.4\textwidth]{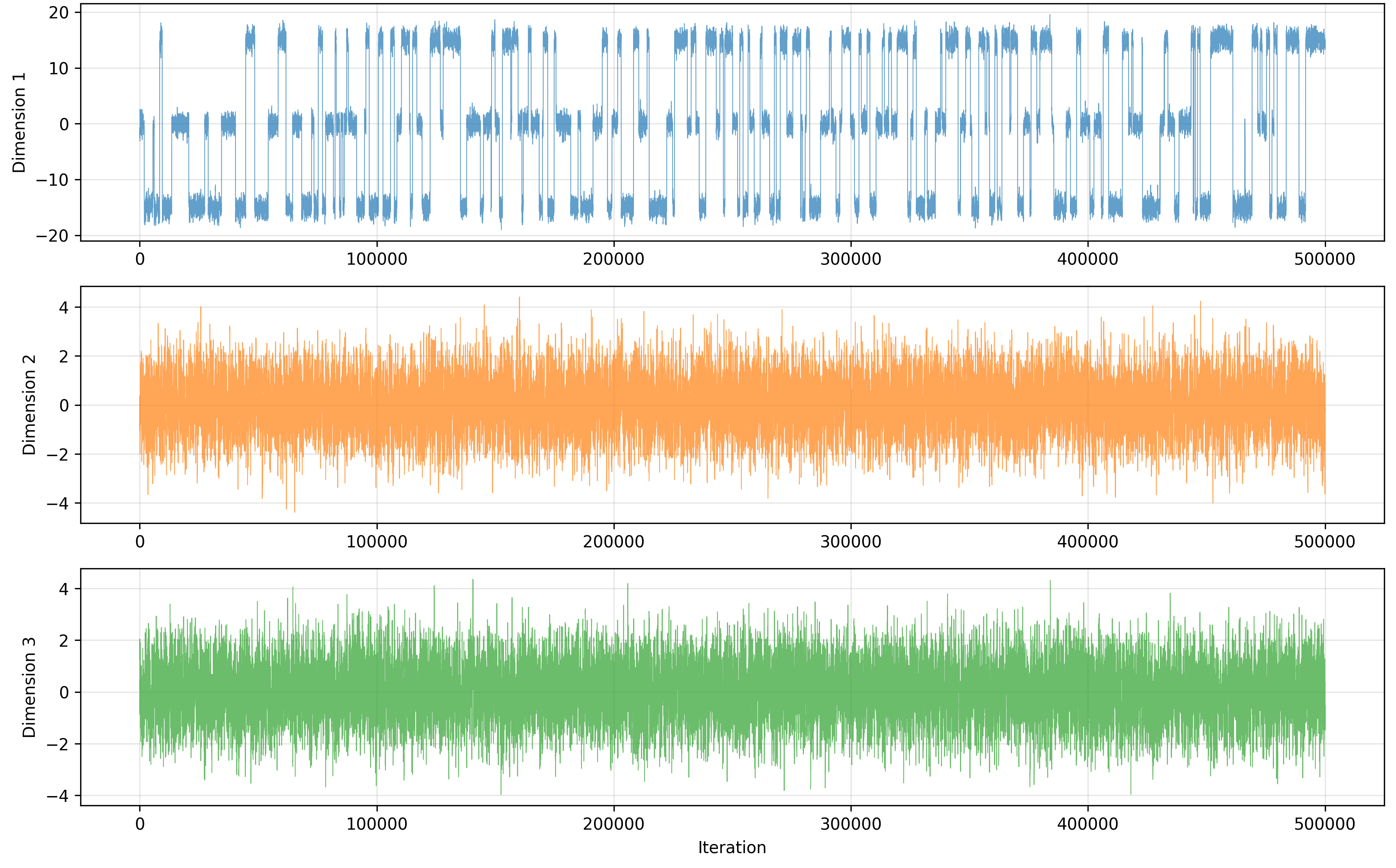}
    \caption{Traceplot of the first three dimensions of the cold chain from one of the parallel tempering simulations on the three-mixture target distribution $\pi_6$ (Eq \ref{eq:3mode}).}
    \label{fig:pt-trace-3mix}
\end{figure}

\end{document}